\documentclass[useAMS,usenatbib]{mn2e}
\usepackage{txfonts}
\usepackage{graphicx}
\usepackage{natbib}
\usepackage{amssymb}

\def\del#1{{}}

\defcitealias{2007MNRAS...378..385P}{Paper I}
\defcitealias{2007PfrommerIII}{Paper III}

\del{
\bibitem[\protect\citeauthoryear{{Miniati}, {Ryu}, {Kang} \& {Jones}}{{Miniati}
  et~al.}{2001a}]{2001ApJ...559...59M}
{Miniati} F.,  {Ryu} D.,  {Kang} H.,    {Jones} T.~W.,  2001a, \apj, 559, 59
\bibitem[\protect\citeauthoryear{{Miniati}, {Jones}, {Kang} \& {Ryu}}{{Miniati}
  et~al.}{2001b}]{2001ApJ...562..233M}
{Miniati} F.,  {Jones} T.~W.,  {Kang} H.,    {Ryu} D.,  2001b, \apj, 562, 233
}

\newcommand{\apj}{ApJ}

\newcommand{\dd}{\mathrm{d}}
\newcommand{\bra}{\langle}
\newcommand{\ket}{\rangle}
\newcommand{\ltsima}{$\; \buildrel < \over \sim \;$}
\newcommand{\lsim}{\lower.5ex\hbox{\ltsima}}
\newcommand{\gtsima}{$\; \buildrel > \over \sim \;$}
\newcommand{\gsim}{\lower.5ex\hbox{\gtsima}}

\newcommand{\e}{{\rm e}}
\newcommand{\p}{{\rm p}}
\newcommand{\inj}{{\rm inj}}
\newcommand{\lin}{{\rm lin}}
\newcommand{\me}{m_\e}
\renewcommand{\mp}{m_\p}
\newcommand{\CR}{{\rm CR}}
\newcommand{\CRe}{\rmn{CRe}}
\renewcommand{\th}{{\rm th}}

\newcommand{\IC}{\rmn{IC}}
\newcommand{\T}{\rmn{T}}

\newcommand{\B}{{\mathcal B}}
\newcommand{\M}{{\mathcal M}}
\newcommand{\cre}{\mathrm{CRe}}

\newcommand{\vir}{\mathrm{vir}}
\newcommand{\eps}{\varepsilon}
\newcommand{\vecbf}{\mathbfit}
\newcommand{\vel}{\upsilon}

\title[Cosmic rays in clusters of galaxies -- II. Radio halos, relics, and
$\gamma$-ray emission] {Simulating cosmic rays in clusters of galaxies -- II. A
unified scheme for radio halos and relics with predictions of the $\gamma$-ray
emission}

\author[C.~Pfrommer, T.~A.~En{\ss}lin, V.~Springel]
  {Christoph~Pfrommer,$^1$\thanks{e-mail: pfrommer@cita.utoronto.ca (CP);
  ensslin@mpa-garching.mpg.de (TAE); volker@mpa-garching.mpg.de (VS)}
  Torsten~A.~En{\ss}lin,$^2$\footnotemark[1]
  Volker~Springel$^2$\footnotemark[1] \\ 
  $^1$Canadian Institute for Theoretical Astrophysics, University of Toronto,
  60 St. George Street, Toronto, Ontario, M5S 3H8, Canada \\
  $^2$Max-Planck-Institut f\"ur Astrophysik, Karl-Schwarzschild-Stra{\ss}e 1,
  Postfach 1317, 85741 Garching, Germany }

\begin{document}
\pagerange{\pageref{firstpage}--\pageref{lastpage}} \pubyear{2003}
\maketitle
\label{firstpage}

\begin{abstract}
  The thermal plasma of galaxy clusters lost most of its information on how
  structure formation proceeded as a result of dissipative processes. In
  contrast, non-equilibrium distributions of cosmic rays (CR) preserve the
  information about their injection and transport processes and provide thus a
  unique window of current and past structure formation processes.  This
  information can be unveiled by observations of non-thermal radiative
  processes, including radio synchrotron, hard X-ray, and $\gamma$-ray
  emission.  To explore this, we use high-resolution simulations of a sample of
  galaxy clusters spanning a mass range of about two orders of magnitudes, and
  follow self-consistent CR physics on top of the radiative hydrodynamics.  We
  model relativistic electrons that are accelerated at cosmological structure
  formation shocks and those that are produced in hadronic interactions of CRs
  with ambient gas protons.  We find that the CR proton pressure traces the
  time integrated non-equilibrium activities of clusters and is modulated by
  the recent dynamical activities.  In contrast, the pressure of primary
  shock-accelerated CR electrons resembles current accretion and merging shock
  waves that break at the shallow cluster potential in the virial regions. The
  resulting synchrotron emission is predicted to be polarised and has an
  inhomogeneous and aspherical spatial distribution which matches the
  properties of observed radio relics. We propose a unified scheme for the
  generation of giant radio halos as well as radio mini-halos that naturally
  arises from our simulated synchrotron surface brightness maps and emission
  profiles. Giant radio halos are dominated in the centre by secondary
  synchrotron emission with a transition to the radio synchrotron radiation
  emitted from primary, shock-accelerated electrons in the cluster
  periphery. This model is able to explain the regular structure of radio halos
  by the dominant contribution of hadronically produced electrons. At the same
  time, it is able to account for the observed correlation of mergers with
  radio halos, the larger peripheral variation of the spectral index, and the
  large scatter in the scaling relation between cluster mass and synchrotron
  emission.  Future low-frequency radio telescopes (LOFAR, GMRT, MWA, LWA) are
  expected to probe the accretion shock regions of clusters and the warm-hot
  intergalactic medium, depending on the adopted model for the magnetic fields.
  The hadronic origin of radio halos can be scrutinised by the detection of
  pion-decay induced $\gamma$-rays following hadronic CR interactions.  The
  high-energy $\gamma$-ray emission depends only weakly on whether radiative or
  non-radiative gas physics is simulated due to the self-regulated nature of
  the CR cooling processes. Our models predict a $\gamma$-ray emission level
  that should be observable with the GLAST satellite.
\end{abstract}

\begin{keywords}
  cosmology: large-scale structure of Universe, galaxies: cluster: general,
  magnetic fields, cosmic rays, radiation mechanisms: non-thermal
\end{keywords}

\section{Introduction}

\subsection{Motivation}

A substantial number of luminous X-ray clusters show diffuse large-scale radio
emission. Generally these radio phenomena can be divided into two categories
that differ morphologically, in their degree of polarisation, as well as in
their characteristic emission regions with respect to the cluster halo.  The
large-scale ``radio relic'' or ``radio gischt'' emission
\citep{2004rcfg.proc..335K}, that has a high degree of polarisation, is
irregularly shaped and occurs at peripheral cluster regions, can be attributed
to merging or accretion shock waves as proposed by \citet{1998A&A...332..395E}.
Prominent examples for large scale ``radio relic'' emission have been observed
in Abell~3667 \citep{1997MNRAS.290..577R}, Abell~3376
\citep{2006Sci...314..791B}, and Abell~2256 \citep{1976A&A....52..107B,
  1978MNRAS.185..607M, 1979A&A....80..201B, 1994ApJ...436..654R,
  2006AJ....131.2900C}.  In contrast, the origin of ``cluster radio halos''
that resemble the regular morphology of the X-ray emitting intra-cluster
plasma is not understood to date. Prominent examples for ``radio halo''
emission can be obtained from \citet{1999NewA....4..141G} and include the Coma
cluster \citep{1989Natur.341..720K, 1997A&A...321...55D} and the galaxy
cluster 1E 0657-56 \citep{2000ApJ...544..686L}.  In principle, observations of
non-thermal cluster phenomena could provide an independent and complementary
way of studying the growth of structure in our Universe and could shed light
on the existence and the properties of the warm-hot intergalactic medium
(WHIM), provided the underlying processes are understood.  Sheets and
filaments are predicted to host this WHIM with temperatures in the range
$10^5\,\mbox{K}<T<10^7\,\mbox{K}$ whose evolution is primarily driven by shock
heating from gravitational perturbations breaking on mildly nonlinear,
non-equilibrium structures \citep{1998ApJ...509...56H, 1999ApJ...514....1C,
  2001ApJ...552..473D, 2004ApJ...611..642F, 2005ApJ...620...21K}.  Once a
cluster has virialised, the thermal plasma lost most information on how the
formation proceeded due to the dissipative processes driving the plasma
towards a Maxwell-Boltzmann momentum distribution that is characterised by its
temperature only. In contrast, non-equilibrium distributions of cosmic rays
preserve the information about their injection and transport processes much
better, and thus provide a unique window of current and past structure
formation processes.

The information about these non-equilibrium processes is encoded in the
spectral and spatial distribution of cosmic ray electrons and protons.
Radiative loss processes of these non-thermal particle distributions produce
characteristic radio synchrotron, hard X-ray inverse Compton, and hadronically
induced $\gamma$-ray emission. Suitably combining various non-thermal emission
processes will allow us to infer the underlying non-equilibrium processes of
clusters as well as to improve our knowledge about fundamental plasma physics.
The upcoming generation of radio, hard X-ray, and $\gamma$-ray instruments
opens up the extragalactic sky in unexplored wavelength ranges: low-frequency
radio arrays ({\em GMRT}\footnote{{\bf G}iant {\bf M}eterwave {\bf R}adio {\bf
    T}elescope}, {\em LOFAR}\footnote{{\bf LO}w {\bf F}requency {\bf AR}ray},
{\em MWA}\footnote{{\bf M}ileura {\bf W}idefield {\bf A}rray}, {\em
  LWA}\footnote{{\bf L}ong {\bf W}avelength {\bf A}rray}), the future hard
X-ray satellite missions {\em NuSTAR}\footnote{{\bf Nu}clear {\bf
    S}pectroscopic {\bf T}elescope {\bf Ar}ray} and {\em Simbol-X}, and
$\gamma$-ray instruments (the {\em GLAST}\footnote{{\bf G}amma-ray {\bf L}arge
  {\bf A}rea {\bf S}pace {\bf T}elescope} satellite and imaging atmospheric
\v{C}e\-ren\-kov telescopes {\em H.E.S.S.}\footnote{{\bf H}igh {\bf E}nergy
  {\bf S}tereoscopic {\bf S}ystem}, {\em MAGIC}\footnote{{\bf M}ajor {\bf
    A}tmospheric {\bf G}amma {\bf I}maging {\bf C}erenkov {\bf T}elescope},
  {\em VERITAS}\footnote{{\bf V}ery {\bf E}nergetic {\bf R}adiation {\bf
      I}maging {\bf T}elescope {\bf A}rray {\bf S}ystem}, and {\em
    CANGAROO}\footnote{{\bf C}ollaboration of {\bf A}ustralia and {\bf N}ippon
    for a {\bf GA}mma {\bf R}ay {\bf O}bservatory in the {\bf O}utback}) will
  allow us to probe non-thermal cluster physics with a multi-faceted approach.
  There have been pioneering efforts to simulate the non-thermal emission from
  clusters by numerically modelling discretised cosmic ray (CR) energy
  spectra on top of Eulerian grid-based cosmological simulations
  \citep{2001CoPhC.141...17M, 2001ApJ...559...59M, 2001ApJ...562..233M,
    2002MNRAS.337..199M, 2003MNRAS.342.1009M}.  However, these approaches
  neglected the hydrodynamic pressure of the CR component, were unable to
  resolve the observationally accessible, dense central regions of clusters,
  and they neglected dissipative gas physics including radiative cooling, star
  formation, and supernova feedback.  To allow studies of the dynamical effects
  of CR protons in radiatively cooling galactic and cluster environments, we
  have developed a CR proton formalism that is based on smoothed particle
  hydrodynamical representation of the equations of motion. The emphasis is
  given to the dynamical impact of CR protons on hydrodynamics, while
  simultaneously allowing for the important CR proton injection and loss
  processes in a cosmological setting \citep{2007A&A...473...41E,
    2006...Jubelgas, 2006MNRAS.367..113P}.  This enables us to account for the
  pressure support provided by CR protons to the plasma of clusters of
  galaxies.  A substantial CR proton pressure contribution might have a major
  impact on the properties of the intra-cluster medium (ICM) and could modify
  thermal cluster observables such as the X-ray emission and the
  Sunyaev-Zel'dovich (SZ) effect \citep{2007MNRAS...378..385P}.

\begin{figure}
\resizebox{\hsize}{!}{\includegraphics{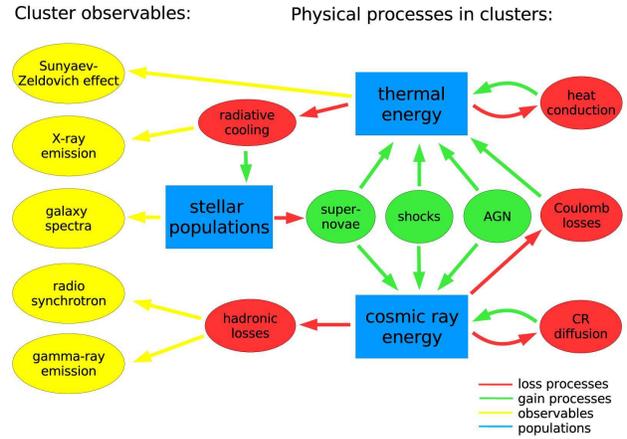}}
\caption{Overview over the relevant physical processes in galaxy clusters. The
  right side shows the interplay of different physical processes
  highlighting the interplay of the energy reservoirs of the thermal plasma and
  cosmic ray protons (shown in blue) while the left side shows observables
  that inform about the properties of clusters and their dynamical state. Gain
  processes are denoted in green, while loss or redistribution processes are
  denoted in red.}
\label{fig:CR_flowchart}
\end{figure}

\subsection{Cosmic ray physics}

We give a short and simplified overview over the relevant physical processes in
galaxy clusters in Fig.~\ref{fig:CR_flowchart} before introducing different CR
populations that are relevant for the non-thermal emission from clusters.  The
{\em upper central part} of Fig.~\ref{fig:CR_flowchart} shows standard
processes which are usually considered in simulations.  Radiative cooling of
the gas leads eventually to star formation in the densest regions that exceed a
certain density threshold. This happens in the central cluster regions and
within substructures leading to individual galaxies.  Once the nuclear energy
has been used up, massive stars explode in supernovae that drive strong shock
waves into the ambient interstellar medium (ISM) which resupply thermal and
turbulent energy. On larger scales, structure formation shock waves dissipate
gravitational energy associated with hierarchical clustering into thermal
energy of the gas, thus supplying the ICM with entropy and thermal pressure
support.  There are three main observables associated with these processes: the
hot ICM emits thermal bremsstrahlung radiation with an X-ray luminosity that
depends on the square of the electron number density.  The amplitude of the
Sunyaev-Zel'dovich effect \citep[][]{1972CoASP...4..173S} depends on the
pressure of the thermal electron population integrated along the line-of-sight
through the cluster.  Finally, galaxy spectra probe directly the stellar
populations of intra-cluster galaxies and indirectly the cluster's potential
through their velocity dispersion \citep[for reviews
  see][]{1988xrec.book.....S, 2005RMP...77...207V}.

The {\em lower part} of Fig.~\ref{fig:CR_flowchart} sketches the cosmic ray
physics within clusters. CR protons behave differently compared to the thermal
gas. Their equation of state is softer, they are able to travel actively over
macroscopic distances, and their energy loss time-scales are typically larger
than the thermal ones.  Besides thermalization, collisionless shocks are also
able to accelerate ions of the high-energy tail of the Maxwellian through
diffusive shock acceleration \citep[for reviews see][]{1983RPPh...46..973D,
1987PhR...154....1B, 2001RPPh...64..429M}. These energetic ions are reflected
at magnetic irregularities through magnetic resonances between the gyro-motion
and waves in the magnetised plasma and are able to gain energy in moving back
and forth through the shock front. This acceleration process typically yields a
CR proton population with a power-law distribution of the particle momenta. CRs
are accelerated on galactic scales through supernova shocks while they are
injected by structure formation shock waves on even larger scales up to tens of
Mpc.  So far, we have neglected feedback from active galactic nuclei (AGN) in
our simulations despite its importance \citep[for first numerical simulations
of thermal `radio-mode' feedback within cosmological simulations,
see][]{2006MNRAS.366..397S}.  Gravitational energy associated with the
accretion onto super-massive black holes is converted into large-scale jets and
eventually dissipated into thermal and CR energy.

\begin{figure}
\resizebox{\hsize}{!}{\includegraphics{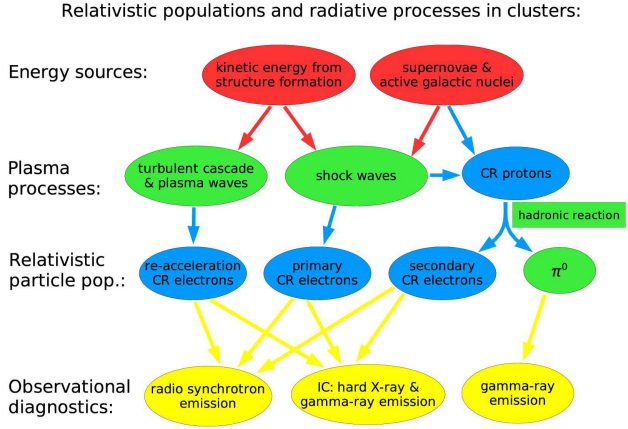}}
\caption{Schematic overview over non-thermal radiative processes in galaxy
  clusters. Various gravitational and non-gravitational energy sources (shown
  in red) are able to accelerate relativistic particle populations (shown in
  blue) by means of different plasma processes (shown in green). Non-thermal
  cluster observables (shown in yellow) are tracers of these cosmic ray
  populations: any cosmic ray electron population can emit radio synchrotron
  radiation as well as inverse Compton emission that extends from the X-ray
  into the $\gamma$-ray regime. In contrast, the characteristic spectral
  signature accompanying $\gamma$-ray emission from hadronic cosmic ray
  interactions is a unique sign of a cosmic ray proton population in the
  intra-cluster plasma.}
\label{fig:Radiative_processes}
\end{figure}

The ubiquitous cosmic magnetic fields couple the otherwise dynamically
independent ingredients like the ICM plasma, and the CR gas into a single,
however complex fluid.  Magnetic fields prevent charged relativistic particles
to travel macroscopic distances with their intrinsic velocity close to the
speed of light. Instead, the particles gyrate around, and travel slowly along
magnetic field lines. Occasionally, they get scattered on magnetic
irregularities.  On macroscopic scales, the transport can often be described as
a diffusion process that redistributes the CR energy density macroscopically
provided the gyro-radius of charged relativistic particles can be regarded to
be small. Thus, the diffusive CR transport in tangled magnetic fields
  effectively confines the CRs with energies $E<2\times 10^7$~GeV to clusters
  and yields to CR proton lifetimes of the order of the Hubble time
\citep{1996SSRv...75..279V, 1997ApJ...477..560E, 1997ApJ...487..529B,
  1998APh.....9..227C}, long enough to diffuse away from the production site
and to maintain a space-filling distribution over the cluster volume.  Thermal
heat conduction is an analogous process that reallocates the thermal energy of
the ICM.

The CR energy reservoir suffers two main loss processes: (1) CR energy is
transferred into the thermal energy reservoir through individual electron
scatterings in the Coulomb field of the CR particle as well as by small
momentum transfers through excitations of quantised plasma oscillations. We
refer to the sum of both effects as Coulomb losses
\citep{1972Phy....58..379G}. (2) Provided the CR momentum exceeds the threshold
$p \simeq 0.8\,\mbox{GeV} / c$ for the hadronic reaction with ambient protons,
they produce pions which decay into secondary electrons, positrons, neutrinos,
and $\gamma$-rays:
\begin{eqnarray}
  \pi^\pm &\rightarrow& \mu^\pm + \nu_{\mu}/\bar{\nu}_{\mu} \rightarrow
  e^\pm + \nu_{e}/\bar{\nu}_{e} + \nu_{\mu} + \bar{\nu}_{\mu}\nonumber\\
  \pi^0 &\rightarrow& 2 \gamma \,.\nonumber
\end{eqnarray}
Only CR protons above this kinematic threshold are therefore visible
through their decay products via radiative processes, making them
directly observationally detectable. As shown in
Fig.~\ref{fig:Radiative_processes}, these {\em secondary relativistic
  electrons and positrons} can emit a halo of radio synchrotron
emission in the presence of ubiquitous intra-cluster magnetic fields
\citep{1980ApJ...239L..93D, 1982AJ.....87.1266V, 1999APh....12..169B,
  2000A&A...362..151D, 2001ApJ...562..233M, 2003A&A...407L..73P,
  2004A&A...413...17P, 2004MNRAS.352...76P, 2007A&A...465...41M} as
well as inverse Compton emission by scattering photons from the cosmic
microwave background into the hard X-ray and
$\gamma$-regime.\footnote{In the following, we use the term secondary
  CR electrons synonymously for the likewise produced electrons and
  positrons.}  Future $\gamma$-ray satellites should be able to detect
the associated hadronically induced $\gamma$-ray emission resulting
from neutral pion decay and allow unambiguous conclusions on the
parent CR population in clusters.

Structure formation shocks can also directly accelerate so-called {\em primary
CR electrons} giving rise to an irregularly shaped radio and inverse Compton
morphology due to the comparatively short synchrotron lifetimes of CR electrons
of $\tau \simeq 10^8$~yr.  To complicate this picture even more, there are
other processes that accelerate relativistic electrons.  Re-acceleration
processes of `mildly' relativistic electrons ($\gamma\simeq 100-300$) that are
being injected over cosmological timescales into the ICM by sources like radio
galaxies, supernova remnants, merger shocks, or galactic winds can provide an
efficient supply of highly-energetic CR electrons.  Owing to their long
lifetimes of a few times $10^9$ years these `mildly' relativistic electrons can
accumulate within the ICM \citep{2002mpgc.book....1S}, until they experience
continuous in-situ acceleration either via interactions with
magneto-hydrodynamic waves, or through turbulent spectra
\citep{1977ApJ...212....1J, 1987A&A...182...21S, 2001MNRAS.320..365B,
2002ApJ...577..658O, 2004MNRAS.350.1174B,2007MNRAS.378..245B}.  This gives rise
to a third population of {\em re-accelerated CR electrons} that also
contributes to the observed radio and inverse Compton emission.  Since the
distribution of magnetic field strengths with cluster radius is also not well
known, radio synchrotron emission alone has limited predictive power.
Unfortunately, the conceptually simpler inverse Compton emission is hard to
observe because of the strong radiation background in the soft and hard X-ray
regime. 

Nevertheless, there seems to be growing evidence for an excess of hard X-ray
emission compared to the expected thermal bremsstrahlung in a number of
clusters that is based on observations with instruments on board five different
X-ray satellites. Prominent examples include the Coma cluster
(\citealt{1999ApJ...511L..21R, 1999ApJ...513L..21F, 2002ApJ...579..587R,
  2004ApJ...602L..73F, 2007ApJ...654L...9F};\footnote{The results of these
  papers have been challenged by an analysis that takes into account all
  systematic uncertainties in the critical parameters including the choice of a
  source-free background field and the modelling of the thermal model for the
  ICM \citep{2004A&A...414L..41R, 2007astro.ph..2417R}.}
\citealt{2007A&A...470..835E}; using the {\em Rossi X-ray Timing Explorer}
({\em RXTE}), {\em BeppoSAX}, and {\em INTEGRAL}) , the ``Bullet'' cluster 1ES
0657--558 \citep[][ using {\em RXTE}]{2006ApJ...652..948P}, Abell 2256
\citep[][ using {\em RXTE}]{2003ApJ...595..137R}, the Ophiuchus cluster
\citep[][ using {\em INTEGRAL}]{2007arXiv0712.2326E}, and the Perseus cluster
\citep[][ using {\em Chandra} and {\em XMM-Newton}]{2005MNRAS.360..133S,
  2007Molendi}.  The currently favoured theoretical explanation of inverse
Compton radiation by CR electrons faces problems since the magnetic field
estimates inferred by combining synchrotron and inverse Compton emission are
much smaller than those derived from Faraday rotation measurements \citep[cf.][
  for an extensive discussion]{2007PfrommerIII}.  It has been proposed that a
fraction of the diffuse cosmological $\gamma$-ray background radiation
originates from the same processes \citep{2000Natur.405..156L,
  2002MNRAS.337..199M, 2003ApJ...588..155R, 2003ApJ...594..709B,
  2005ApJ...618..675K}.

This paper studies directly the CR related multi-frequency observables aiming
at understanding the cluster radio halo emission.  In a companion paper, we
study the interplay of thermal gas and CRs and their effect on the observables
of the thermal gas such as X-ray emission and the Sunyaev-Zel'dovich effect
\citep[][ hereafter Paper I]{2007MNRAS...378..385P}.  For consistency reasons
with that paper, we scale cluster masses and length units with the
dimensionless Hubble constant, $h$, where $H_0 = 100\, h\mbox{ km s}^{-1}
\mbox{ Mpc}^{-1}$. However, non-thermal surface brightness and luminosities (for
radio, hard X-ray, and $\gamma$-ray emission) are reported in units of the
currently favoured Hubble constant, $h_{70}$, where $H_0 = 70\, h_{70}\mbox{ km
s}^{-1} \mbox{ Mpc}^{-1}$ since primary and secondary emission processes have a
different scaling with $h$.  We derive cluster scaling relations for
non-thermal observables and compare our results to observations in our
follow-up paper \citep[][ hereafter Paper III]{2007PfrommerIII}.

The outline of the paper is as follows. Section~\ref{sec:methodology}
describes our methodology including the general setup of the simulations, our
cluster sample, the different physical processes which we simulated, and
highlights important properties of radiative processes considered in
this work.  In Sect.~\ref{sec:results}, we present and interpret the results
on the line-of-sight projections and emission profiles of the different
non-thermal emission mechanisms, and correlations of various non-thermal
emission processes with the thermal X-ray emission. We compare our results to
previous findings in the literature and point out future theoretical work that
is needed to complement this work (cf.{\ }Sect.~\ref{sec:discussion}). The
conclusions are drawn in Sect.~\ref{sec:conclusions}.
Appendix~\ref{sec:electron_pop} describes the modelling of the primary and
secondary CR electron population while Appendix~\ref{sec:rad_proc_formulae}
describes the formulae of the non-thermal emission mechanisms ranging from
radio synchrotron radiation, inverse Compton emission, as well as hadronically
induced $\gamma$-ray emission.

\section{Methodology}
\label{sec:methodology}

\subsection{General approach}
\label{sec:approach}

We follow the CR proton pressure dynamically in our simulations while taking
into account all relevant CR injection and loss terms in the ICM, except for a
possible production of CR protons by AGN. In contrast, we model the CR electron
population in a post-processing step since it does not modify the hydrodynamics
owing to its negligible pressure contribution.  In this paper, we concentrate
on three observationally motivated wave-bands.  (1) Radio synchrotron emission
between 15~MHz and $1.4$~GHz, (2) non-thermal hard X-ray emission at energies
$E_\gamma>10$~keV, and (3) $\gamma$-ray emission at energies
$E_\gamma>100$~MeV. Studying our simulated radio synchrotron maps and emission
profiles, we develop a {\em unified scheme} for the generation of the diffuse
large scale radio emission of clusters such as giant radio halos, mini-halos,
as well as the radio relic emission.\footnote{Our `unified scheme' unifies
  apparently different diffuse radio phenomena in clusters (giant relics,
  halos, and mini-halos) with a simple and physically motivated model. The
  `unified scheme' should not be confused with a `complete model' and we want
  to point out that we have not accounted for all possible CR processes that
  could be of interest in the context of cluster physics.}

\subsection{The simulations}
\label{sec:sims}

\subsubsection{General setup and cluster sample}

This section provides a short overview of the simulations and physical models
used. Further details can be found in \citetalias{2007MNRAS...378..385P}.
All simulations were performed using the ``concordance'' cosmological cold dark
matter model with a cosmological constant ($\Lambda$CDM).  The cosmological
parameters of our model are: $\Omega_\rmn{m} = \Omega_\rmn{DM} + \Omega_\rmn{b}
= 0.3$, $\Omega_\rmn{b} = 0.039$, $\Omega_\Lambda = 0.7$, $h = 0.7$, $n = 1$,
and $\sigma_8 = 0.9$.  Here, $\Omega_\rmn{m}$ denotes the total matter density
in units of the critical density for geometrical closure today,
$\rho_\rmn{crit} = 3 H_0^2 / (8 \upi G)$. $\Omega_\rmn{b}$ and $\Omega_\Lambda$
denote the densities of baryons and the cosmological constant at the present
day. The Hubble constant at the present day is parametrised as $H_0 = 100\,h
\mbox{ km s}^{-1} \mbox{Mpc}^{-1}$, while $n$ denotes the spectral index of the
primordial power-spectrum, and $\sigma_8$ is the {\em rms} linear mass
fluctuation within a sphere of radius $8\,h^{-1}$Mpc extrapolated to $z=0$.

Our simulations were carried out with an updated and extended version of the
distributed-memory parallel TreeSPH code GADGET-2 \citep{2005MNRAS.364.1105S,
2001NewA....6...79S} that includes self-consistent cosmic ray physics
\citep{2007A&A...473...41E, 2006...Jubelgas, 2006MNRAS.367..113P}. Gravitational
forces were computed using a combination of particle-mesh and tree algorithms.
Hydrodynamic forces are computed with a variant of the smoothed particle
hydrodynamics (SPH) algorithm that conserves energy and entropy where
appropriate, i.e. outside of shocked regions \citep{2002MNRAS.333..649S}.

\begin{table}
\caption{\scshape: Cluster sample}
\begin{tabular}{l l l l r r}
\hline
\hline
Cluster & sim.'s & dyn. state$^{(1)}$ & $M_{200}^{(2)}$ & $R_{200}^{(2)}$ & $kT_{200}^{(3)}$ \\
& & & [$h^{-1}\,\rmn{M}_\odot$] & [$h^{-1}\,$Mpc] & [keV] \\
\hline
1  & g8a  & CC    & $1.8\times 10^{15}$ & 2.0  & 13.1 \\
2  & g1a  & CC    & $1.3\times 10^{15}$ & 1.8  & 10.6 \\
3  & g72a & PostM & $1.1\times 10^{15}$ & 1.7  & 9.4  \\
4  & g51  & CC    & $1.1\times 10^{15}$ & 1.7  & 9.4  \\
                                                    
5  & g1b  & M     & $3.7\times 10^{14}$ & 1.2  & 4.7  \\
6  & g72b & M     & $1.5\times 10^{14}$ & 0.87 & 2.4  \\
7  & g1c  & M     & $1.4\times 10^{14}$ & 0.84 & 2.3  \\
8  & g8b  & M     & $1.0\times 10^{14}$ & 0.76 & 1.9  \\
9  & g1d  & M     & $9.2\times 10^{13}$ & 0.73 & 1.7  \\
                                                    
10 & g676 & CC    & $8.8\times 10^{13}$ & 0.72 & 1.7  \\
11 & g914 & CC    & $8.5\times 10^{13}$ & 0.71 & 1.6  \\
12 & g1e  & M     & $6.4\times 10^{13}$ & 0.65 & 1.3  \\
13 & g8c  & M     & $5.9\times 10^{13}$ & 0.63 & 1.3  \\
14 & g8d  & PreM  & $5.4\times 10^{13}$ & 0.61 & 1.2  \\
\hline
\end{tabular}   
\begin{quote} 
  {\scshape Notes:}\\
  (1) The dynamical state has been classified through a combined criterion
  invoking a merger tree study and the visual inspection of the X-ray
  brightness maps. The labels for the clusters are M--merger, PostM--post
  merger (slightly elongated X-ray contours, weak cool core region
  developing), PreM--pre-merger (sub-cluster already within the virial
  radius), CC--cool core cluster with extended cooling region (smooth X-ray
  profile).\\
  (2) The virial mass and radius are related by $M_\Delta(z) = \frac{4}{3}
  \pi\, \Delta\, \rho_\rmn{crit}(z) R_\Delta^3 $, where $\Delta=200$ denotes a
  multiple of the critical overdensity $\rho_\rmn{crit}(z) = 3 H (z)^2/ (8\pi
  G)$. \\  
  (3) The virial temperature is defined by $kT_\Delta = G M_\Delta \, \mu\,
  m_\p / (2 R_\Delta)$, where $\mu$ denotes the mean molecular weight.
\end{quote}
\label{tab:sample}
\end{table} 

We have performed high-resolution hydrodynamic simulations of the formation of
14 galaxy clusters. The clusters span a mass range from $5 \times 10^{13}\,
h^{-1}\, \rmn{M}_\odot$ to $2 \times 10^{15}\, h^{-1}\, \rmn{M}_\odot$ and show
a variety of dynamical states ranging from relaxed cool core clusters to
violent merging clusters (cf. Table~\ref{tab:sample}). The clusters have
originally been selected from a low-resolution dark-matter-only simulation
\citep{2001MNRAS.328..669Y}. Using the `zoomed initial conditions' technique
\citep{1993ApJ...412..455K}, the clusters have been re-simulated with higher
mass and force resolution by adding short-wavelength modes within the
Lagrangian regions in the initial conditions that will evolve later-on into the
structures of interest \citepalias{2007MNRAS...378..385P}.  We re-simulated
three isolated clusters (cluster 4, 10, and 11) and three super-cluster regions
which are each dominated by a large cluster (cluster 1, 2, and 3) and
surrounded by satellite clusters (cluster 5 to 9 and 12 to 14).  In
high-resolution regions, the dark matter particles had masses of $m_\rmn{DM} =
1.13 \times 10^9\,h^{-1}\,\rmn{M}_\odot$ and SPH particles had $m_\rmn{gas} =
1.7\times 10^8\,h^{-1}\,\rmn{M}_\odot$ so each individual cluster is resolved
by $8 \times 10^4$ to $4\times 10^6$ particles, depending on its final mass.
The SPH densities were computed from 48 neighbours, allowing the SPH smoothing
length to drop at most to half of the value of the gravitational softening
length of the gas particles. This choice of the SPH smoothing length leads to
our minimum gas resolution of approximately $8\times
10^{9}\,h^{-1}\,\rmn{M}_\odot$.  For the initial redshift we chose
$1+z_\rmn{init}=60$.  The gravitational force softening was of a spline form
\citep[e.g.,][]{1989ApJS...70..419H} with a Plummer equivalent softening length
that is assumed to have a constant comoving scale down to $z = 5$, and a
constant value of $5\,h^{-1}$kpc in physical units at later epochs.

We analysed the clusters with a halo-finder based on spherical overdensity
followed by a merger tree analysis in order to get the mass accretion history
of the main progenitor. We also produced projections of the X-ray emissivity at
redshift $z=0$ in order to get a visual impression of the cluster
morphology. The dynamical state of a cluster is defined by a combined
criterion: (i) if the cluster did not experience a major merger with a
progenitor mass ratio 1:3 or larger after $z=0.8$ (corresponding to a look-back
time of $\sim 5\, h^{-1}\,$Gyr) and (ii) if the visual impression of the
cluster's X-ray morphology is relaxed, it was defined to be a cool core
cluster.  The spherical overdensity definition of the virial mass of the
cluster is given by the material lying within a sphere centred on a local
density maximum, whose radial extend $R_\Delta$ is defined by the enclosed
threshold density condition $M (< R_\Delta) / (4 \pi R_\Delta^3 / 3) =
\rho_\rmn{thres}$. We chose the threshold density $\rho_\rmn{thres}(z) =
\Delta\, \rho_\rmn{crit} (z)$ to be a multiple $\Delta=200$ of the critical
density of the universe $\rho_\rmn{crit} (z) = 3 H (z)^2/ (8\pi G)$. We assume a
constant $\Delta=200$ although some treatments employ a time-varying $\Delta$
in cosmologies with $\Omega_\rmn{m} \ne 1$ \citep{1996MNRAS.282..263E}. In the
reminder of the paper, we use the terminology $R_\rmn{vir}$ instead of
$R_{200}$.

\subsubsection{The models}
\label{sec:models}

\begin{table}
\caption{\scshape: Different physical processes in our simulations:}
\begin{center}
\begin{tabular}{l | c c c c c}
\hline
\hline
Simulated physics$^{(1)}$ & \multicolumn{5}{c}{simulation models$^{(1)}$:}\\
& S1 & & S2 & & S3 \\ 
\hline
thermal shock heating & \checkmark & & \checkmark & & \checkmark \\
radiative cooling     &            & & \checkmark & & \checkmark \\
star formation        &            & & \checkmark & & \checkmark \\
Coulomb CR losses     & \checkmark & & \checkmark & & \checkmark \\
hadronic CR losses    & \checkmark & & \checkmark & & \checkmark \\
shock-CRs             & \checkmark & & \checkmark & & \checkmark \\
supernova-CRs         &            & &            & & \checkmark \\
\hline
\end{tabular}   
\end{center}
\begin{quote} 
  {\scshape Notes:}\\
  (1) This table serves as an overview over our simulated models. The first
  column shows the simulated physics and the following three columns show our
  different simulation models with varying gas and cosmic ray physics. Model
  S1 models the thermal gas non-radiatively and includes CR physics, while the
  models S2 and S3 use radiative gas physics with different variants of
  CR physics.\\
\end{quote}
\label{tab:models}
\end{table} 

For each galaxy cluster we ran three different simulations with varying gas and
cosmic ray physics (cf.~Table~\ref{tab:models}).  The first set of simulations
used non-radiative gas physics only, i.e.~the gas is transported adiabatically
unless it experiences structure formation shock waves that supply the gas with
entropy and thermal pressure support.  Additionally we follow cosmic ray (CR)
physics including adiabatic CR transport processes, injection by cosmological
structure formation shocks with a Mach number dependent acceleration scheme, as
well as CR thermalization by Coulomb interaction and catastrophic losses by
hadronic interactions (model S1).  The second set of simulations follows the
radiative cooling of the gas, star formation, supernova feedback, and a
photo-ionising background \citepalias[details can be found
  in][]{2007MNRAS...378..385P}.  As before in model S1, we account for CR
acceleration at structure formation shocks and allow for all CR loss processes
(model S2). The last set of simulations additionally assumes that a constant
fraction $\zeta_\rmn{SN} = \eps_\rmn{CR,inj}/\eps_\rmn{diss} = 0.3$ of the
kinetic energy of a supernova ends up in the CR population (model S3), which is
motivated by TeV $\gamma$-ray observations of a supernova remnant that find an
energy fraction of $\zeta_\rmn{SN} \simeq 0.1 - 0.3$ when extrapolating the CR
distribution function \citep{2006Natur.439..695A}. We choose a maximum value
for this supernova energy efficiency owing to the large uncertainties and our
aim to bracket the realistic case with the two radiative CR
simulations. Generally, we use model S2 as our standard case which is a
conservative choice for the CR pressure and explore how the physics of the
other models change the resulting non-thermal cluster observables. In this
work, we don't account for AGN sources of cosmic rays, but that this will be
addressed in upcoming work \citep{2007Sijacki}.

Radiative cooling was computed assuming an optically thin gas of primordial
composition (mass-fraction of $X_\rmn{H} = 0.76$ for hydrogen and $1-X_\rmn{H}
= 0.24$ for helium) in collisional ionisation equilibrium, following
\citet{1996ApJS..105...19K}. We have also included heating by a photo-ionising,
time-dependent, uniform ultraviolet (UV) background expected from a population
of quasars \citep{1996ApJ...461...20H}, which reionises the Universe at $z
\simeq 6$.  Star formation is treated using the hybrid multiphase model for the
interstellar medium introduced by \citet{2003MNRAS.339..289S}. In short, the
ISM is pictured as a two-phase fluid consisting of cold clouds that are
embedded at pressure equilibrium in an ambient hot medium.  The clouds form
from the cooling of high density gas, and represent the reservoir of baryons
available for star formation. When stars form, the energy released by
supernovae heats the ambient hot phase of the ISM, and in addition, clouds in
supernova remnants are evaporated. These effects establish a tightly
self-regulated sub-resolution model for star formation in the ISM.

Cosmic ray physics was computed by using a new formulation that follows the
most important injection and loss processes self-consistently while accounting
for the CR pressure in the equations of motion \citep{2007A&A...473...41E,
  2006...Jubelgas, 2006MNRAS.367..113P}. We refer to these papers for a
detailed description of the formalism, providing here only a short summary of
the model.  In our methodology, the non-thermal cosmic ray population of each
gaseous fluid element is approximated by a simple power law spectrum in
particle momentum, characterised by an amplitude, a low-momentum cut-off, and
a fixed slope $\alpha = 2.3$. This choice is justified by taking the mean of
the Mach number distribution weighted by the dissipated energy at shocks which
is closely related to the spectral index of a CR power-law distribution
\citepalias{2007MNRAS...378..385P}.  Adiabatic CR transport processes such as
compression and rarefaction, and a number of physical source and sink terms
which modify the cosmic ray pressure of each particle are modelled. The most
important sources considered are injection by supernovae (in our radiative
simulations) and diffusive shock acceleration at cosmological structure
formation shocks, while the primary sinks are thermalization by Coulomb
interactions, and catastrophic losses by hadronic interactions.

\subsection{Essentials of radiative processes}
\label{sec:rad_proc}

We are interested in the non-thermal emission of the inter-galactic medium at
radio frequencies ($\nu > 10$~MHz) as well as at hard X-ray/$\gamma$-ray energies
($E_\gamma > 10$~keV).  This non-thermal emission is generated by CR electrons
with energies $E_\e > \mbox{GeV}$ as can be readily inferred from the classical
synchrotron and inverse Compton formulae,
\begin{eqnarray}
  \label{eq:synchrotron}
  \nu_\rmn{synch} &=& \frac{3 e B}{2\pi\, m_\rmn{e} c}\,\gamma^2 \simeq 
  1 \mbox{ GHz}\, \frac{B}{\umu\mbox{G}}\, 
  \left(\frac{\gamma}{10^4}\right)^2,\\
  \label{eq:ICphoton}
  h\nu_\rmn{IC} &=& \frac{4}{3}\, h\nu_\rmn{init}\,\gamma^2 \simeq 
  90 \mbox{ keV}\, \frac{\nu_\rmn{init}}{\nu_\rmn{CMB}}\, 
  \left(\frac{\gamma}{10^4}\right)^2,
\end{eqnarray}
where $e$ denotes the elementary charge, $h$ the Planck constant, $c$ the
speed of light, $m_\e$ the electron mass, the particle kinetic energy $E /
(m_\e c^2)= \gamma-1$ is defined in terms of the Lorentz factor $\gamma$, and
$B = \sqrt{\bra \vecbf{B}^2\ket}$ is the {\em rms} of the magnetic vector
field $\vecbf{B}$. We chose CMB photons $h\nu_\rmn{CMB}\simeq 0.66$~meV as
source for the inverse Compton emission using Wien's displacement law. Thus,
the same CR electron population seen in the radio band via synchrotron
emission can be observed in the hard X-ray regime through the IC process.

\subsubsection{Synchrotron and IC emission from equilibrium spectra}

\begin{figure}
\resizebox{\hsize}{!}{\includegraphics{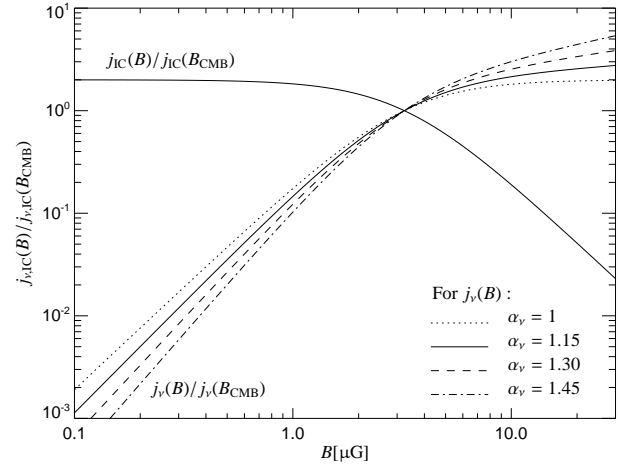}}
  \caption{The synchrotron and inverse Compton (IC) emissivity of an
  equilibrium distribution of CR electrons is shown for various spectral
  indices $\alpha_\nu$. The normalisation is given by the respective
  emissivities at the equivalent magnetic field strength of the energy density
  of the CMB, $B_\rmn{CMB} = 3.24\,\umu\mbox{G}$.  The weak field regime on the
  left-hand side is characterised by the dominant inverse Compton (IC) emission
  while the strong field regime on the right side has the synchrotron
  emission as the dominant electron cooling channel.  }
  \label{fig:IC-synchro_B}
\end{figure}

The synchrotron and IC emissivities of an equilibrium distribution of CR
electrons is characterised by two distinctive regimes.  The weak field regime
is characterised by the dominating IC emission while the strong field regime
has the synchrotron emission as the dominant electron cooling channel. Using
the formulae for the equilibrium distribution of CR electrons
(Eqns.~(\ref{eq:C_e_prim}) or (\ref{eq:C_e})), the synchrotron and IC
emissivities depend on the magnetic field strength as follows:
\begin{equation}
  \label{eq:IC-synchro_B}
  j_\nu \propto \frac{B^{\alpha\nu+1}}{\eps_B + \eps_\rmn{CMB}}
  \quad\mbox{and}\quad
  j_\rmn{IC} \propto \frac{1}{\eps_B + \eps_\rmn{CMB}},
\end{equation}
where $\alpha_\nu$ denotes the synchrotron spectral index that is defined by
$j_\nu \propto \nu^{-\alpha_\nu}$, and $\eps_B = B^2 / (8\, \upi)$.
Figure~\ref{fig:IC-synchro_B} shows these emissivities, normalised by
respective emissivities at the equivalent magnetic field strength of the CMB
energy density at $z=0$, $B_\rmn{CMB} = 3.24\,\umu\mbox{G}$.  In the IC
dominated electron cooling regime leftwards of $B_\rmn{CMB}$, the synchrotron
emissivity quickly decreases relative to the IC emissivity, showing thus a
strong dependence on the magnetic field strength. There is an interesting
twist associated with the different spectral indices of the synchrotron
emission in clusters. \citetalias{2007MNRAS...378..385P} shows that the
characteristic shock strength increases as one moves outwards from the cluster
centre due to the decrease of the sound velocity in combination with the
shallower peripheral cluster potential. CR acceleration crucially depends on
the shock strength according to Eqn.~(\ref{eq:ainj}) predicting steep CR
spectra at the centre that flatten on average towards the cluster periphery.
Since the magnetic field has a decreasing profile with radius
(Eqn.~\ref{eq:magnetic_scaling}), the synchrotron emission of clusters should
qualitatively be given by the upper envelope of the family of emissivity
curves labelled by different spectral indices of Fig.~\ref{fig:IC-synchro_B}.
This simplified picture assumes that the electron spectra are dominated by
injection and are neither qualitatively modified by CR transport processes
such as CR diffusion nor by the hadronically injected electron spectra.

\subsubsection{Comparison of inverse Compton and $\bgamma$-ray spectra}

\begin{figure}
\resizebox{\hsize}{!}{\includegraphics{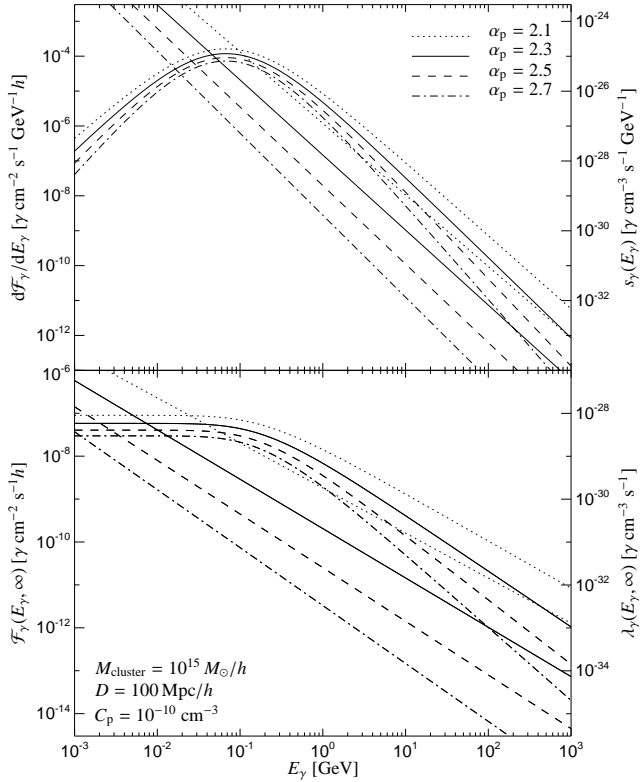}}
  \caption{Spectral distribution of the differential $\gamma$-ray flux
  $\dd\mathcal{F}_\gamma / \dd E_\gamma$ (upper panel) and the integrated
  $\gamma$-ray flux $\mathcal{F}_\gamma(E_\gamma, \infty)$ (bottom panel) for
  different spectral indices. Shown are inverse Compton spectra from a
  secondary CR electron population and pion decay induced $\gamma$-ray spectra
  (broken power-laws) both resulting from hadronic CR proton interactions. The
  model calculations assume a normalisation for the CR proton distribution of
  $C_\p = 10^{-10}\mbox{ cm}^{-3}$, a nucleon density of $n_\rmn{N} =
  10^{-3}\mbox{ cm}^{-3}$ (for $\lambda_\gamma$), and put the fiducial cluster
  with mass $M_\rmn{cluster} = 10^{15}\,M_\odot / h$ at a distance $D = 100\,
  \mbox{Mpc} / h$ to derive $\mathcal{F}_\gamma$.}
  \label{fig:spectrum}
\end{figure}

Figure~\ref{fig:spectrum} compares the spectral distribution of the pion-decay
induced $\gamma$-ray emission (broken power-laws) with the inverse Compton
emission both resulting from hadronic CR proton interactions.  Note, that the
relative normalisation of both emission components is governed by hadronic
physics and does not depend on the gas and CR proton number densities.  For our
choice of the CR proton spectral index of $\alpha_\p=2.3$, the ratio of
pion-decay to secondary IC emission in the energy range $E_\gamma > 100\mbox{
MeV}$ can be readily inferred to be $\mathcal{F}_\gamma/\mathcal{F}_\rmn{IC} =
20$.  The asymptotic behaviour for the $\gamma$-ray number flux of both
emission components at high energies is given by
\begin{equation}
  \label{eq:asymptotics}
  \mathcal{F}_\gamma \propto E_\gamma^{-\alpha_\p+1}
  \quad\mbox{and}\quad
  \mathcal{F}_\rmn{IC} \propto E_\gamma^{-\alpha_\nu}
   = E_\gamma^{-\alpha_\p/2}.
\end{equation}
Assuming a spectral CR index of $\alpha_\p=2$ yields the same asymptotic
behaviour while increasing $\alpha_\p$ results in a shallower decline for
$\mathcal{F}_\rmn{IC}$ with energy such that eventually the IC component will
surpass the pion decay emission. This however is well above the energy range
$E_\gamma \gg 10\mbox{ TeV}$ that is of interest for imaging air \v{C}erenkov
telescopes. For an energy range $E_\gamma < 1\mbox{ MeV}$ the secondary IC
emission dominates the hadronically induced channel. In contrast to the
secondary IC emission, the IC emission level of primary CR electrons depends
on the dynamical activity of the region, in particular on the shock strength
and the associated amount of dissipated energy. Comparing the primary to the
secondary IC emission will be one goal of this paper.


\section{Results}
\label{sec:results}

\subsection{Cluster environment and cosmic ray pressure}
\label{sec:CRpressure}

\begin{figure*}
\begin{center}
  \begin{minipage}[t]{0.495\textwidth}
    \centering{\it \large Gas density:}
  \end{minipage}
  \hfill
  \begin{minipage}[t]{0.495\textwidth}
    \centering{\it \large Gas temperature:}
  \end{minipage}
\resizebox{0.5\hsize}{!}{\includegraphics{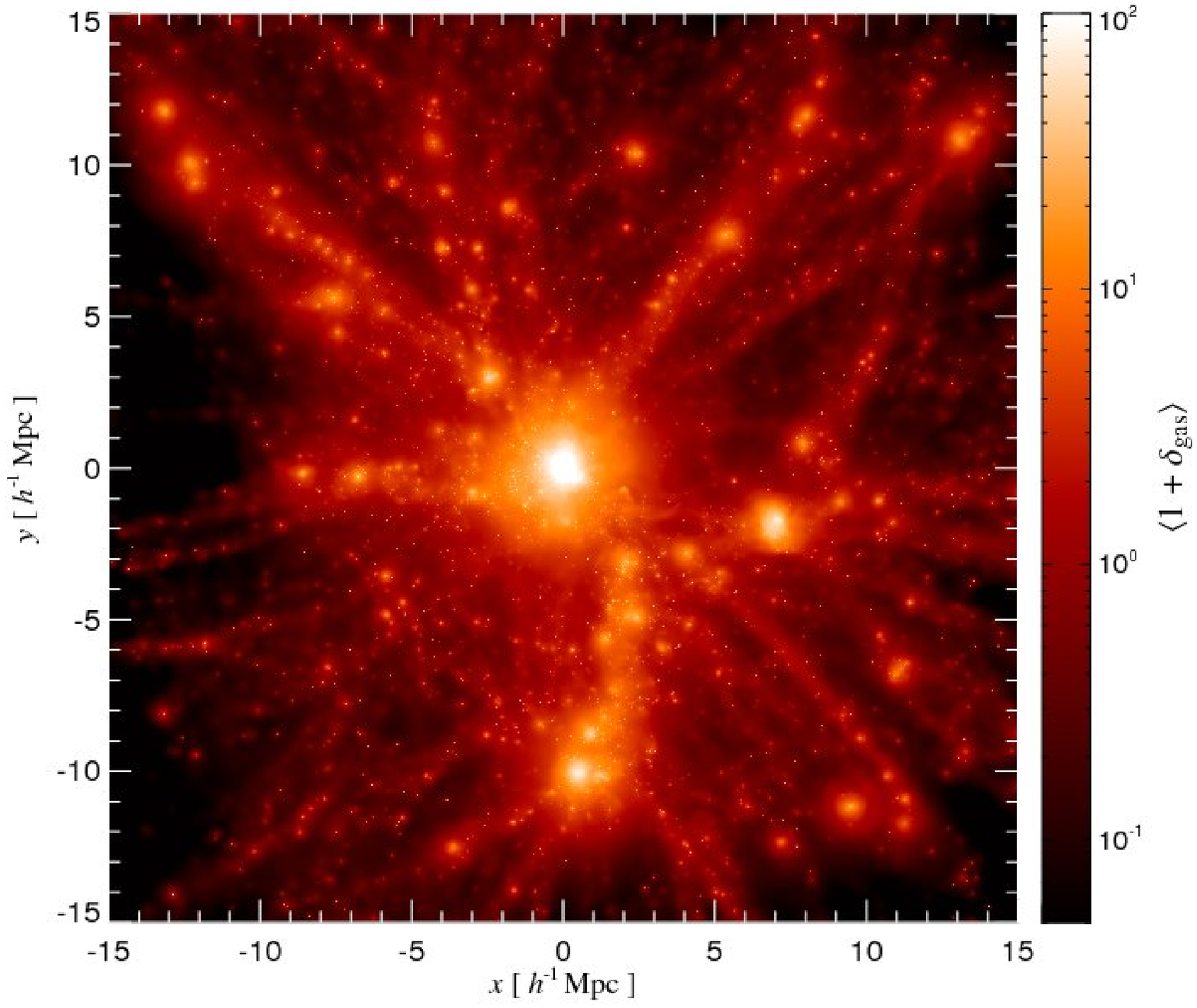}}%
\resizebox{0.5\hsize}{!}{\includegraphics{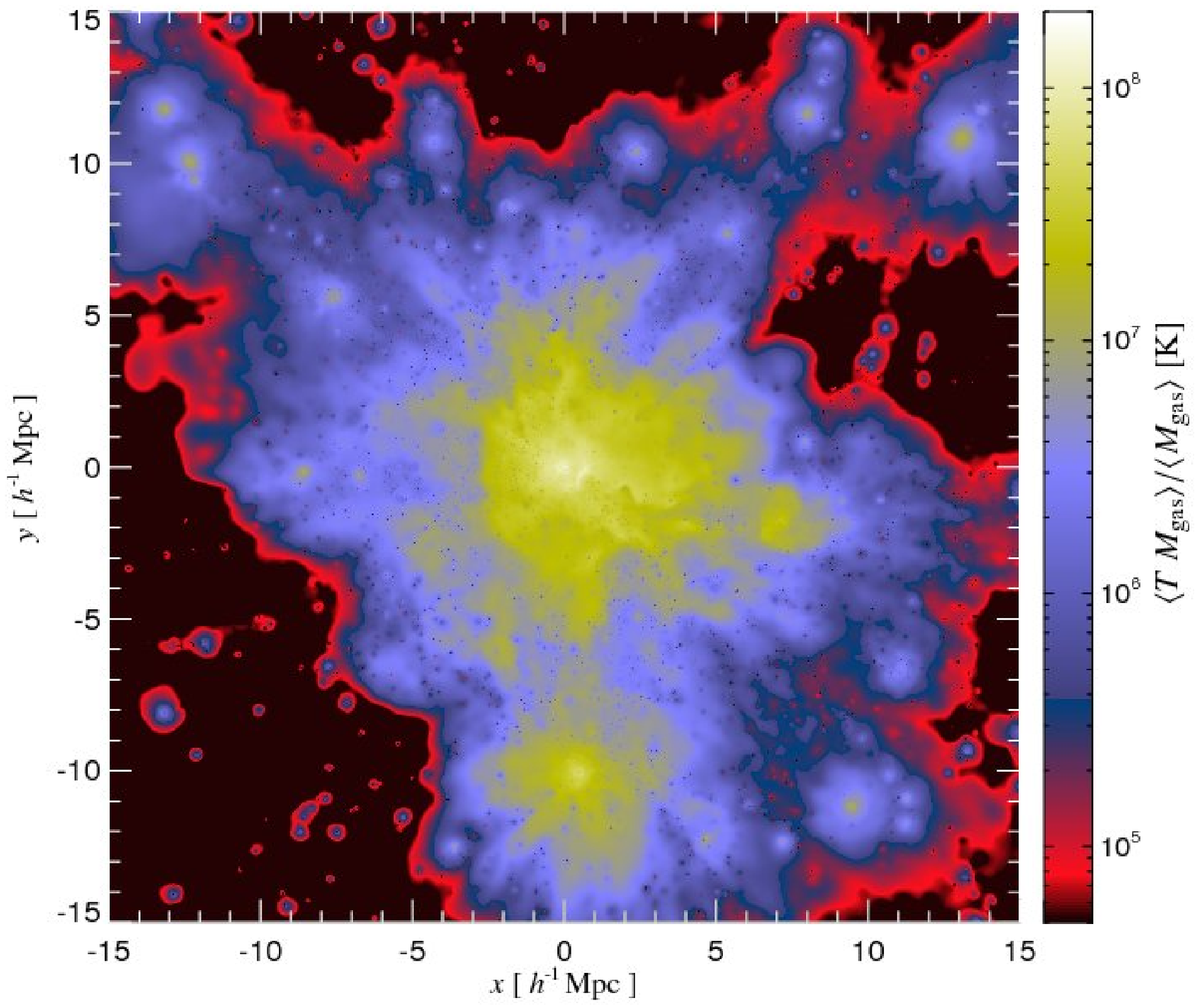}}\\
  \begin{minipage}[t]{0.495\textwidth}
    \centering{\it \large Shock Mach numbers weighted by $\dot{\eps}_\rmn{diss}$:}
  \end{minipage}
  \hfill
  \begin{minipage}[t]{0.495\textwidth}
    \centering{\it \large Shock Mach numbers weighted by $\dot{\eps}_\rmn{CR,inj}$:}
  \end{minipage}
\resizebox{0.5\hsize}{!}{\includegraphics{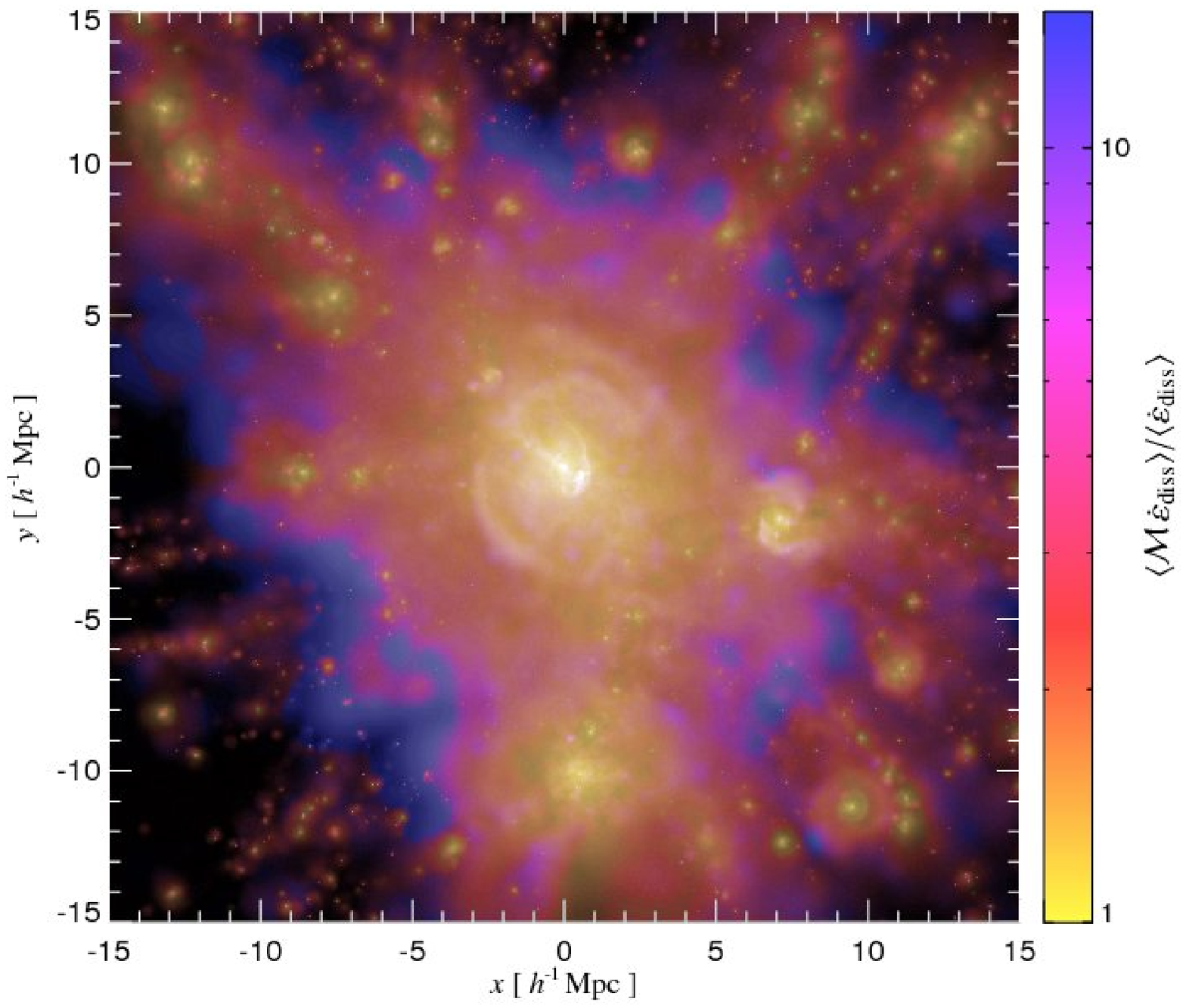}}%
\resizebox{0.5\hsize}{!}{\includegraphics{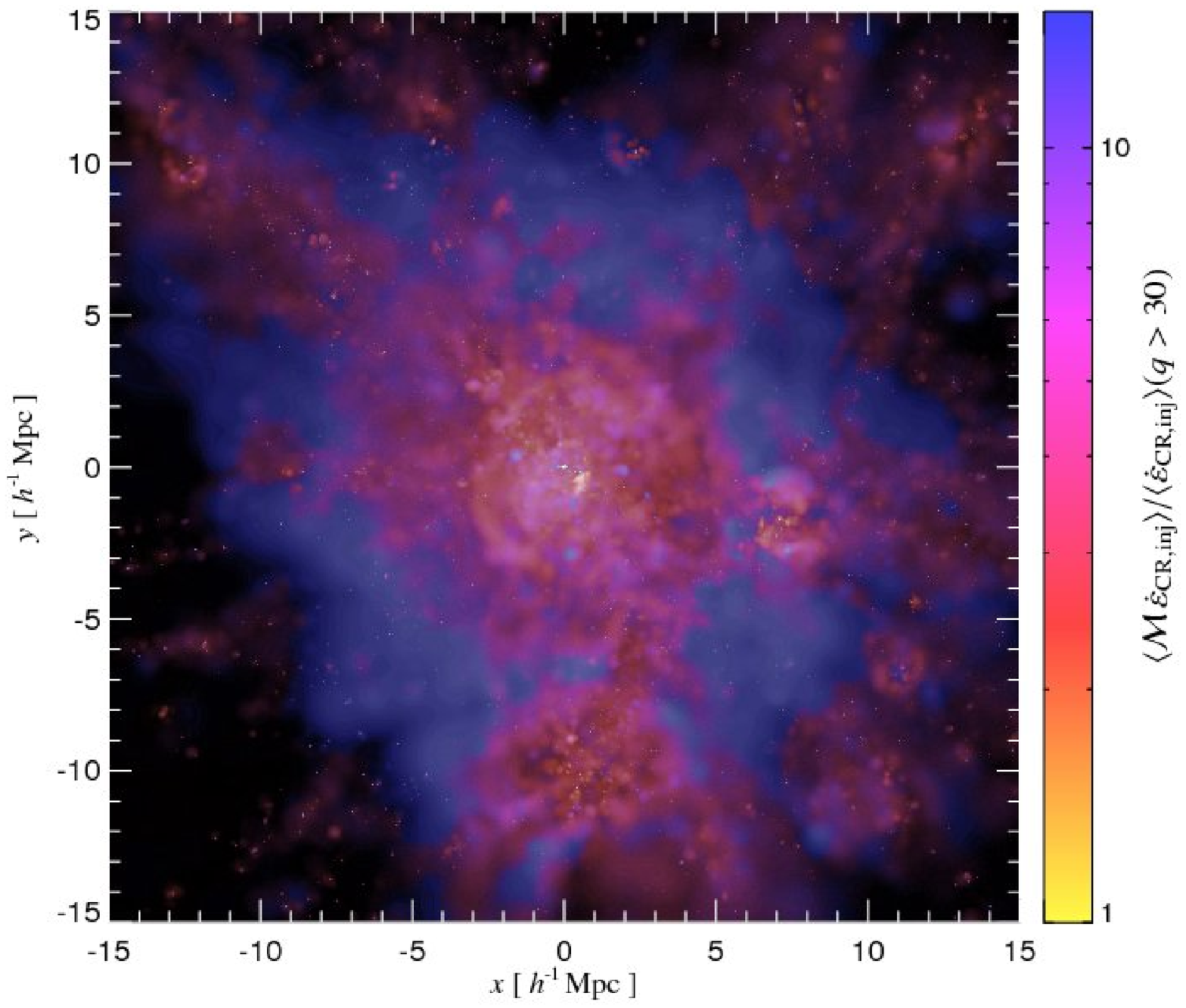}}\\
\end{center}
  \caption{The environment of a large post-merging galaxy cluster ($M\simeq
    10^{15} h^{-1}\,\rmn{M}_\odot$) in our simulation with radiative gas
    physics and star formation including CRs from structure formation shocks
    only (model S2). Shown are the line-of-sight averaged density (top
    left side), the mass weighted temperature (top right side), the
    Mach number of shocks weighted by the energy dissipation rate in colour
    (while the brightness displays the logarithm of the dissipation rate,
    bottom left side), and the Mach number of shocks weighted by the
    energy injection rate of CR protons in colour (while the brightness
    displays the logarithm of the CR proton energy injection rate, bottom
    right side). For better comparison, we used the same colour and
    brightness scale in the bottom plots.  Only CR protons with a dimensionless
    momentum $q = \beta\gamma>30$ have been considered for calculating the CR
    energy density since lower energetic CR protons are not detectable at radio
    frequencies $\nu>120 \mbox{ MHz}$ by means of hadronically produced
    secondary electrons.}
  \label{fig:g72}
\end{figure*}
\begin{figure*}
\begin{center}
  \begin{minipage}[t]{0.495\textwidth}
    \centering{\it \large CR proton pressure:}
  \end{minipage}
  \hfill
  \begin{minipage}[t]{0.495\textwidth}
    \centering{\it \large Relative pressure of CR protons:}
  \end{minipage}
\resizebox{0.5\hsize}{!}{\includegraphics{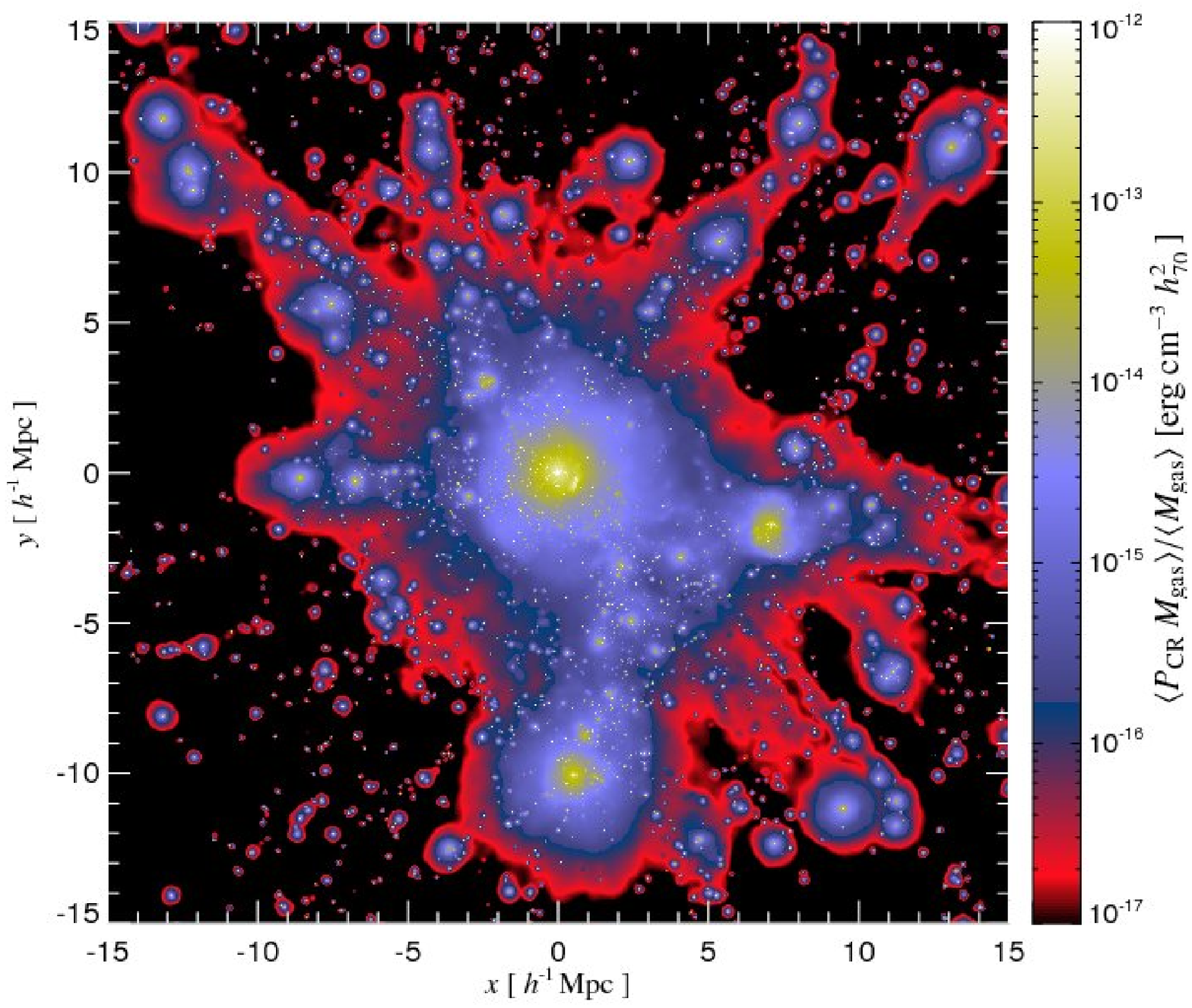}}%
\resizebox{0.5\hsize}{!}{\includegraphics{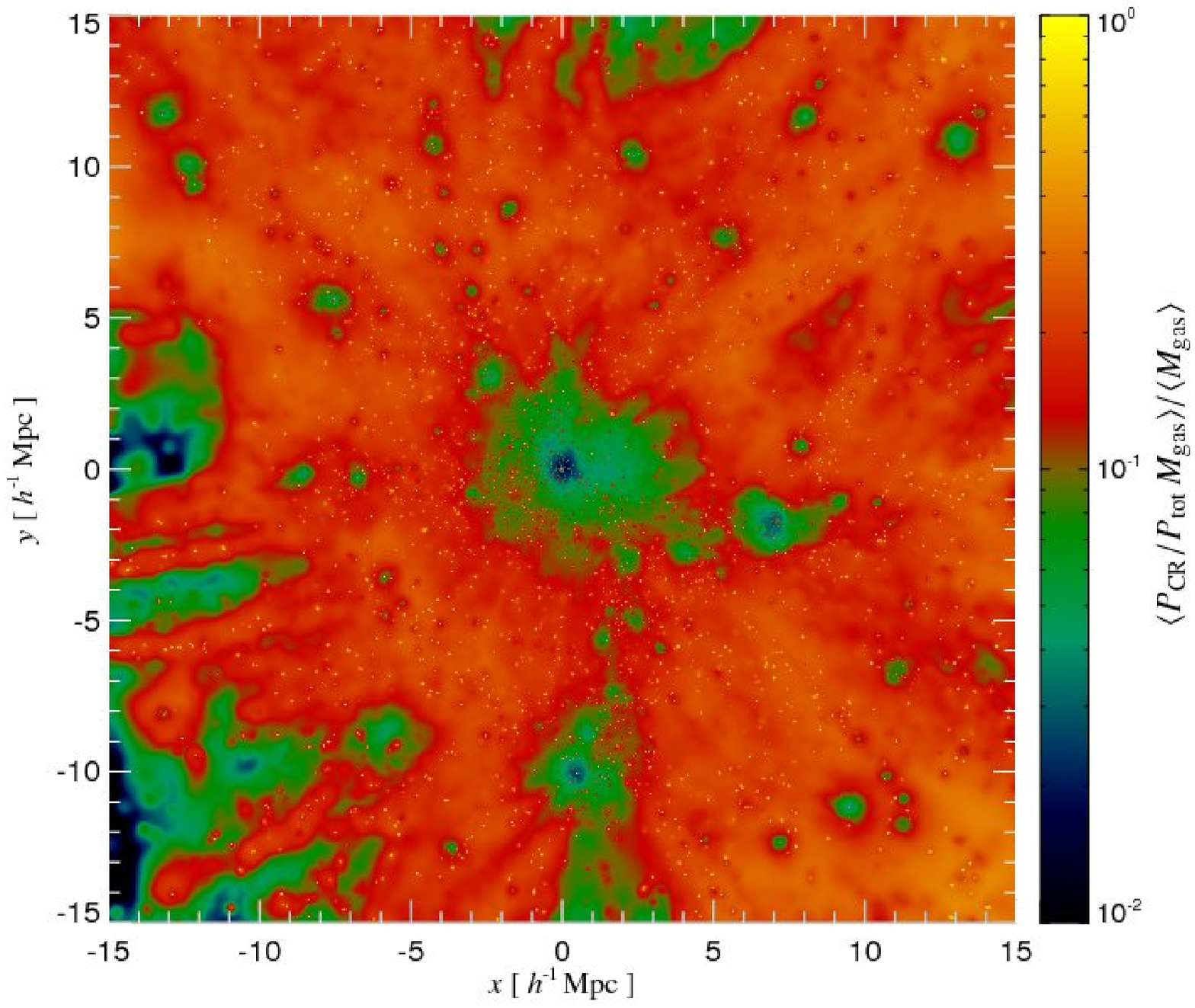}}\\
  \begin{minipage}[t]{0.495\textwidth}
    \centering{\it \large Relative pressure of primary CR electrons:}
  \end{minipage}
  \hfill
  \begin{minipage}[t]{0.495\textwidth}
    \centering{\it \large Relative pressure of secondary CR electrons:}
  \end{minipage}
\resizebox{0.5\hsize}{!}{\includegraphics{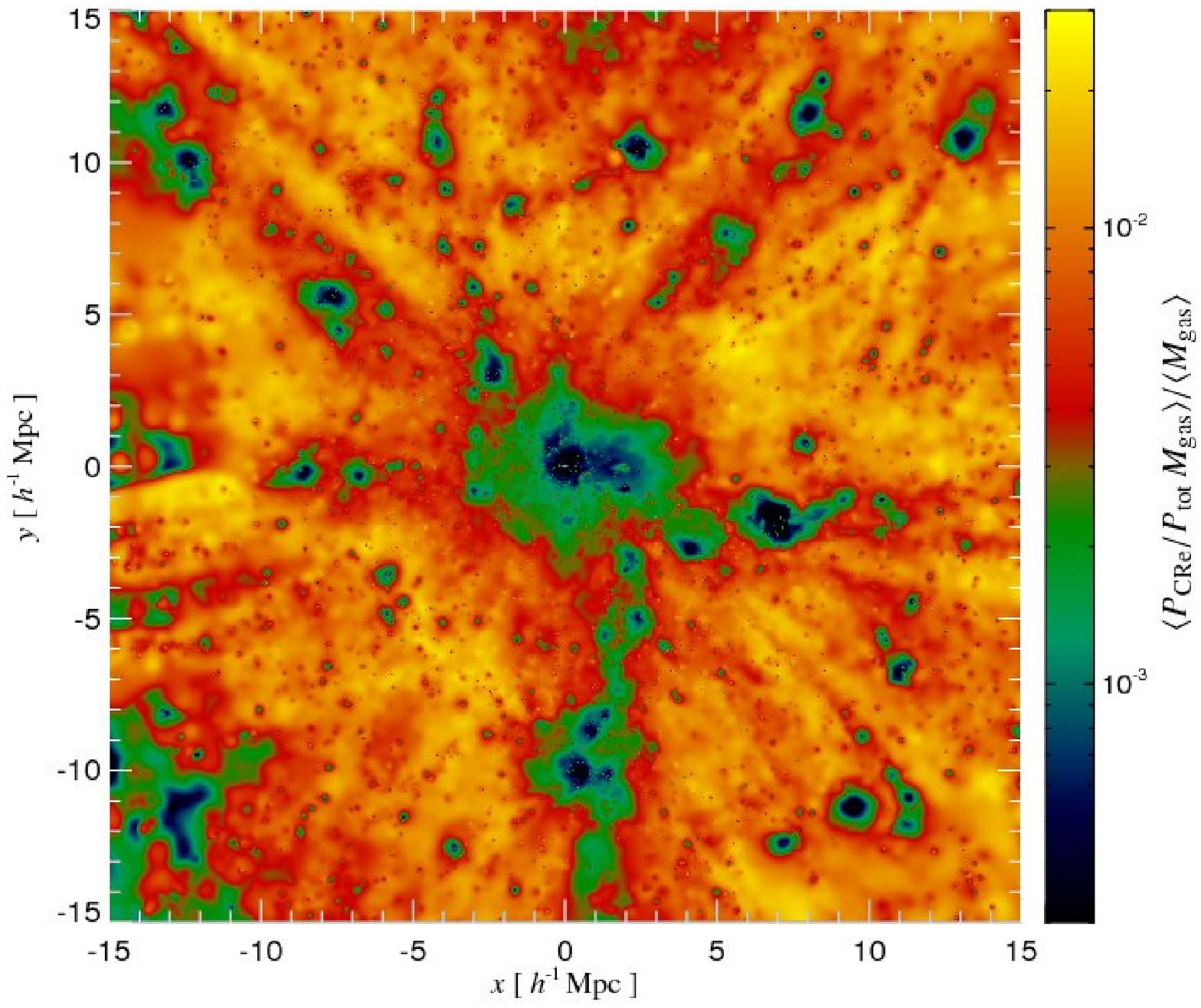}}%
\resizebox{0.5\hsize}{!}{\includegraphics{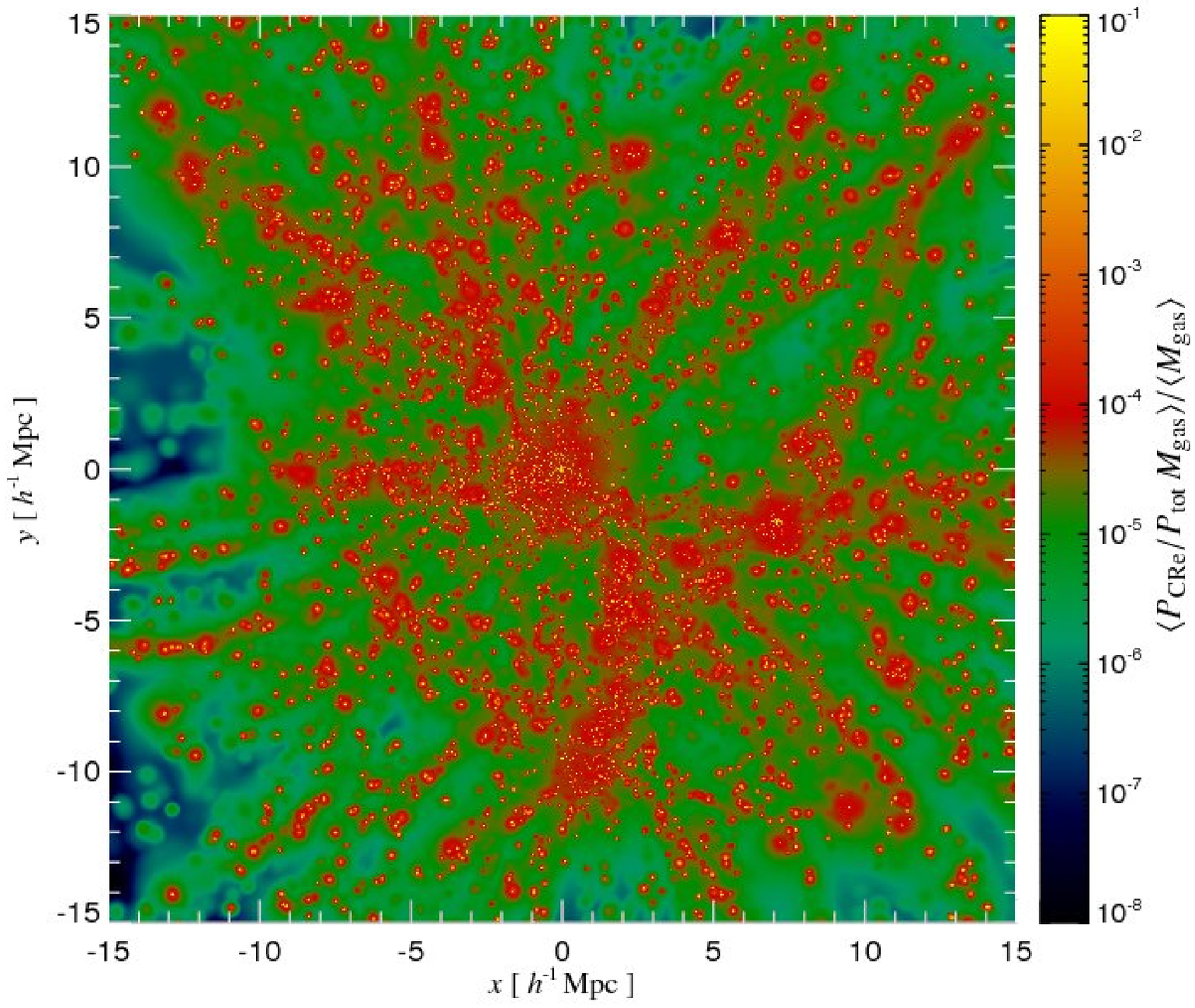}}\\
\end{center}
  \caption{CR proton and electron pressure maps in the super-cluster
    environment of a large post-merging galaxy cluster ($M\simeq 10^{15}
    h^{-1}\,\rmn{M}_\odot$) in our radiative simulation (model S2). The mass
    weighted CR pressure (top left side) is contrasted with the mass
    weighted CR pressure {\em relative} to the total gas pressure (top
    right side). Since the CR proton pressure decreases less steeply than
    the gas pressure as a function of cluster radius this results in an
    increasing relative CR pressure profile towards the periphery.  In the
    bottom panels, we show the mass weighted CR electron pressure relative to
    the total pressure for primary, shock-accelerated electrons (left
    side) and secondary electrons resulting from hadronic CR proton
    interactions (right side). The CR electron pressure derives from the
    respective equilibrium distribution functions that balance acceleration and
    cooling processes.}
  \label{fig:g72_CR}
\end{figure*}

Figure~\ref{fig:g72} shows the region around the largest cluster with merging
activity in our sample ($M\simeq 10^{15} h^{-1}\,\rmn{M}_\odot$, with the
identifier g72a) in our simulation with radiative gas physics, star formation,
and CRs from structure formation shocks only (model S2).  This galaxy cluster
experienced a large merger with a mass ratio of
$m_\rmn{merger}/m_\rmn{progenitor}=0.3$ at redshift $z=0.25$ preceded by a
minor merger mass ratio of $m_\rmn{merger}/m_\rmn{progenitor}=0.1$ at $z=0.4$.
The environment is dominated by the large central post-merging cluster and
surrounded by smaller satellite clusters and groups. The line-of sight average
of the density shows the suitably normalised quantity $1+\delta_\rmn{gas} =
\rho / (\Omega_\rmn{b}\, \rho_\rmn{crit})$.  While the ICM of the central massive
cluster reaches central temperatures above that of the virial temperature of
$kT_\vir = 9.4 \mbox{ keV}$, the surrounding warm-hot intergalactic medium
(WHIM) acquires temperatures of $kT \sim (10^{-2} - 10^{-1}) \mbox{ keV}$. The
spatial distribution of shock strengths can be studied best by looking at the
Mach numbers weighted by the energy dissipation rate at structure formation
shocks (represented by the colour hue in the bottom left panel of
Fig.~\ref{fig:g72}).  The brightness scales logarithmically with the projected
dissipation rate $\bra \dot{\eps}_\rmn{diss}\ket_\rmn{los}$.  Within this
super-cluster region most of the energy is dissipated in weak internal shocks
with Mach numbers $\M\lesssim 2$ which are predominantly central flow shocks or
merger shock waves traversing the cluster centre.  Collapsed cosmological
structures are surrounded by several quasi-spherical shells of external shocks
with successively higher Mach numbers, but they play only a minor role in the
energy balance of thermalization as can be inferred by its dim brightness.
Clearly visible are spherical shells of shocks at different radii from the
cluster centre. Two distinct outgoing shock waves at distances of 2 and
$3\,h^{-1}\mbox{ Mpc}$ to the cluster centre are visible that are triggered by
the merger, followed by shells of stronger shocks further outwards. This
picture is dramatically changed if we perform the weighting of the Mach numbers
with the energy injection rate of CR protons, $\dot{\eps}_\CR$ (shown in colour
while the brightness displays the logarithm of the CR proton energy injection
rate, bottom right side).  Only CR protons with a dimensionless momentum
$q = \beta\gamma>30$ ($E\gtrsim30$~GeV) have been considered for calculating
the CR energy density since lower energetic CR protons are not detectable at
radio frequencies $\nu>120 \mbox{ MHz}$ by means of hadronically produced
secondary electrons (assuming $B=1\,\umu\mbox{G}$). As expected, weak shocks
with Mach numbers $1<\M<2$ almost disappear in this representation due to their
small acceleration efficiency. This effect is amplified by considering only
energetic CR protons with $q >30$. Comparing the Mach numbers weighted by
$\dot{\eps}_\CR$ to those weighted by $\dot{\eps}_\rmn{diss}$ uncovers the
three-dimensional picture of these shock surfaces. The powerful (in terms of
energy dissipation rate) but weak (in terms of shock strength) internal shock
waves are surrounded by shock surfaces of successively increasing Mach numbers
that are only revealed in these projections if one disfavours these internal
shocks in the weighting function.

What are the consequences for the CR proton pressure within galaxy clusters?
Similar to the thermal pressure, it peaks in the centre and falls of with
radius. The CR pressure is additionally enhanced at strong shock waves, that
have a modulating effect on the underlying CR pressure distribution.  The
latter can be seen by looking at the strong tangential shock wave at $r\simeq
600 \,h^{-1}\mbox{ kpc}$ southwards from the cluster centre in the mass
weighted CR proton pressure map (top left panel in
Fig.~\ref{fig:g72_CR}). The CR pressure peaks roughly at $P_\CR\simeq 10^{-12}
\mbox{erg cm}^{-3}\,h_{70}^{2}$ at the cluster centre.  Even more revealing
is the mass-weighted CR proton pressure relative to the total pressure $X_\CR =
P_\CR / P_\rmn{tot}$, where $P_\rmn{tot} = P_\CR + P_\th$ (top right panel
in Fig.~\ref{fig:g72_CR}).  The relative CR pressure $X_\CR$ acquires
comparatively high values within the WHIM that are hydrodynamically important,
their importance decreases (on average) as we move inwards due to a combination
of the following reasons: (1) weak central flow shocks are inefficient in
accelerating CRs \citepalias[e.g.,][]{2007MNRAS...378..385P} and (2) adiabatic
compression of a composite of CRs and thermal gas disfavours the CR pressure
relative to the thermal pressure due to the softer equation of state of CRs.
Within each individual galaxy as well as within the cluster centre, the CR
pressure reaches equipartition or dominates the thermal pressure as can be seen
by the numerous yellow points sprinkled over the map, each corresponding to a
galaxy.  This is due to the long CR cooling time scales compared to those of
the thermal gas, an effect that diminishes the thermal gas pressure relative to
that of CRs \citepalias{2007MNRAS...378..385P}.

It is very instructive to compare the CR proton to the CR electron pressure
since protons and electrons are subject to different cooling mechanisms due to
their large mass difference. The CR proton cooling timescale is generally
larger than that of CR electrons such that protons accumulate within the ICM on
a Hubble timescale and maintain a comparatively smooth distribution over the
cluster volume (top panels of Fig.~\ref{fig:g72_CR}). This implies that the CR
proton pressure traces the time integrated non-equilibrium activities of a
cluster and is only modulated by recent dynamical activities \citepalias[see
also][ for average values of the relative CR energy in different dynamical
cluster environments]{2007MNRAS...378..385P}.  In contrast, the pressure of
primary CR electrons resembles the current dynamical, non-equilibrium activity
of the forming structure and results in an inhomogeneous and aspherical spatial
distribution. To underpin this argument, in the bottom panels of
Fig.~\ref{fig:g72_CR}, we show the mass-weighted CR electron pressure relative
to the total pressure $X_\CRe = P_\CRe / P_\rmn{tot}$.  On the left side,
we show the relative pressure of primary, shock-accelerated electrons while the
relative pressure of secondary electrons resulting from hadronic CR proton
interactions is shown on the right side. The CR electron pressure derives
from the respective equilibrium distribution functions that balance
acceleration and cooling processes as laid out in Appendices~\ref{sec:fe_eq}
and \ref{sec:fe_had,eq}. Note that the colour scale for the relative pressure
of {\em primary CR electrons} spans exactly two orders of magnitude (like in
the case of CR protons), peaking in the dilute WHIM at roughly 3 per cent
rather than 30 per cent as in the case of CR protons. The relative CR electron
pressure $X_{\CRe,\rmn{prim}}$ decreases towards clusters and groups due to
larger Coulomb losses and smaller shock acceleration efficiencies within
collapsed objects as in the case of CR protons.  Interestingly,
$X_{\CRe,\rmn{prim}}$ within galaxies is suppressed by roughly one order of
magnitude with respect to the ambient intergalactic medium in which the galaxy
resides due to large Coulomb losses. This is quite different from CR protons
that acquire equipartition with the thermal gas inside galaxies.  In general,
the spatial variations of $X_{\CRe,\rmn{prim}}$ are larger than in the case of
protons, showing that the CR electron pressure indeed reflects the active
dynamical structure formation activities mediated by shock waves.  In contrast,
the mass weighted relative pressure of {\em secondary CR electrons} is shown on
a colour scale that spans seven orders of magnitude, due to the high dynamic
range of this quantity. The CR electron pressure is proportional to the number
densities of CR protons and of the gas $P_{\CRe,\rmn{sec}} \propto n_\CR
n_\rmn{N}$, causing $X_{\CRe,\rmn{sec}}$ to peak towards densest structures and
thus filling in the diminishing primary CR electron pressure inside dense
structures such as galaxies. As a word of caution, we do not account for the
re-acceleration of CR electrons e.g.~via resonant pitch angle scattering by
compressible magneto-{hy\-dro\-dy\-nam\-ical} (MHD) modes neither do we account
for a previously injected and aged electron population which could change the
presented picture. Further work is required to elucidate these electron
components in simulations.

\subsection{Radio synchrotron emission}
\label{sec:Radio_synchrotron}

\subsubsection{Projected radio maps}
\label{sec:radio_proj}

\begin{figure*}
\begin{center}
  \begin{minipage}[t]{0.495\textwidth}
    \centering{\it \large Radio web (primary CRes, 150 MHz):}
  \end{minipage}
  \hfill
  \begin{minipage}[t]{0.495\textwidth}
    \centering{\it \large Central radio halos (secondary CRes, 150 MHz):}
  \end{minipage}
\resizebox{0.5\hsize}{!}
{\includegraphics{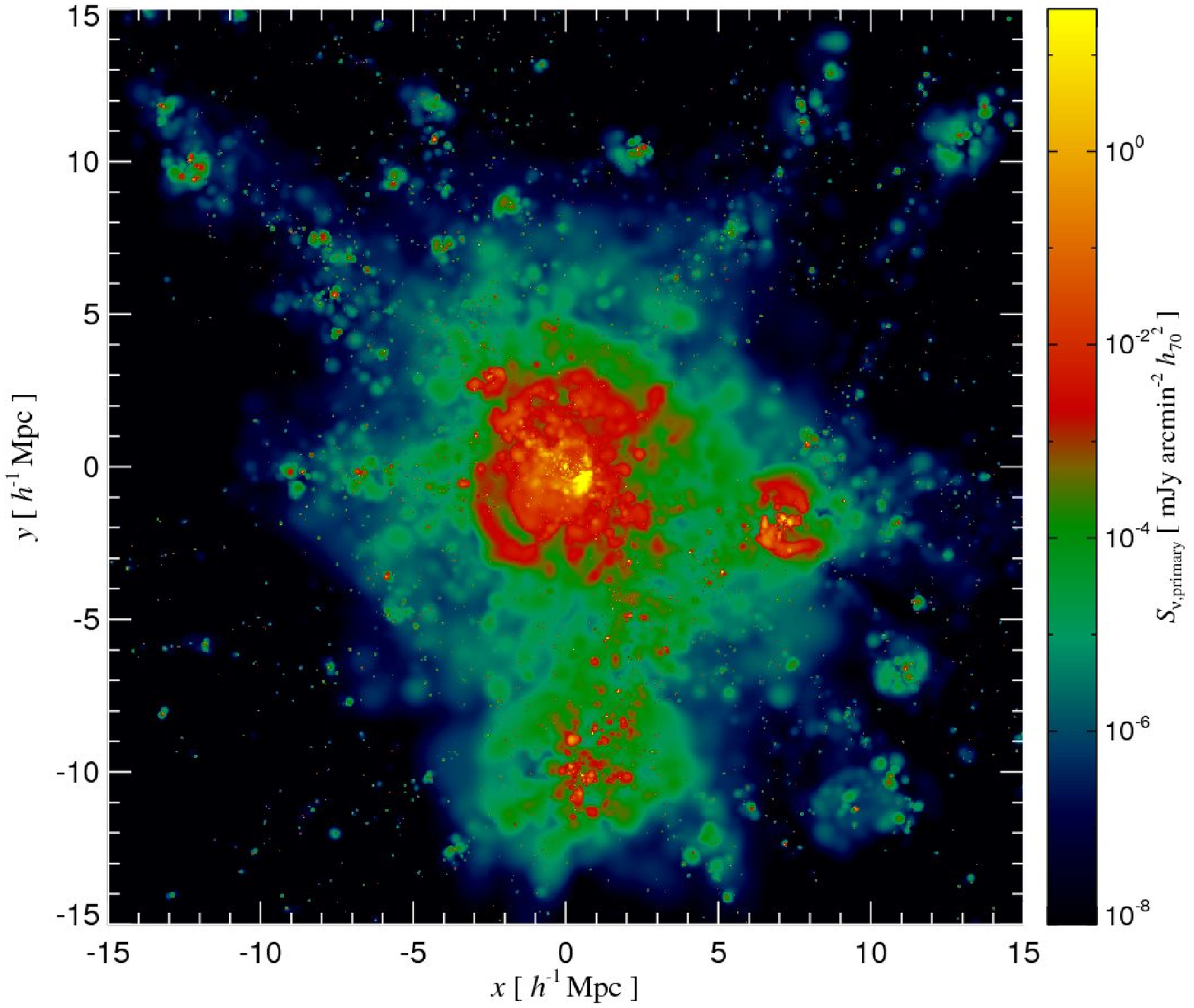}}%
\resizebox{0.5\hsize}{!}
{\includegraphics{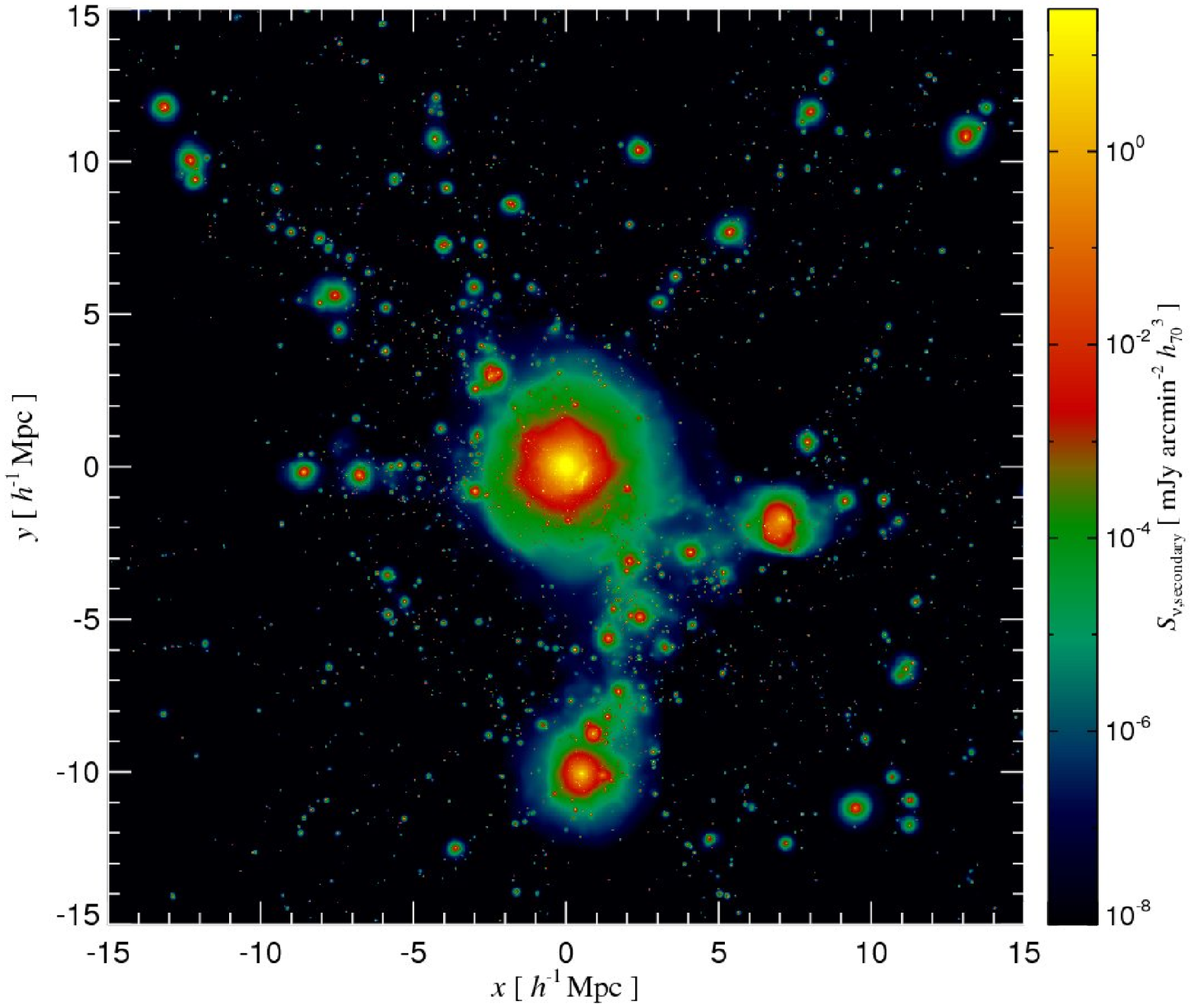}}\\
  \begin{minipage}[t]{0.495\textwidth}
    \centering{\it \large Total radio synchrotron emission, 150 MHz:}
  \end{minipage}
  \hfill
  \begin{minipage}[t]{0.495\textwidth}
    \centering{\it \large Radio spectral index map:}
  \end{minipage}
\resizebox{0.5\hsize}{!}
{\includegraphics{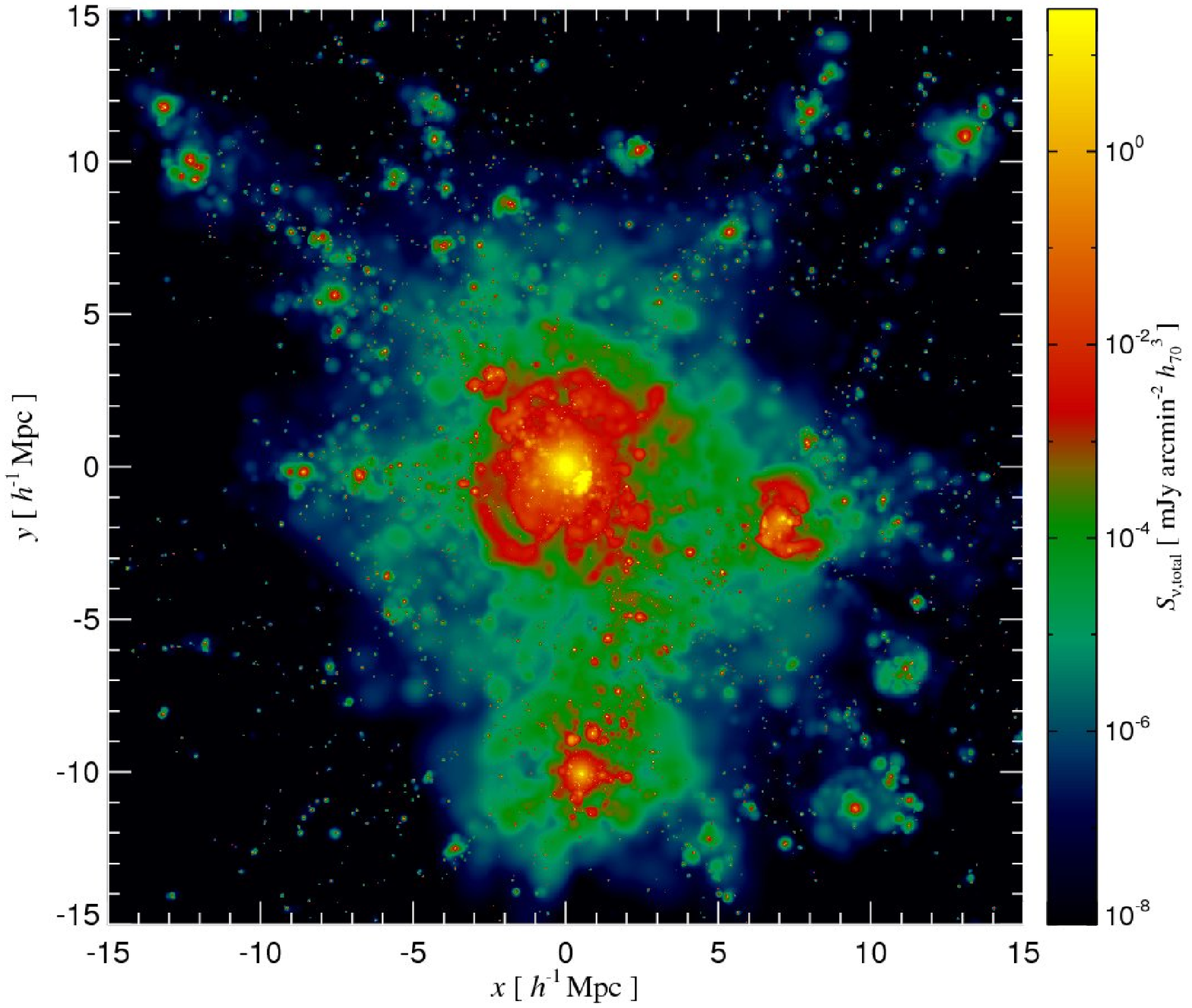}}%
\resizebox{0.5\hsize}{!}
{\includegraphics{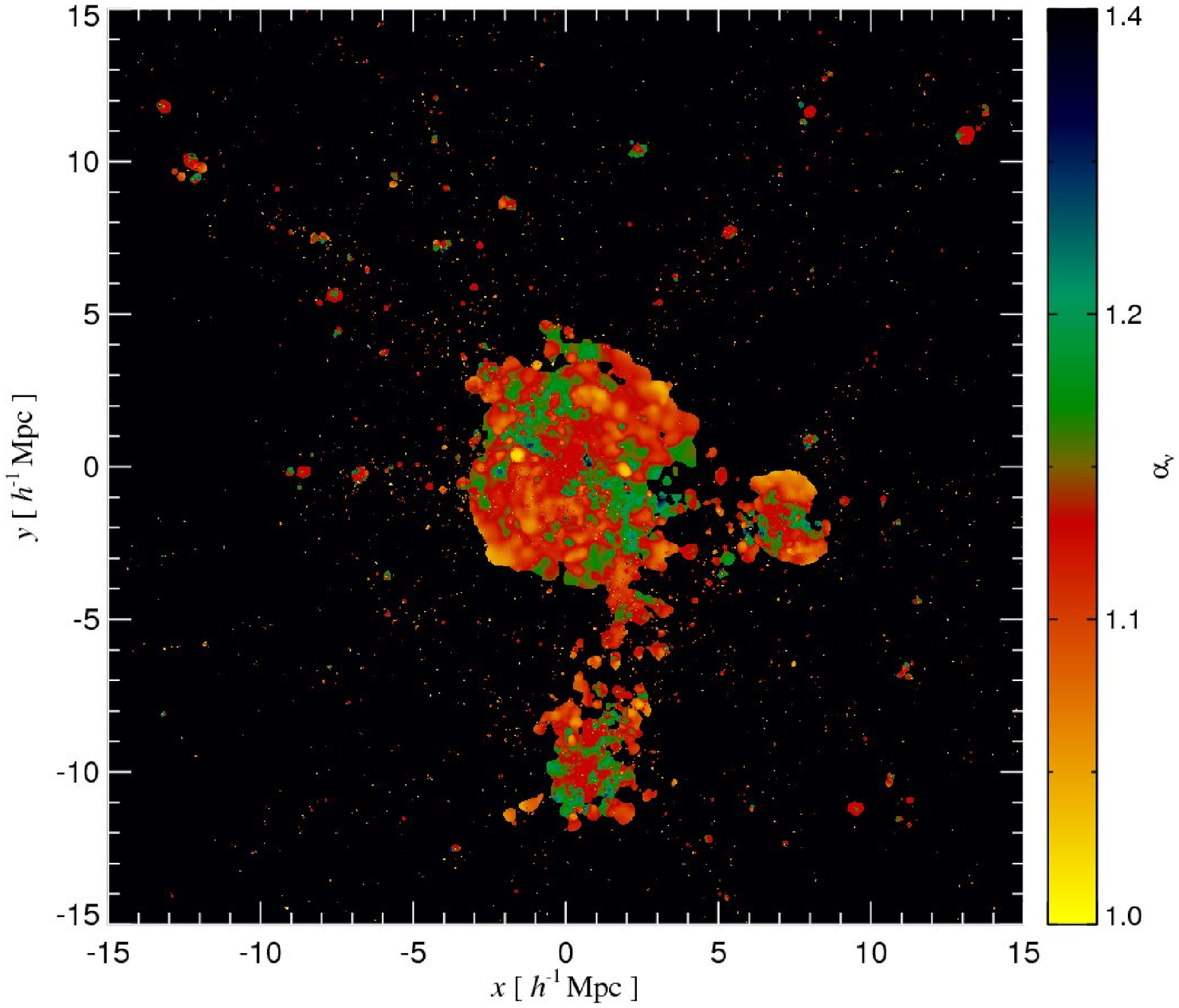}}\\
\end{center}
  \caption{The large-scale ``radio web'' at 150 MHz of the super-cluster region
    of our Coma-like cluster that experienced a recent merger (g72a) in our
    radiative simulation (model S2). We show the synchrotron emission of {\em
      primary CR electrons} that were accelerated directly at structure
    formation shocks (top left side) as well as the radio emission of {\em
      secondary CR electrons} that results from hadronic CR proton
    interactions with ambient gas protons (top right side).  The bottom left
    panel shows the {\em giant radio halo} emission of this cluster that is
    characterised in the centre by the regular smooth morphology of the
    secondary radio emission. At larger radii, we observe a transition to the
    the irregularly shaped primary radio ``gischt'' emission with a prominent
    radio relic to the lower right of the cluster.  The radio spectral index
    between 150 MHz and 1.4 GHz (bottom right panel) shows larger variations
    in the peripheral cluster regions. These are caused by projecting the
    radio emission from inhomogeneously accelerated primary CR electrons and
    reflect the strong variation of the Mach numbers of structure formation
    shocks at the outer cluster regions. }
  \label{fig:radio-processes}
\end{figure*}

\begin{figure*}
\begin{center}
  \begin{minipage}[t]{0.495\textwidth}
    \centering{\it \large Radio emission (1.4 GHz, radiative sim., S2):}
  \end{minipage}
  \hfill
  \begin{minipage}[t]{0.495\textwidth}
    \centering{\it \large Radio emission (15 MHz, radiative sim., S2):}
  \end{minipage}
\resizebox{0.5\hsize}{!}
{\includegraphics{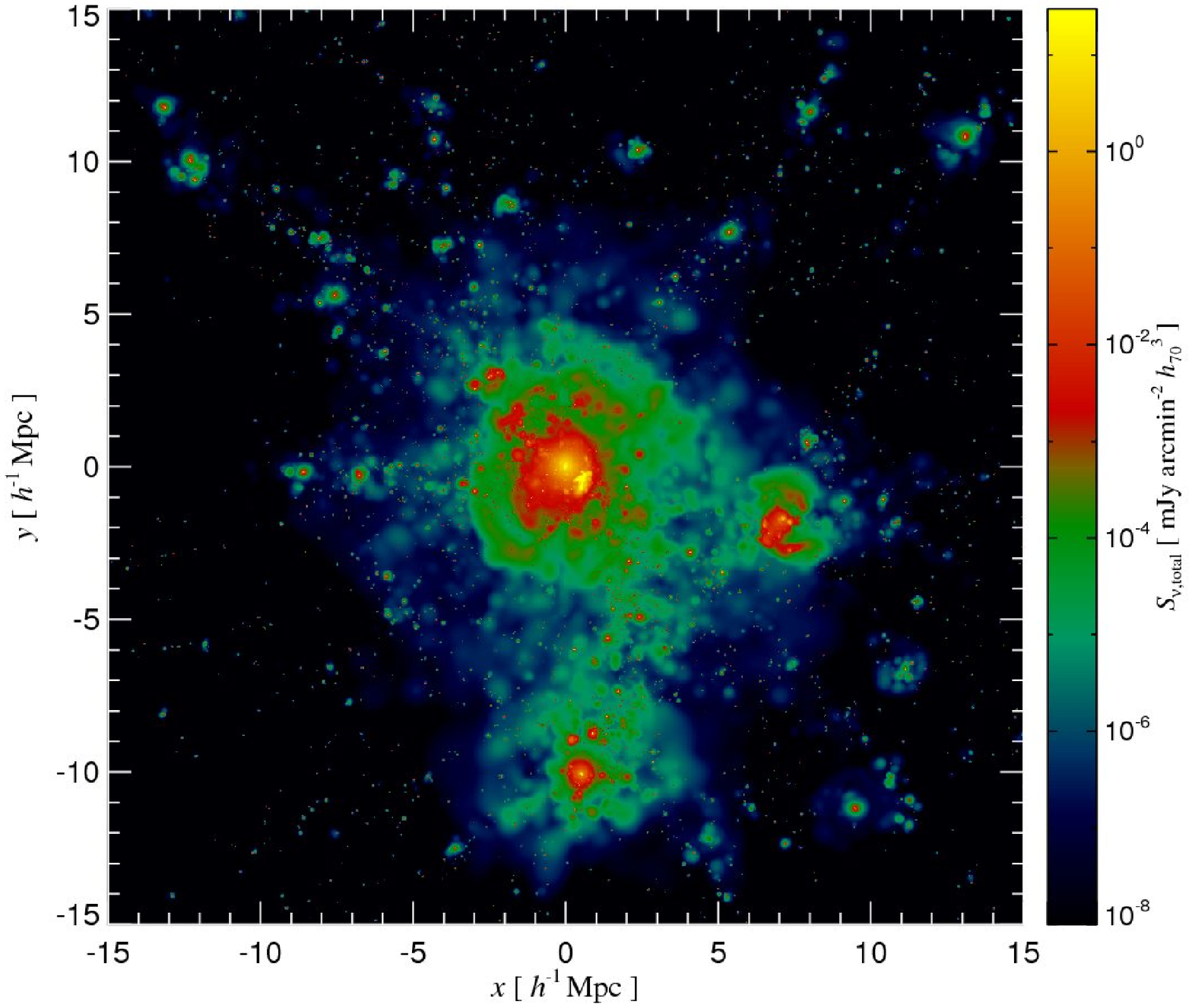}}%
\resizebox{0.5\hsize}{!}
{\includegraphics{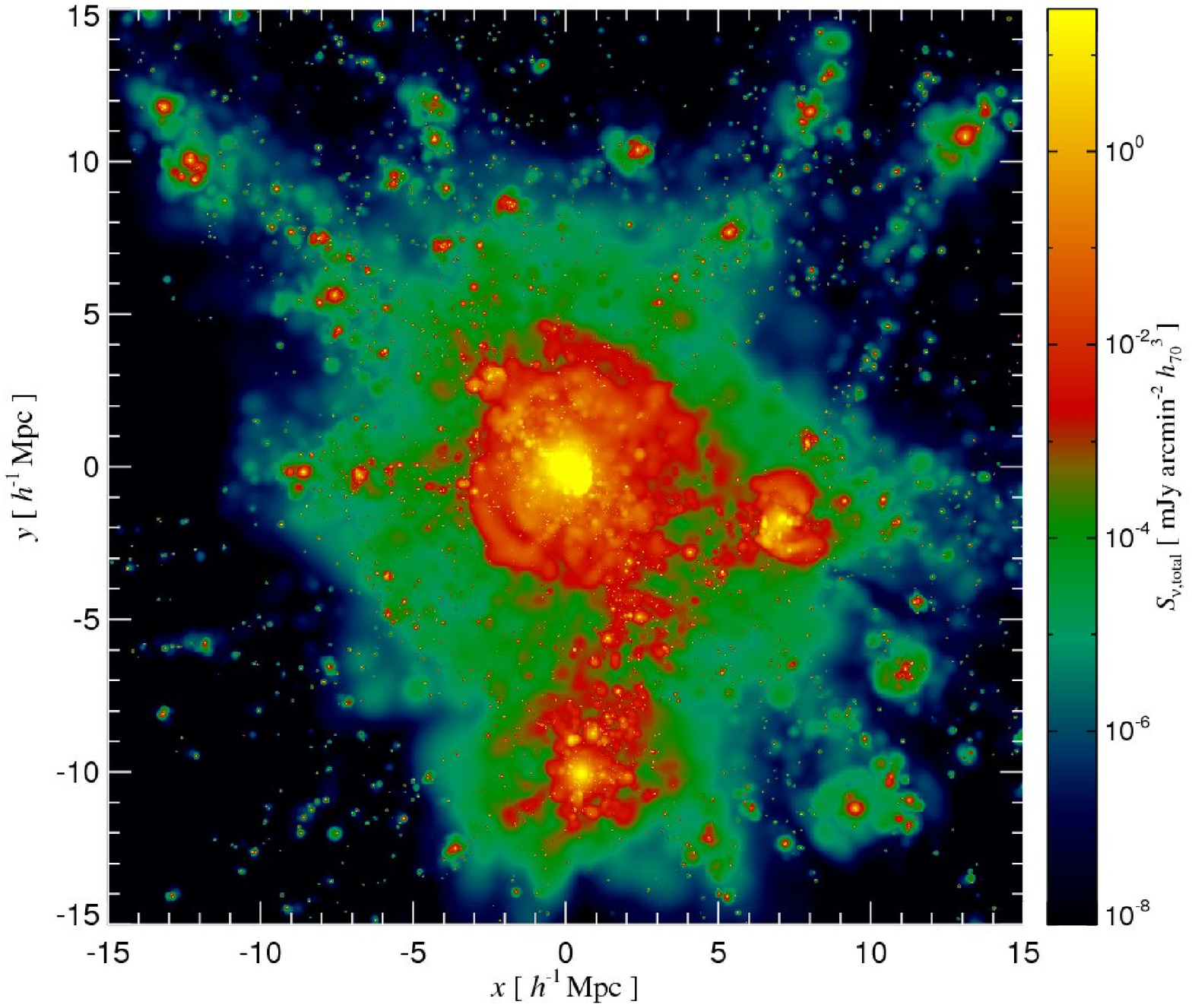}}\\
  \begin{minipage}[t]{0.495\textwidth}
    \centering{\it \large Radio emission (150 MHz, non-radiative sim., S1):}
  \end{minipage}
  \hfill
  \begin{minipage}[t]{0.495\textwidth}
    \centering{\it \large Radio emission (150 MHz, weaker B-decay, S1):}
  \end{minipage}
\resizebox{0.5\hsize}{!}
{\includegraphics{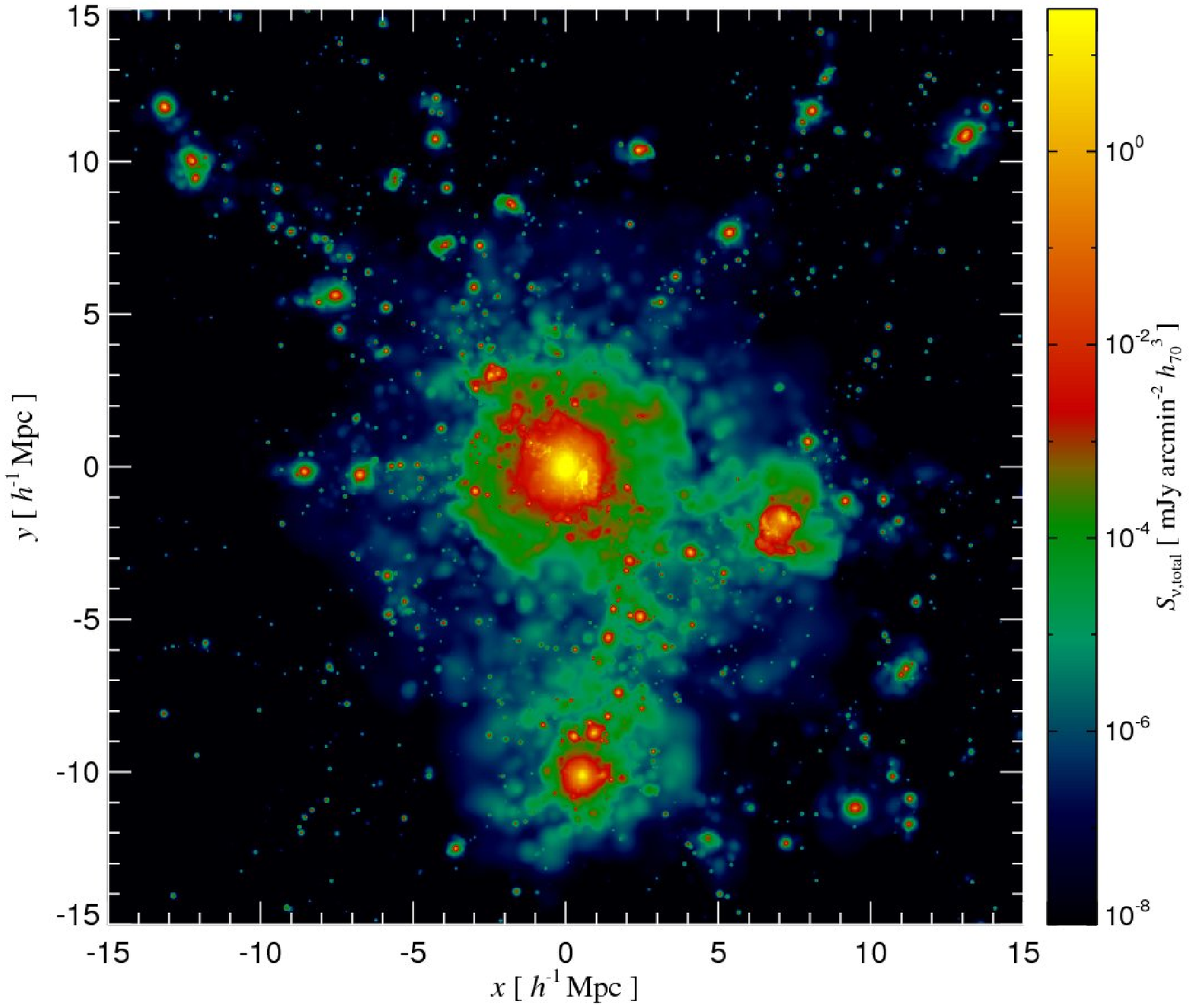}}%
\resizebox{0.5\hsize}{!}
{\includegraphics{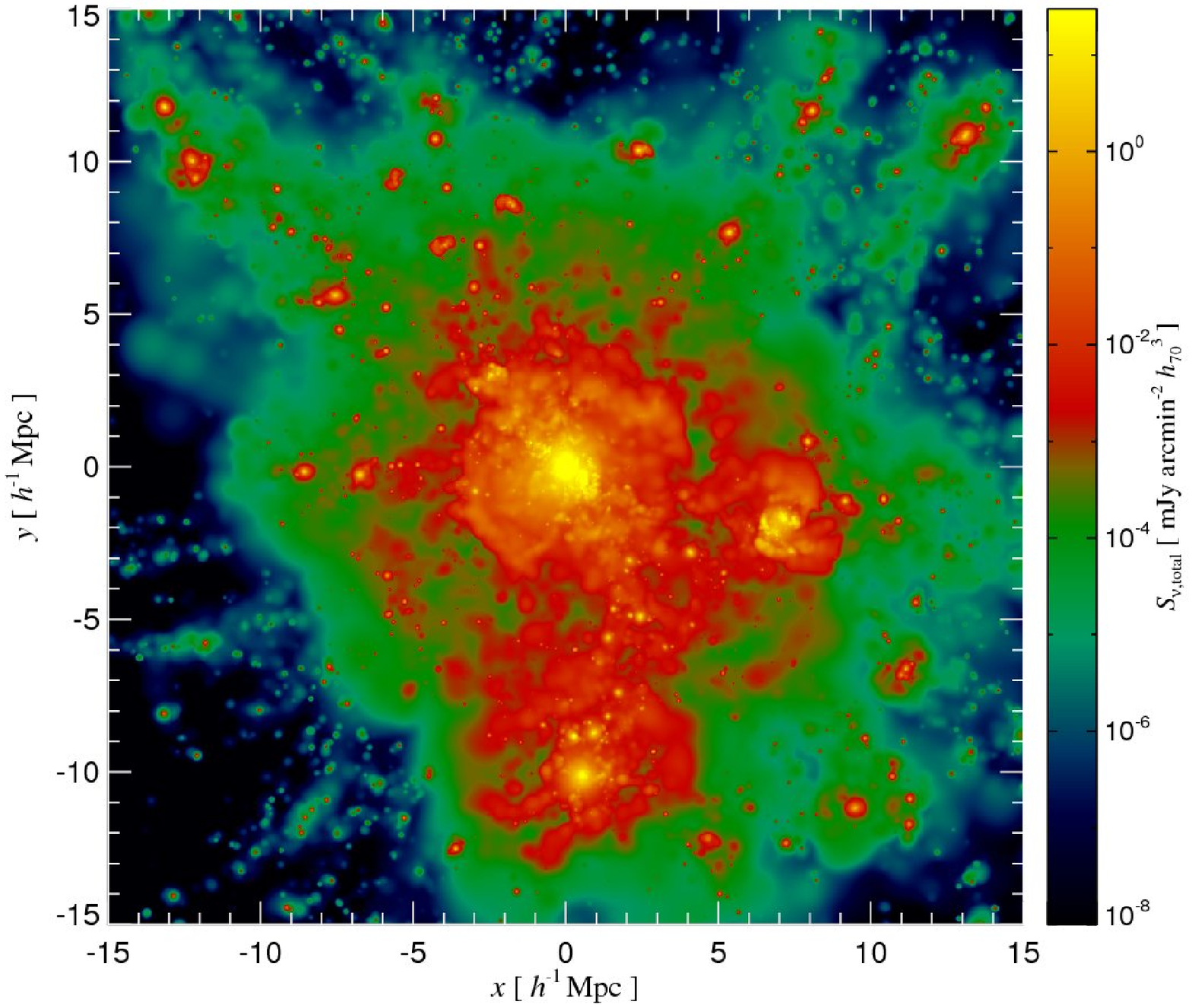}}\\
\end{center}
  \caption{Dependence of the primary and secondary radio synchrotron emission
    on the observing frequency, the model for the magnetic field, as well as
    the type of simulated gas physics (radiative versus non-radiative). The
    top panels show the synchrotron radiation at 1.4 GHz and 15 MHz in our
    radiative simulation (S2) assuming a simple scaling of the magnetic energy
    density with the thermal energy density, $\eps_B \propto \eps_\th$. This
    demonstrates the potential of low-frequency radio arrays in studying
    non-thermal properties of the inter-galactic medium.  The bottom left
    panel shows the total radio emission at 150 MHz in our non-radiative
    simulation (S1) with the same magnetic model.  The bottom right panel
    again shows the total radio emission at 150 MHz in our non-radiative
    simulation, however, with a shallower magnetic decline, $\eps_B \propto
    \eps_\th^{0.5}$. For convenience, the colour scale is the same in all
    panels such that the emission at 15 MHz in the central cluster region is
    highly saturated.}
  \label{fig:radio-dependencies}
\end{figure*}

Figure~\ref{fig:radio-processes} shows the large-scale ``radio web''of the
super-cluster region of our Coma-like cluster that experienced a recent
merger.  The radio emission is computed assuming a simple scaling model for
the magnetic field of
\begin{equation}
  \label{eq:magnetic_scaling}
  \eps_B = \eps_{B,0} \left(\frac{\eps_\rmn{th}}{\eps_\rmn{th,0}}\right)^{2
  \alpha_B},
\end{equation}
where the central magnetic energy density $\eps_{B,0}$ and $\alpha_B$
are free parameters in our model, and $\eps_\rmn{th,0}$ denotes the
thermal energy density at the cluster centre. It is motivated by
turbulent dynamo theories of the growth of magnetic field strength
that are typically saturating at a level proportional to the turbulent
energy density or the thermal energy density
\citep[e.g.,][]{2003PhRvL..90x5003S, 2006PhPl...13e6501S}.  This
allows us to explore the unknown behaviour of the large scale magnetic
field parametrically (cf. Appendix~\ref{sec:magfield} for more
discussion). Our standard model (also adopted in
Fig.~\ref{fig:radio-processes}) assumes a central magnetic field
strength of $B_0 = 10 \,\umu$G and $\alpha_B = 0.5$ which implies a
constant ratio of thermal-to-magnetic pressure of 20 in our Coma-like
cluster.

In the top panels of Fig.~\ref{fig:radio-processes}, we separately show the
synchrotron emission of {\em primary CR electrons} that were accelerated
directly at structure formation shocks as well as the radio emission of {\em
  secondary CR electrons} that results from hadronic CR proton interactions
with ambient gas protons. The combined radio synchrotron emission (shown in
the bottom left panel) shows that the morphologically smooth secondary
component dominates the radio emission of the central cluster regions. In
contrast, the irregularly shaped primary radio relic emission dominates in the
cluster periphery and the super-cluster region that is believed to host the
warm-hot intergalactic medium (WHIM). These observations are supported by the
radio spectral index map between 150 MHz and 1.4 GHz (bottom right panel)
which shows larger variations in the peripheral cluster regions. These are
caused by projecting the radio emission from inhomogeneously accelerated
primary CR electrons and reflect the strong variation of the Mach numbers of
structure formation shocks at the outer cluster regions. Based on these
findings, we put forward a new {\em unified scheme} for the generation of giant
radio halos as well as radio mini-halos.  The {\em giant radio halo} emission
in merging clusters shows a transition from the secondary radio emission in
the centre to the dominant primary emission component at the outer parts of
radio halos.  Gravitational energy, that is associated with the merger, is
virialised by a morphologically complex network of strong shock waves in the
cluster outskirts. This induces an irregular radio `gischt' emission in the
cluster periphery that represents radio synchrotron radiation emitted from
shock-accelerated electrons.  Our simulated radio emission maps of relaxed
cool core clusters show a significantly reduced level of this primary emission
component such that the diffuse radio emission in these systems is solely
determined by the secondary radio emission, producing a {\em radio mini-halo}.
Note that our simple magnetic model does not take into account the adiabatic
compression of magnetic fields during the formation of a cool core. This
effect should furthermore decrease the emission size of radio mini-halos
making it comparable to the cool core region.

Closer inspection of the primary radio emission map (top left panel in
Fig.~\ref{fig:radio-processes}) shows a bright {\em radio relic} on the lower
right with respect to the cluster centre at a distance of $r\simeq 0.6
h^{-1}\mbox{ Mpc}$. This is caused by an outgoing merger shock wave that
steepens as it reaches the shallower peripheral cluster potential.  Further
outwards at a distance of 2 and $3\,h^{-1}\mbox{ Mpc}$ to the cluster centre,
there is a class of tangentially curved radio relics visible in orange and
red. These are uniquely associated with strong shock waves as can be inferred
by comparing the primary radio emission to the dissipated energy at shock
waves (shown as brightness in the bottom left panel of Fig.~\ref{fig:g72}).
The statistical study of the radio emission at different frequencies and
Faraday rotation of these objects will enable us to investigate
non-equilibrium processes of virialisation including the acceleration of
cosmic rays, the growth of magnetic fields, and kinetic energy in bulk motions
that are expected to source turbulence in clusters.  Comparing the emission
level of the projected radio surface brightness maps to the LOFAR point source
sensitivity of $0.25 \mbox{ mJy / (arcmin hour)}$ at $\nu = 120$~MHz shows
that moderately long exposures of super cluster regions have the potential to
detect the large-scale ``radio web'' and to study the magnetic field on
Mpc-scales that is woven into the web.  Ongoing work that includes simulated
mock observations for radio interferometers studies associated questions in
greater detail (Battaglia et al.{\ }in prep.).

\begin{figure*}
\begin{center}
  \begin{minipage}[t]{0.485\textwidth}
    \centering{\it \large Primary vs. secondary radio emission:}
  \end{minipage}
  \hspace{0.02\textwidth}
  \begin{minipage}[t]{0.485\textwidth}
    \centering{\it \large Influence of sim. physics and magnetic models:}
  \end{minipage}
\resizebox{0.5\hsize}{!}
{\includegraphics{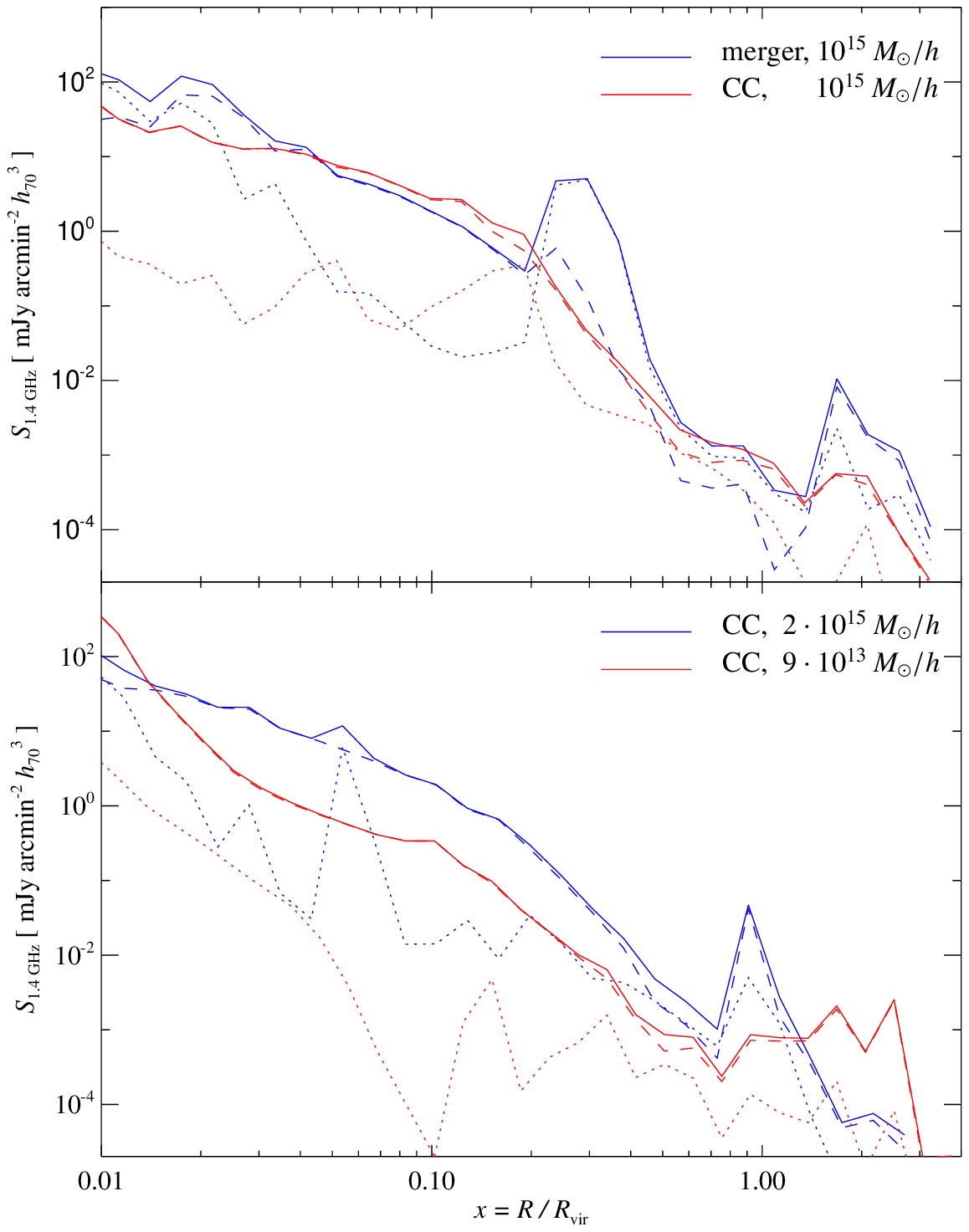}}%
\resizebox{0.5\hsize}{!}
{\includegraphics{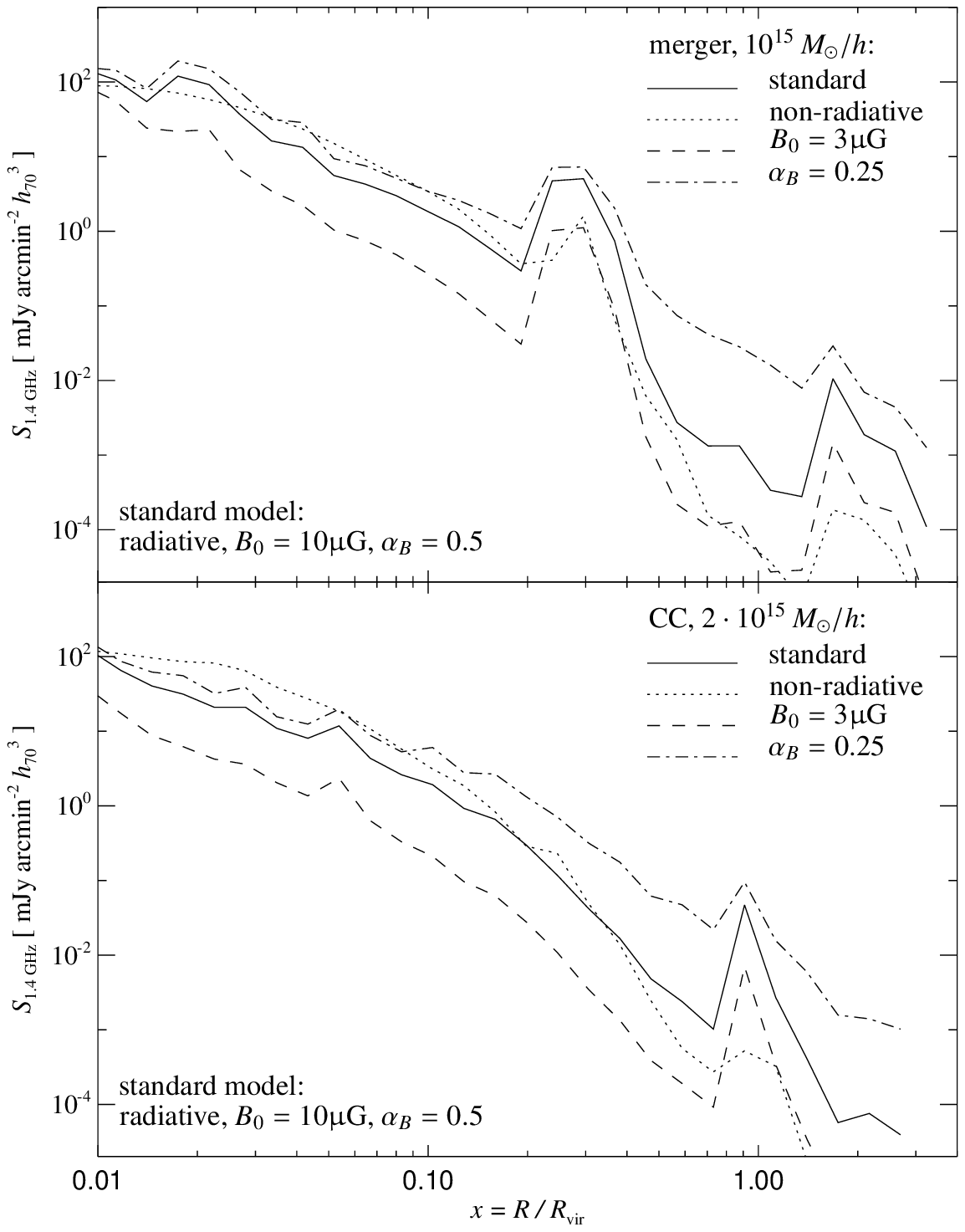}}\\
\end{center}
\caption{Azimuthally averaged radio synchrotron brightness profiles at $\nu =
  1.4$~GHz. The left side compares the synchrotron emission of {\em
    primary CR electrons} that were accelerated directly at structure formation
  shocks (dotted lines) with that of {\em secondary CR electrons} that result
  from hadronic CR proton interactions with ambient gas protons (dashed line)
  while the solid line shows the sum of both emission components. The upper
  panel compares a post-merging cluster (g72a) with a cool core (CC) cluster (g51)
  of the same mass and the bottom panel compare the radio emission of two
  differently sized CC clusters (g8a versus g676).  The right side shows
  the influence of different simulation and magnetic field models on the total
  synchrotron emission profiles for the post-merging cluster g72a (upper panel) and
  the CC cluster g8a (bottom panel).  The standard model uses radiative gas
  physics (S2), has a central magnetic field strength of $B_0 = 10\,\umu$G,
  and scales as $\eps_B^{} \propto \eps_\th^{2 \alpha_B}$ with
  $\alpha_B=0.5$. The other lines result from varying each of these assumptions
  separately leaving the others unchanged.  The radio emission in our
  non-radiative simulation (S1, dotted lines) declines faster.  The emission
  profiles for a lower central magnetic field strength (dashed lines) and with
  a weaker magnetic decline (dash-dotted lines) illustrate the uncertainty in
  the simulated radio profiles.  }
  \label{fig:synchro_profiles}
\end{figure*}

Figure~\ref{fig:radio-dependencies} shows the dependence of the primary and
secondary radio synchrotron emission on the observing frequency, the model for
the magnetic field, as well as the type of simulated gas physics (radiative
versus non-radiative).  The top panels show the high- and low-frequency radio
emission (1.4 GHz and 15 MHz) in our radiative simulation (model S2) using our
standard model for the magnetic field. This demonstrates the potential of
low-frequency radio arrays in studying non-thermal properties of the
inter-galactic medium especially since the associated radio spectrum is
steeper compared to that of the Galactic foreground emission. This will allow
us to address questions such as the existence and properties of the WHIM and
the existence and origin of large scale magnetic fields.  The bottom left
panel shows the total radio emission at 150 MHz in our non-radiative
simulation (model S1) assuming our standard parameters for the magnetic field,
and should be compared to the same panel in Fig.~\ref{fig:radio-processes}.
The level of the primary radio emission in the cluster periphery and the
super-cluster region is reduced in the non-radiative simulation (model S1)
compared to the radiative case (model S2). Some relics in the bottom panel
(model S1) that are at distances of $\gtrsim 2\,h^{-1}\mbox{ Mpc}$ to the
cluster centre even show radio emission at the level that is comparable to
that in the panel above (model S2) despite the lower frequency that should
provide a flux level that is increased by an order of magnitude assuming
$\alpha_\nu \simeq 1$.  The reason for this stems from the larger shock
strength (higher Mach numbers) of characteristic shocks that dissipate
gravitational energy into thermal energy in radiative simulations
\citepalias{2007MNRAS...378..385P}. The enhanced acceleration efficiency of
CRs at stronger shocks leads to the increased primary radio emission in
radiative simulations compared to the non-radiative case.  The bottom right
panel again shows the total radio emission at 150 MHz in our non-radiative
simulation, however, with a shallower magnetic decline, $\alpha_B = 0.25$
which results in $\eps_B \propto \eps_\th^{0.5}$. Although the radial decline
of this model for the magnetic field might be almost too shallow, it serves
for illustrative purposes demonstrating that low-frequency radio arrays in
combination with high-resolution simulations can tightly constrain the large
scale behaviour of the magnetic field.

\subsubsection{Radio emission profiles}
\label{sec:radio_profiles}

\begin{figure*}
\resizebox{0.5\hsize}{!}
{\includegraphics{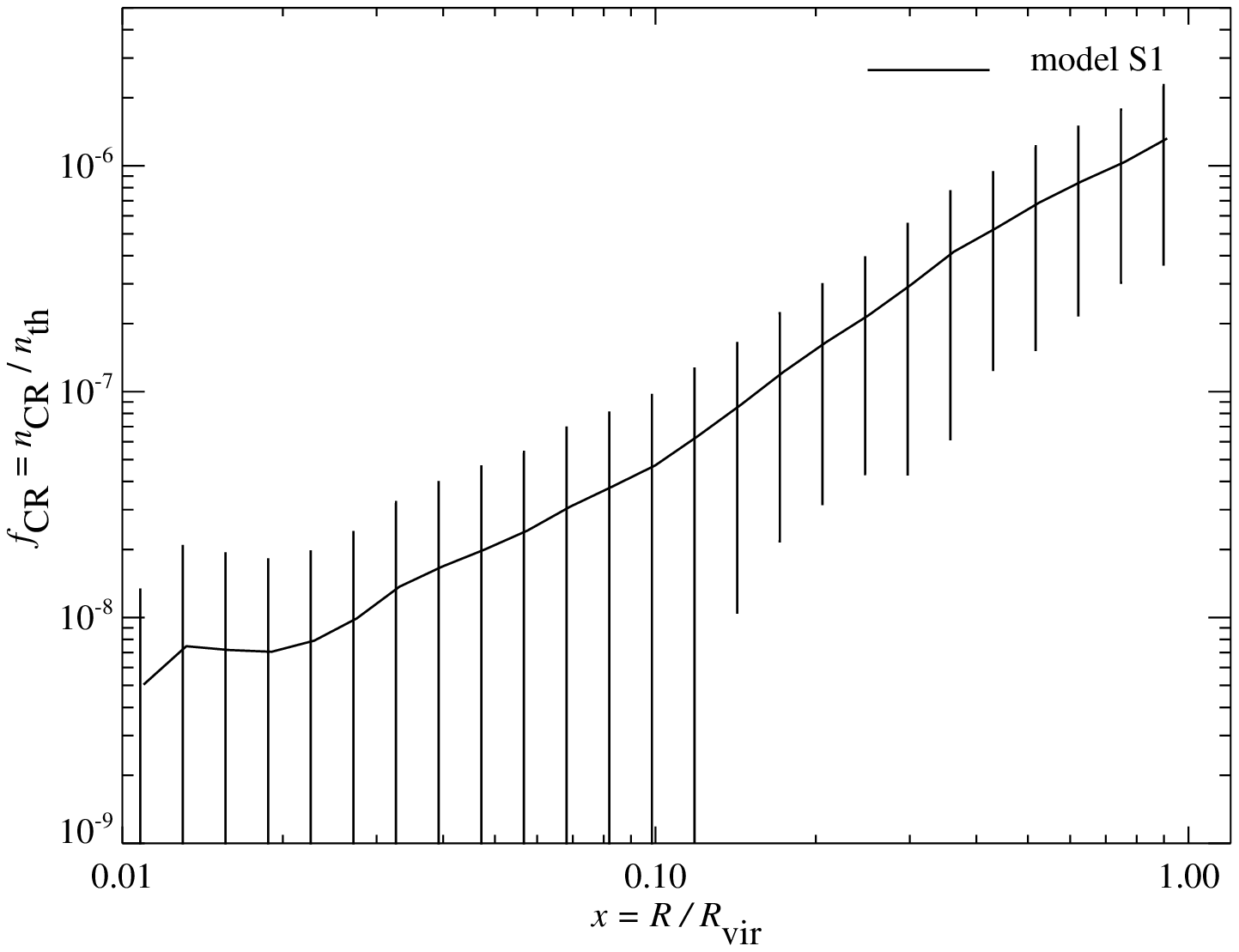}}%
\resizebox{0.5\hsize}{!}
{\includegraphics{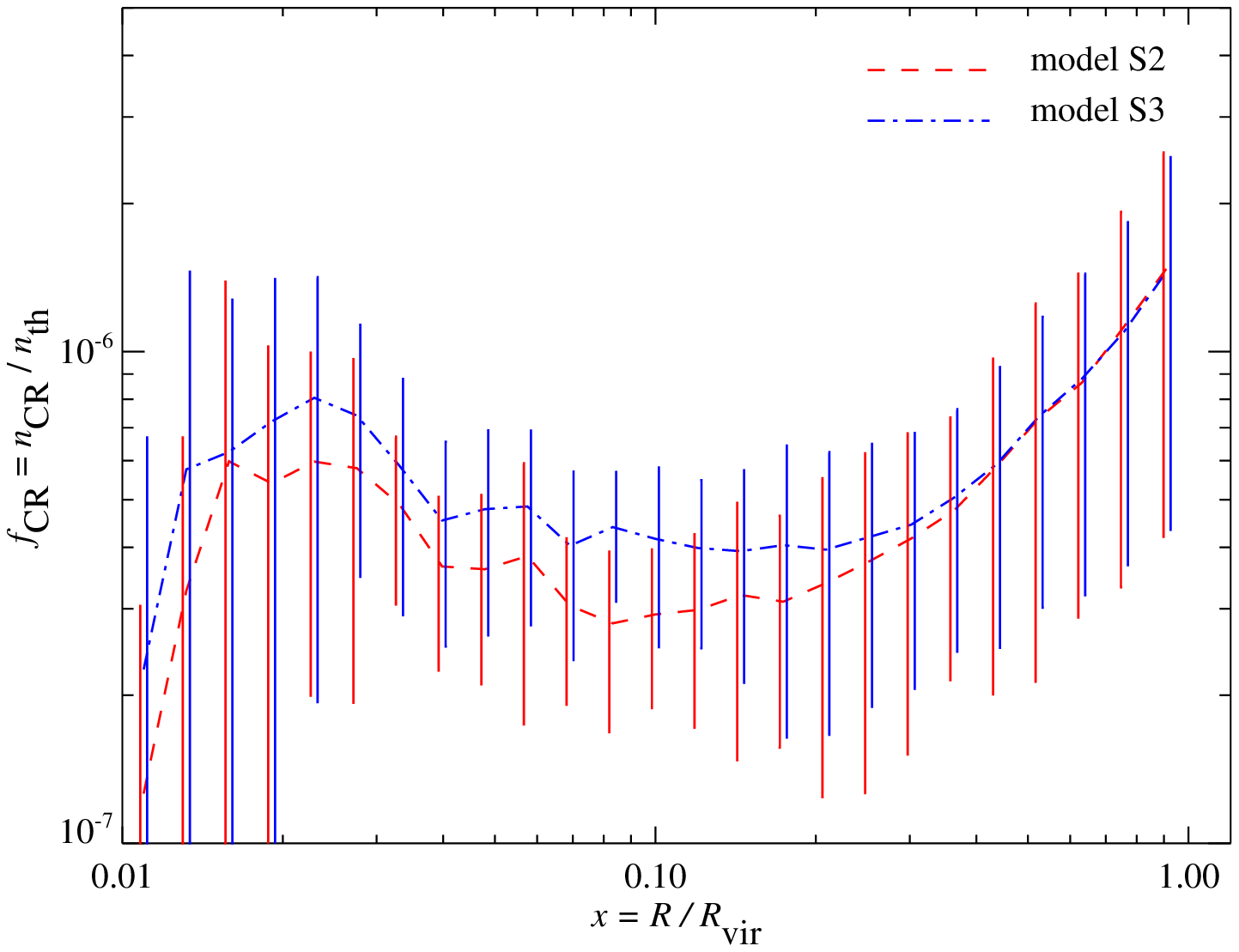}}
\caption{Average profiles for the spherically averaged CR fraction $f_\CR =
  n_\CR / n_\rmn{th}$ of our sample of all 14 clusters at redshift $z=0$.  The
  error bars represent the standard deviation from the sample mean.  The
  increase of $f_\CR$ with radius is independent of the modelled CR physics (S1
  -- solid, S2 -- dashed, S3 -- dash-dotted) and continues beyond the virial
  radius suggesting that this is a generic property of CR physics. The increase
  of $f_\CR$ in our radiative models for small radii $r<0.1 R_\rmn{vir}$ is due
  to the short cooling timescale of the gas compared to that of CR protons
  \citepalias[cf.][]{2007MNRAS...378..385P}. Note the different axes scales in
  both panels. }
  \label{fig:fCR}
\end{figure*}

Radio synchrotron profiles allow us to confirm and quantify the proposed
unified scheme for the cluster radio halo emission.

{\bf Primary versus secondary radio emission:} The left side of
Fig.~\ref{fig:synchro_profiles} compares the synchrotron emission of {\em
  primary CR electrons} that were accelerated directly at structure formation
shocks with that of {\em secondary CR electrons} that result from hadronic CR
proton interactions with ambient gas protons.  In cool core clusters, the
azimuthally averaged secondary radio emission dominates the primary emission
component for radii $r < 3 R_\rmn{vir}$. The smooth secondary component
typically falls of at a radius $r \simeq 0.2 R_\rmn{vir}$ which resembles the
characteristics of observed radio mini halos as observed e.g. in the Perseus
cluster \citep{1990MNRAS.246..477P} or RX~J1347.5-1145
\citep{2007A&A...470L..25G}. Our model predicts diffuse secondary radio
emission in virtually every cool core cluster.

Interestingly, our post-merging cluster g72a shows a transition from the
secondary to the primary radio emission component towards the outer cluster
regions triggered by the dynamical merger activity with strong shock waves
traversing the cluster in order to thermalize the gas. These shock waves
steepen as they break on the shallower peripheral cluster potential.  The
associated increase in the primary radio emission comes hand in hand with a
stronger variation of the radio spectral index towards the cluster periphery
(cf.~Fig.~\ref{fig:radio-processes}).  In the particular case of g72a, the
resulting radio halo profile reaches out to $r \simeq 0.45 R_\rmn{vir}$ which
corresponds to the observed size of $R_\rmn{max} \simeq 1 \mbox{ Mpc} /h_{70}$
of the Coma radio halo at $\nu = 1.4$~GHz \citep{1997A&A...321...55D,
  2002ApJ...567..716R}.  We verified that the transition from the secondary to
the primary radio emission in our simulated giant radio halos is independent
on the chosen projection and a generic prediction for merging clusters.  Our
simple scaling model for the magnetic field of
Eqn.~(\ref{eq:magnetic_scaling}) does clearly not include non-equilibrium
effects related to the growth of the magnetic field.  The enhancement of the
magnetic field strength through turbulent dynamo processes will saturate on a
level which is determined by the strength of the magnetic back-reaction
\citep[e.g.,][]{2003PhRvL..90x5003S} and is typically a fraction of the
turbulent energy density. Thus, in a real cluster, the strong shocks at the
cluster periphery are expected to drive turbulence and strong shear motions
which should in turn lead to a stronger magnetic field amplification. Our
adopted scaling of the magnetic field with the thermal energy density might
partially neglect these effects and should somewhat underestimate the
peripheral radio synchrotron emission.

{\bf Influence of magnetic parametrisation on radio emission:} The right side
of Fig.~\ref{fig:synchro_profiles} shows the influence of different simulations
and magnetic field models on the total synchrotron emission profiles for our
massive post-merging cluster and cool core cluster.  The emission profiles for
lower central magnetic field strength (dashed lines, $B_0 = 3\,\umu$G) shows a
small decrease of the central radio emission by a factor of two while it is
considerably suppressed by an order of magnitude towards larger radii. This is
due to the two distinctive regimes of a synchrotron emitting equilibrium
distribution of CR electrons (cf.~Fig.~\ref{fig:IC-synchro_B}). For field
strengths $B> 3\,\umu$G, the synchrotron flux is almost insensitive to the
field strength while it scales as $j_\nu \propto B^{\alpha\nu+1}$ for weaker
fields $B< 3\,\umu$G which are present at larger cluster radii.  The emission
profile for a weaker magnetic decline (dash-dotted lines, $\alpha_B=0.25$) is
more extended than our standard model, as expected.  Current cosmological MHD
SPH simulations \citep[e.g.,][]{2001A&A...378..777D} may not sufficiently
resolve small scale turbulent dynamo processes and large shear motions that are
thought to amplify the magnetic field in the coarsely sampled super-cluster
regions beyond the accretion shocks. The model with the weaker magnetic decline
is an attempt to parametrise those uncertainties.  Note that despite the
uncertainties in the parametrisation of the magnetic field and thus the overall
radio emission, the conclusions with respect to the different emission
components (primary versus secondary) and their emission characteristics remain
unchanged.

{\bf Influence of simulated physics on radio emission:} The radio emission in
our non-radiative simulations (S1, dotted lines) is much smoother and declines
faster since there is only a very weak transition to the primary component due
to the weaker shocks in the cluster periphery compared to the radiative
simulations (S2). It is surprising that the central radio emission in
simulation models S1 and S2 almost coincide despite the large difference of
the central CR fraction $f_\CR = n_\CR / n_\rmn{th}$ in both models
(cf.~Fig.~\ref{fig:fCR}).  At $r = 0.02 R_\rmn{vir}$ the CR fraction is two
orders of magnitude larger in our radiative simulations compared to our
non-radiative simulations. This can be understood by the self-regulated nature
of CR feedback.  The secondary synchrotron emission scales as $j_\nu \propto
n_\CR n_\rmn{gas} \propto f_\CR n_\rmn{gas}^2$ neglecting the weak additional
density dependence through the magnetic field in the synchrotron regime
(cf.~Fig.~\ref{fig:IC-synchro_B}).  The lower gas density in the radiative
simulations \citepalias[cf.~Figs.~3 and 5 in][]{2007MNRAS...378..385P} almost
exactly balances this difference of the CR fraction such that the resulting
secondary synchrotron emission level in the centre remains only slightly
modified. This is due to a combination of the following reasons. (1) The CR
cooling timescales due to Coulomb and hadronic interactions of CRs,
$\tau_\rmn{pp,Coul}\propto n_\rmn{gas}^{-1}$, is almost an order of magnitude
larger in our non-radiative simulations compared to our radiative case owing
to the central density difference. (2) A second sub-dominant effect is the
reduced depletion of the CR pressure in our radiative simulations due to
adiabatic compression of our composite of CRs and thermal gas which disfavours
the CR pressure relative to the thermal pressure.

Part of this density difference is reinforced in contemporary cosmological
radiative simulations that do not include feedback from AGN. This leads to the
well-known over-cooling problem which results in an overproduction of the
amount of stars, enhanced central gas densities, and too small central
temperatures compared to X-ray observations. The density enhancement at the
very centre and the associated star formation take place at the expense of the
surrounding ICM which ends up being less dense compared to its initial stage
before cooling set in. This hypothetical initial stage is realised by our
non-radiative simulations that does not take into account radiative cooling.
We show that the secondary CR emission (radio synchrotron, inverse Compton,
and pion decay induced $\gamma$-ray emission) within the framework of our CR
model is almost independent of those short-comings in the central cluster
regions. The difference of the radio emission at larger radii between our
models S1 and S2 however is a robust finding and primarily caused by the
difference of the primary radio emission. This difference is due to the on
average stronger shock waves that lead to more efficient CR electron
acceleration in our radiative simulations.

\subsubsection{Discussion of synchrotron polarisation}
\label{sec:radio_discussion}

\begin{figure*}
\begin{center}
  \begin{minipage}[t]{0.495\textwidth}
    \centering{\it \large Thermal X-ray emission:}
  \end{minipage}
  \hfill
  \begin{minipage}[t]{0.495\textwidth}
    \centering{\it \large Pion decay $\gamma$-ray emission ($E_\gamma>100$ MeV):}
  \end{minipage}
\resizebox{0.5\hsize}{!}
{\includegraphics{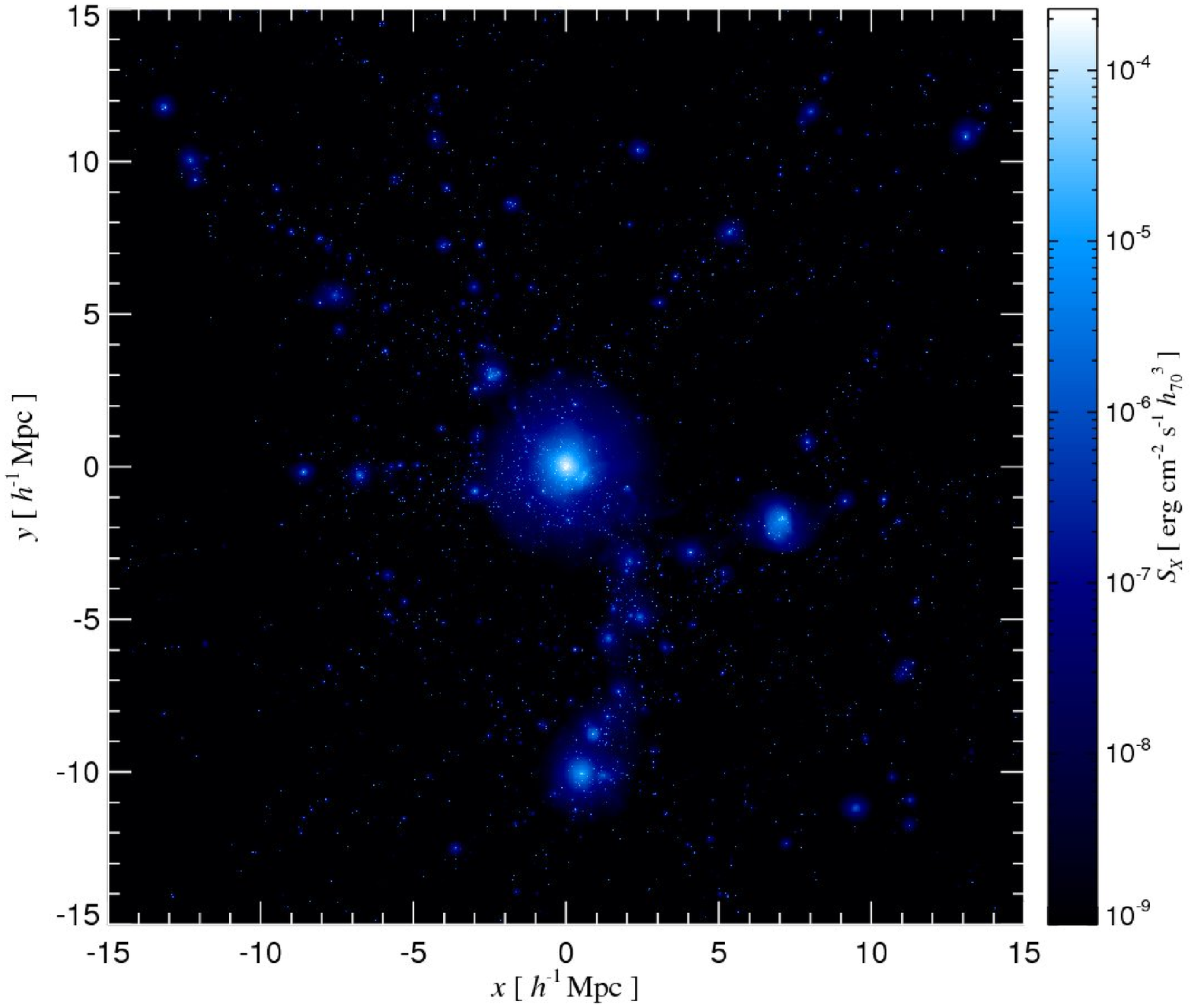}}%
\resizebox{0.5\hsize}{!}
{\includegraphics{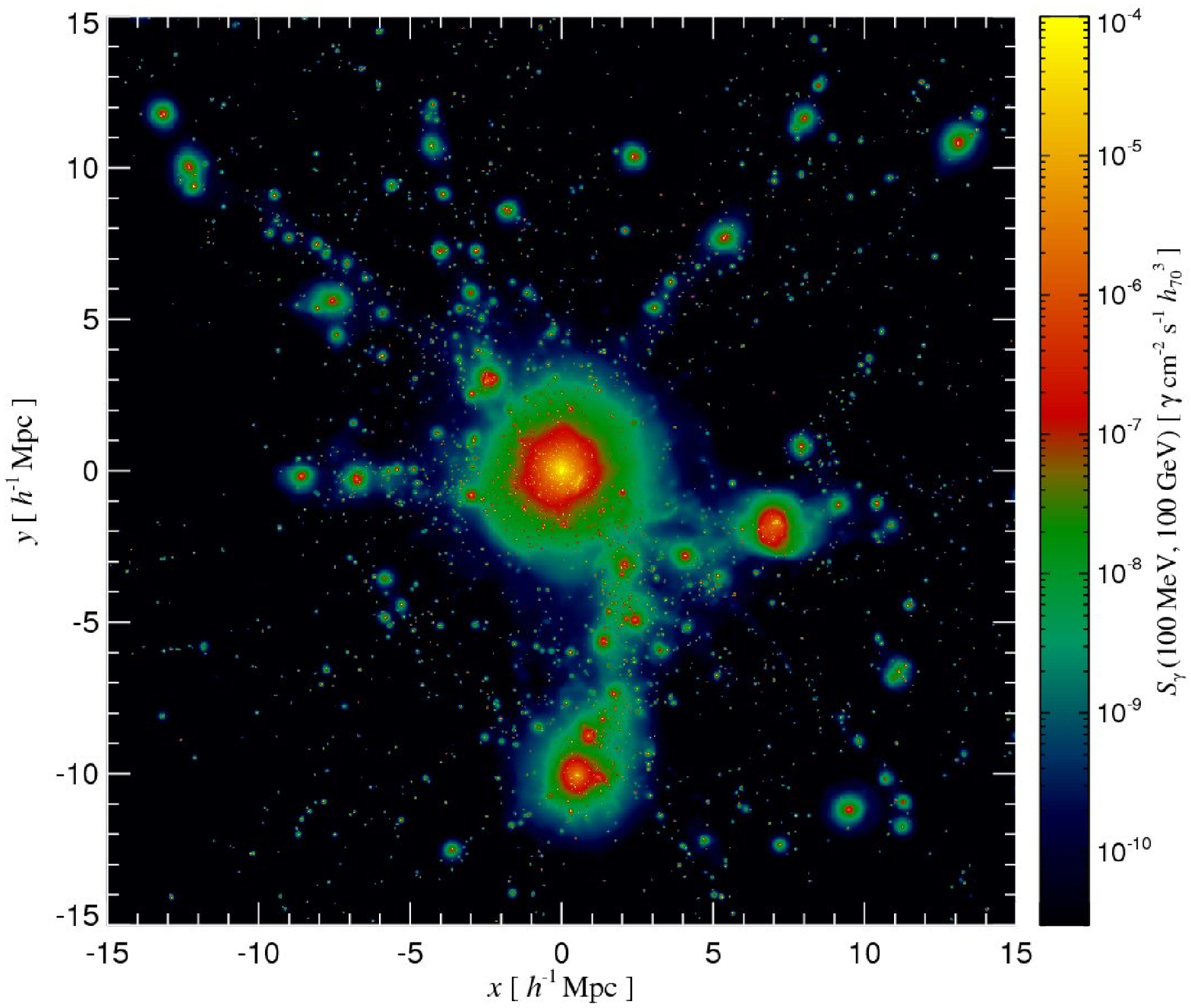}}\\
  \begin{minipage}[t]{0.495\textwidth}
    \centering{\it \large Total inverse Compton emission ($E_\gamma>10$ keV):}
  \end{minipage}
  \hfill
  \begin{minipage}[t]{0.495\textwidth}
    \centering{\it \large Total inverse Compton emission ($E_\gamma>100$ MeV):}
  \end{minipage}
\resizebox{0.5\hsize}{!}
{\includegraphics{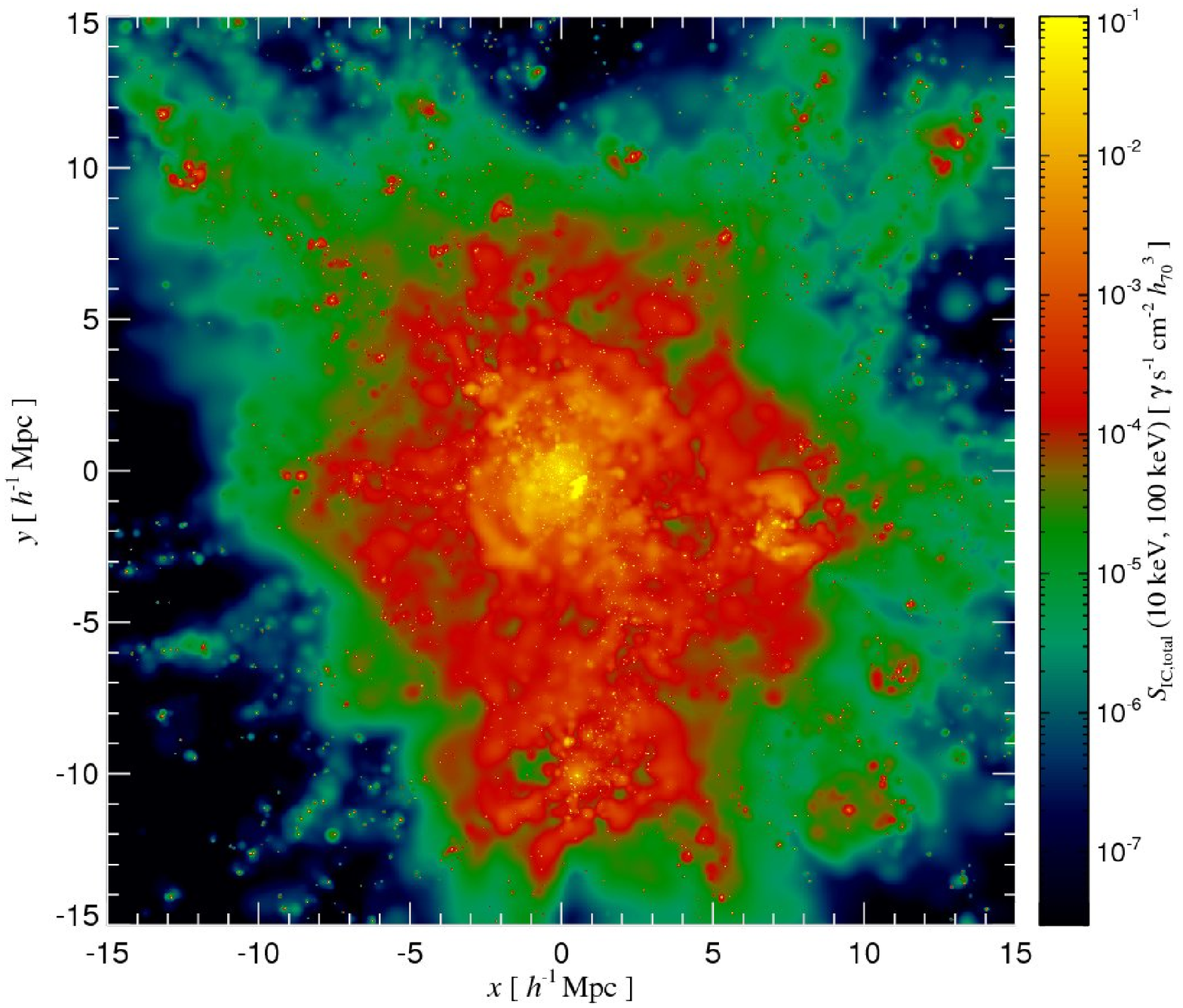}}%
\resizebox{0.5\hsize}{!}
{\includegraphics{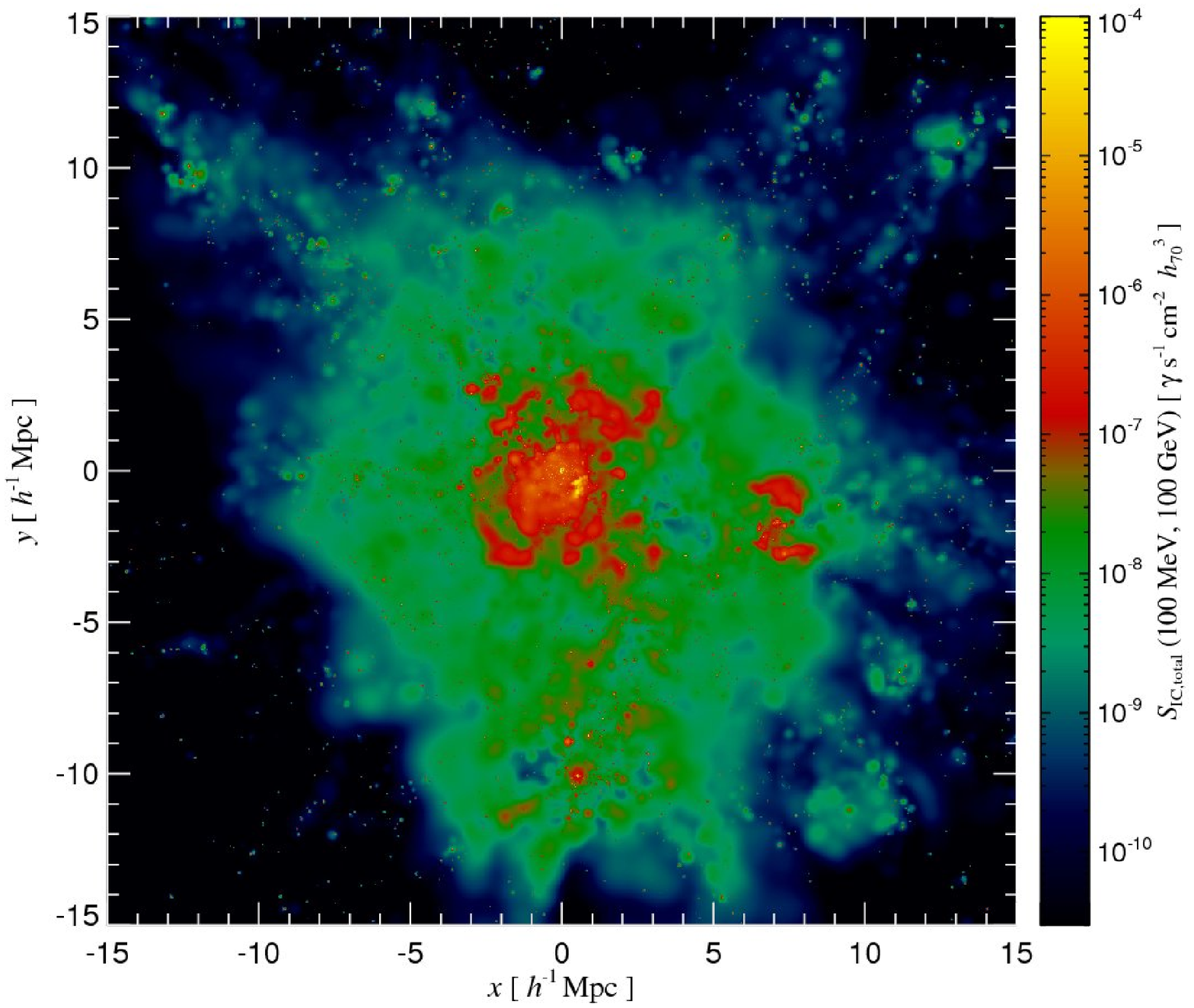}}\\
\end{center}
  \caption{The top panels compare the thermal X-ray emission and the
    $\gamma$-ray emission resulting from hadronic CR interaction with ambient
    gas protons of the super-cluster region of our Coma-like cluster in our
    radiative simulation (model S2, cluster g72a). The hadronic $\gamma$-ray
    emission shows a shallow decline with radius due to the rising
    CR-to-thermal number density profile.  The bottom panels show the inverse
    Compton emission from primary and secondary CR electrons in the hard X-ray
    (left side) as well as the $\gamma$-ray band (right side). The
    primary CR electrons dominate the emission signal on large
    scales. Comparing the $\gamma$-ray emission components (right panels)
    shows that the pion decay $\gamma$-rays exceed the total IC emission at
    energies $E_\gamma>100$~MeV.}
  \label{fig:gamma-rays}
\end{figure*}

Synchrotron emission from primary accelerated electrons is polarised due to a
combination of two effects.  (1) Shock compression aligns unordered magnetic
fields with the shock plane \citep{1998A&A...332..395E}.  Furthermore, shearing
motions induced by oblique shocks stretch these field lines which leads to a
larger magnetic coherence length of two-dimensional field configurations
\citep{2006PhPl...13e6501S}.  If the synchrotron emitting structure that is
energised by the formation shock wave is seen at some angle between the
line-of-sight and the normal of the shock front, the magnetic field structure
projected onto the plane of sky shows a preferential direction which implies a
preferred intrinsic synchrotron polarisation.  (2) The combination of the
localised acceleration site of CR electrons at these shock fronts and the short
synchrotron cooling times (cf.~Fig.~\ref{fig:timescales}) leads to a small
synchrotron emission volume.  Thus, these peripheral radio relics are expected
to show a preferred synchrotron polarisation with the magnetic field aligned
with the shock surface \cite[as observed e.g.~in Abell~3667
  by][]{1997MNRAS.290..577R}.  Superposing many causally unconnected radio
relics in projection leads to a decrease of the degree of polarisation.

Hadronically induced synchrotron emission of the smooth radio halo is virtually
unpolarised assuming statistically isotropic distribution of magnetic field
vectors without a preferred direction. The large emission volume is filled with
magnetised plasma that causes the plane of polarisation to Faraday
rotate. Hadronically generated CR electrons fill the same cluster volume.
Thus, each radio emitting volume element along the line-of-sight, that is
separated by more than the magnetic correlation length or the Faraday depth, if
shorter, radiates causally unconnected intrinsically polarised emission that
averages out to a net unpolarised emission, e.g.~radio halos are Faraday
depolarised.

Combining these considerations with the previously developed model for the
radio halo emission implies a transition from the virtually unpolarised radio
halo emission at small impact parameters to a small degree of polarisation at
the halo periphery, characterised by the dominating primary emission there. In
order to detect this polarisation one might be forced to go out to large impact
parameters with a small resulting synchrotron surface brightness where the
emission is dominated by very few relics along the line-of-sight.  Owing to the
different injection timescales of primary and secondary CR electrons, we
conclude that the secondary halo emission traces the time integrated
non-equilibrium activities of a cluster and is modulated by the recent
dynamical activities \citepalias{2007MNRAS...378..385P}.  In contrast, the
polarised radio relic emission resembles the current dynamical, non-equilibrium
activity of a forming structure and results in an inhomogeneous and aspherical
spatial distribution.

\begin{figure*}
\begin{center}
  \begin{minipage}[t]{0.485\textwidth}
    \centering{\it \large Different $\gamma$-ray emission processes:}
  \end{minipage}
  \hspace{0.02\textwidth}
  \begin{minipage}[t]{0.485\textwidth}
    \centering{\it \large Comparison of different clusters:}
  \end{minipage}
\resizebox{0.5\hsize}{!}
{\includegraphics{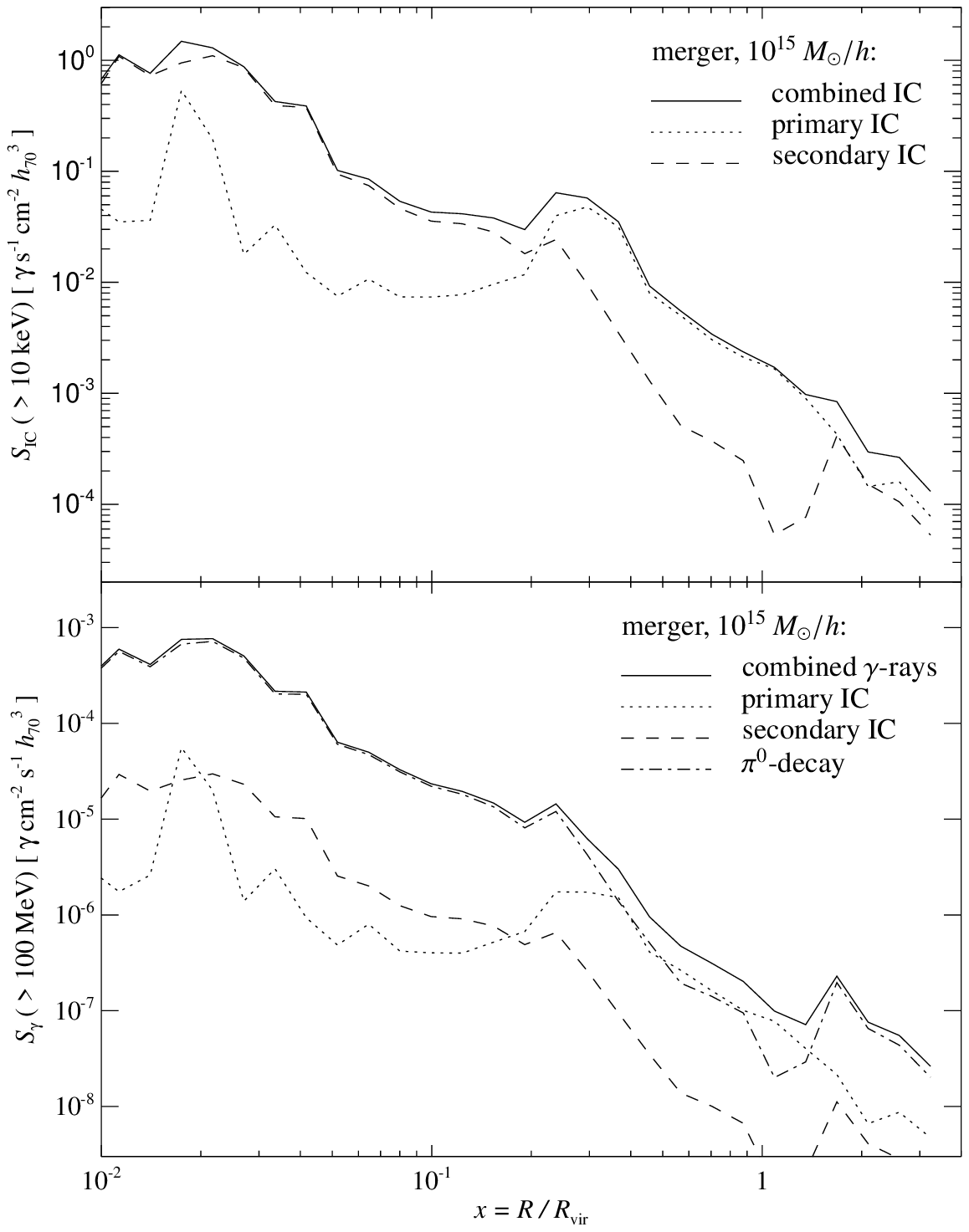}}%
\resizebox{0.5\hsize}{!}
{\includegraphics{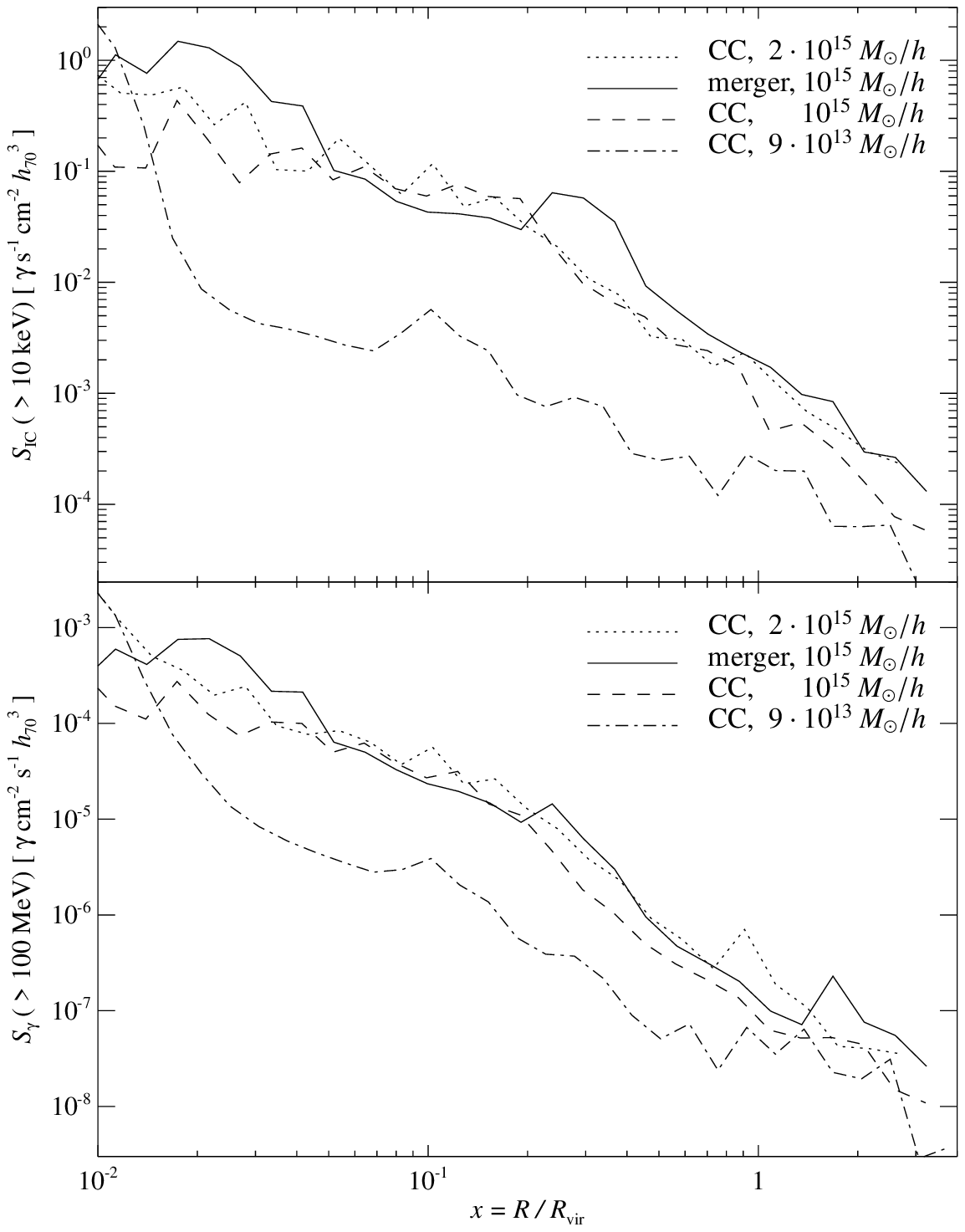}}\\
\end{center}
\caption{Azimuthally averaged profiles of the inverse Compton (IC) surface
  brightness for energies $E_\gamma > 10\mbox{ keV}$ (upper panels) and
  $\gamma$-ray emission profiles for $E_\gamma > 100\mbox{ MeV}$ (bottom
  panels). Using the simulation of the post-merging cluster g72a, the left side
  compares the IC emission of {\em primary CR electrons} that were accelerated
  directly at structure formation shocks (dotted lines) with that of {\em
  secondary CR electrons} that result from hadronic CR proton interactions with
  ambient gas protons (dashed lines) while the solid lines shows the sum of
  both emission components. The dominant emission component at energies
  $E_\gamma > 100\mbox{ MeV}$ is the pion decay induced $\gamma$-ray emission
  (dash-dotted lines).  The right side compares the total hard X-ray and
  $\gamma$-ray emission, respectively, for for the different clusters g8a
  (dotted lines), g72a (solid lines), g51 (dashed lines), and g676
  (dashed-dotted lines). While the hard X-ray/$\gamma$-ray emission clearly
  scales with the cluster mass, the dynamical state of the cluster is equally
  important and can even reverse the mass trend as can be inferred from the IC
  emission for $E_\gamma > 10\mbox{ keV}$ comparing the top two clusters in the
  figure legend.  }
  \label{fig:gamma_profiles}
\end{figure*}

\subsection{Hard X-ray and $\gamma$-ray emission}
\label{sec:X-ray_gamma}

\subsubsection{Projected X-ray and $\gamma$-ray maps}
\label{sec:gamma_proj}

Different non-thermal $\gamma$-ray emission processes are compared to the
thermal X-ray emission in Fig.~\ref{fig:gamma-rays}.  The top panels compare
the morphology of the thermal X-ray surface brightness \citepalias[for details
of the projection, cf.][]{2007MNRAS...378..385P} to that of the $\gamma$-ray
emission resulting from hadronic CR interactions with ambient gas protons of
our Coma cluster region in our radiative simulation.  Although they resemble
each other, the pion decay induced $\gamma$-ray emission declines slower with
radius and makes the $\gamma$-ray halo more extended. The bremsstrahlung
emission of the thermal gas scales as $n_\rmn{th}^2$ (neglecting the radial
dependence of the cooling function) while the hadronically induced $\gamma$-ray
emission scales as $n_\rmn{gas} n_\CR$. The discrepancy between the radial
behaviour of the thermal and non-thermal emission can be understood by looking
at the right side of Fig.~\ref{fig:fCR} which shows average profiles of
the spherically averaged CR fraction $f_\CR = n_\CR / n_\rmn{th}$ of our sample
of all clusters. There is an increase of $f_\CR$ with radius that is
independent of the modelled CR physics and continues beyond the virial
radius. This generic prediction of CR physics is due to the more efficient CR
acceleration at the peripheral strong accretion shocks compared to weak central
flow shocks.

The bottom panels of Fig.~\ref{fig:gamma-rays} show the inverse Compton
emission from primary and secondary CR electrons in the hard X-ray (left
side) as well as the $\gamma$-ray band (right side). The primary CR
electrons dominate the hard X-ray emission signal on large scales. The primary
IC emission directly reflects the inhomogeneous virialisation process that
manifests itself through a filigree web spun by shocks (cf.~Fig.~\ref{fig:g72}).
In principle, IC emission is the cleanest way of probing structure formation
shock waves since the inverse Compton emission is not weighted by the
magnetic energy density as it is the case for synchrotron emission. Visually
comparing the right panels of Fig.~\ref{fig:gamma-rays} implies that the
pion decay induced $\gamma$-ray emission dominates the total IC emission in the
energy range that is of interest to GLAST.

\subsubsection{Hard X-ray IC and $\gamma$-ray emission profiles}
\label{sec:gamma_profiles}

Figure~\ref{fig:gamma_profiles} shows azimuthally averaged profiles of the IC
surface brightness for energies $E_\gamma > 10\mbox{ keV}$ (upper panels) and
$\gamma$-ray emission profiles for $E_\gamma > 100\mbox{ MeV}$ (bottom
panels). Using the simulation of the post-merging cluster g72a, the left
side compares the IC emission of {\em primary CR electrons} that were
accelerated directly at structure formation shocks with that of {\em secondary
CR electrons} that result from hadronic CR proton interactions with ambient gas
protons. The primary and especially the secondary IC emission is generally more
inhomogeneous compared to both radio emission components.  Secondly, the IC
emission declines less steeply with radius and shows almost a power-law profile
compared with the $\beta$-profile in the case of synchrotron emission.  Both
effects are due to the weighting by the magnetic field which in the latter case
tends to smooth out the inhomogeneous CR electron distribution and causes a
steeper decrease of the radio synchrotron emission (at least in our model of
the magnetic field).

At energies $E_\gamma > 100\mbox{ MeV}$, the pion decay induced $\gamma$-ray
emission is the dominant emission component everywhere except in the peripheral
regions of the post-merging cluster g72a where the primary IC emission is of
similar strength (cf.{\ }bottom left panel of
Fig.~\ref{fig:gamma_profiles}).  As a word of caution it should be added, that
our simulation assumes a CR proton spectral index of $\alpha_\p=2.3$. Lowering
this value would result in a larger secondary IC component
(cf. Fig.~\ref{fig:spectrum}). Future work is required to study such a scenario
which might find applications in the outer parts of clusters.  However, the
pion decay component at $E_\gamma > 100\mbox{ MeV}$ should be robust with
respect to variations of $\alpha_\p$ since it samples the pion bump that is
sensitive to threshold effects of the hadronic cross-section but not
to the spectrum of the parent CR distribution.  Despite the fact that the
secondary IC emission still dominates the primary IC emission at $E_\gamma >
100\mbox{ MeV}$, the primary IC component is increased by a factor of two
compared to the hard X-ray emission at $E_\gamma > 10\mbox{ keV}$. We conclude
that the mean IC spectral index is obviously smaller than that of the secondary
emission, $\alpha_\nu=1.15$.  This can be understood by combining the facts
that a superposition of different power-law spectra produces a concave spectrum
and that IC emission at $E_\gamma > 100\mbox{ MeV}$ results from CR electrons
with a Lorentz factor $\gamma\gtrsim 3 \times10^5$ according to
Eqn.~(\ref{eq:ICphoton}). This again stresses the importance of correctly
modelling the peripheral regions of a cluster since they show predominantly the
conditions for strong shock waves that are able to accelerate such flat CR
electron populations.

We move on to compare the hard X-ray and $\gamma$-ray emission for clusters of
different masses and dynamical states.  While the hard X-ray/$\gamma$-ray
emission clearly scales with the cluster mass, the dynamical state of the
cluster is equally important and can even reverse the mass trend (cf.{\ }the
right side of Fig.~\ref{fig:gamma_profiles}).  The reason for this lies in
the enhanced CR pressure in merging clusters \citepalias{2007MNRAS...378..385P}
as well as in the primary IC emission that sensitively traces current
non-equilibrium or merging activities of clusters.

\begin{figure}
  \centering{\it \large  Influence of simulated physics on $\gamma$-ray emission:}
\resizebox{\hsize}{!}
{\includegraphics{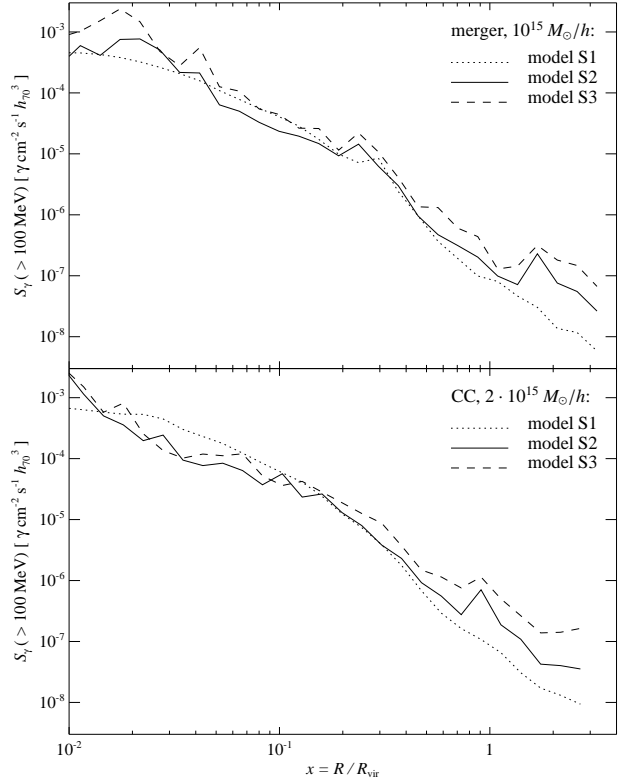}}%
\caption{Influence of simulated physics on azimuthally averaged $\gamma$-ray
  emission profiles for $E_\gamma > 100\mbox{ MeV}$. The top panel shows the
  emission profile of the post-merging cluster g72a while the bottom panel that
  of the large cool core cluster g8a. Shown are our non-radiative simulations
  (dotted lines, model S1), our radiative simulations with CR protons injected
  at structure formation shocks (solid lines, model S2), and the complete CR
  model where we additionally take CR acceleration at SNe shocks into account
  (dashed lines, model S3).  Despite the different physics in these
  simulations, the $\gamma$-ray emission level is similar.  }
  \label{fig:gamma_physics}
\end{figure}
\begin{figure*}
\begin{center}
  \begin{minipage}[t]{0.495\textwidth}
    \centering{\it \large Radiative simulation, $\gamma$-rays ($E_\gamma>100$ MeV):}
  \end{minipage}
  \hfill
  \begin{minipage}[t]{0.495\textwidth}
    \centering{\it \large Non-radiative simulation, $\gamma$-rays ($E_\gamma>100$ MeV):}
  \end{minipage}
\resizebox{0.5\hsize}{!}
{\includegraphics{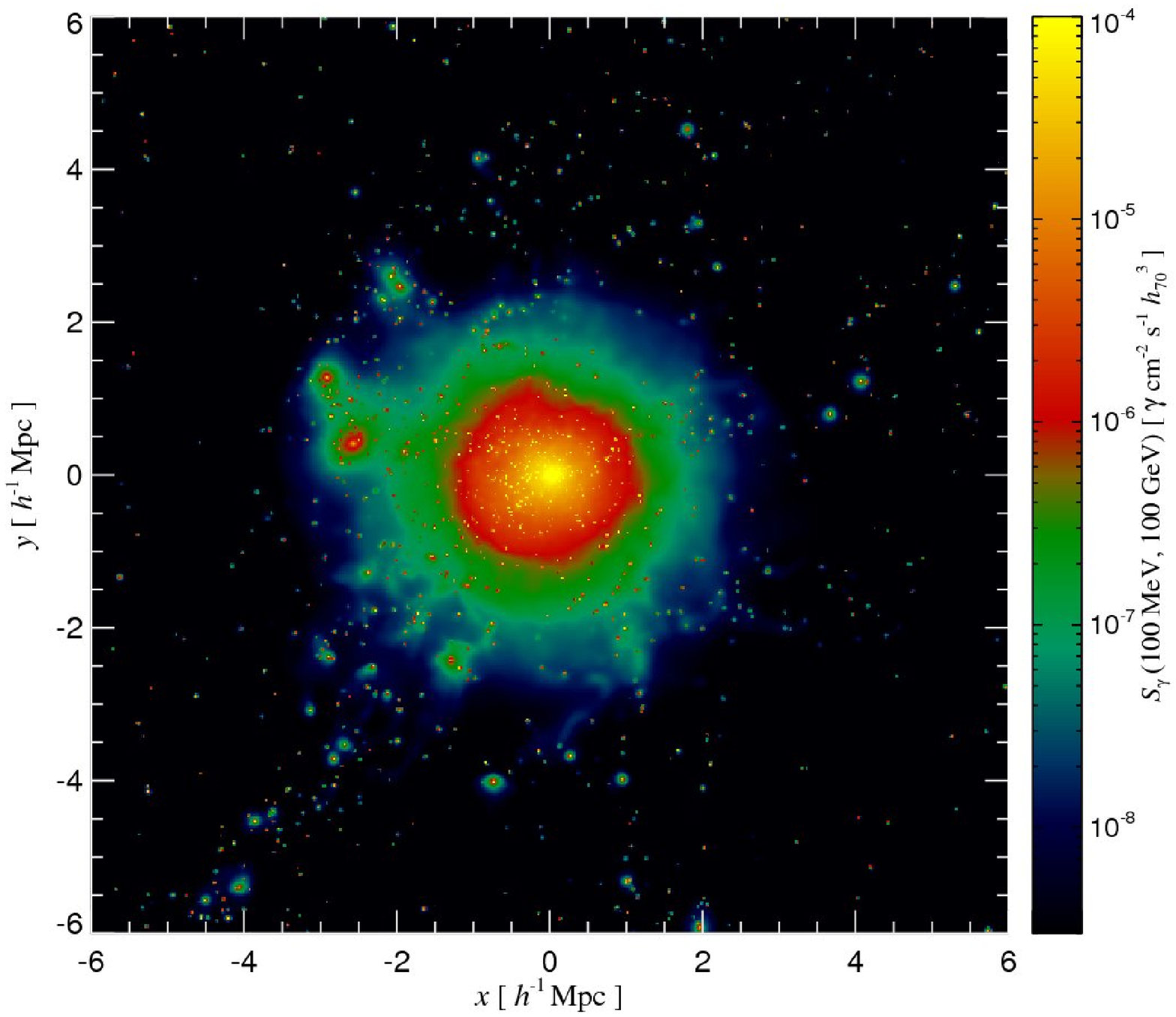}}%
\resizebox{0.5\hsize}{!}
{\includegraphics{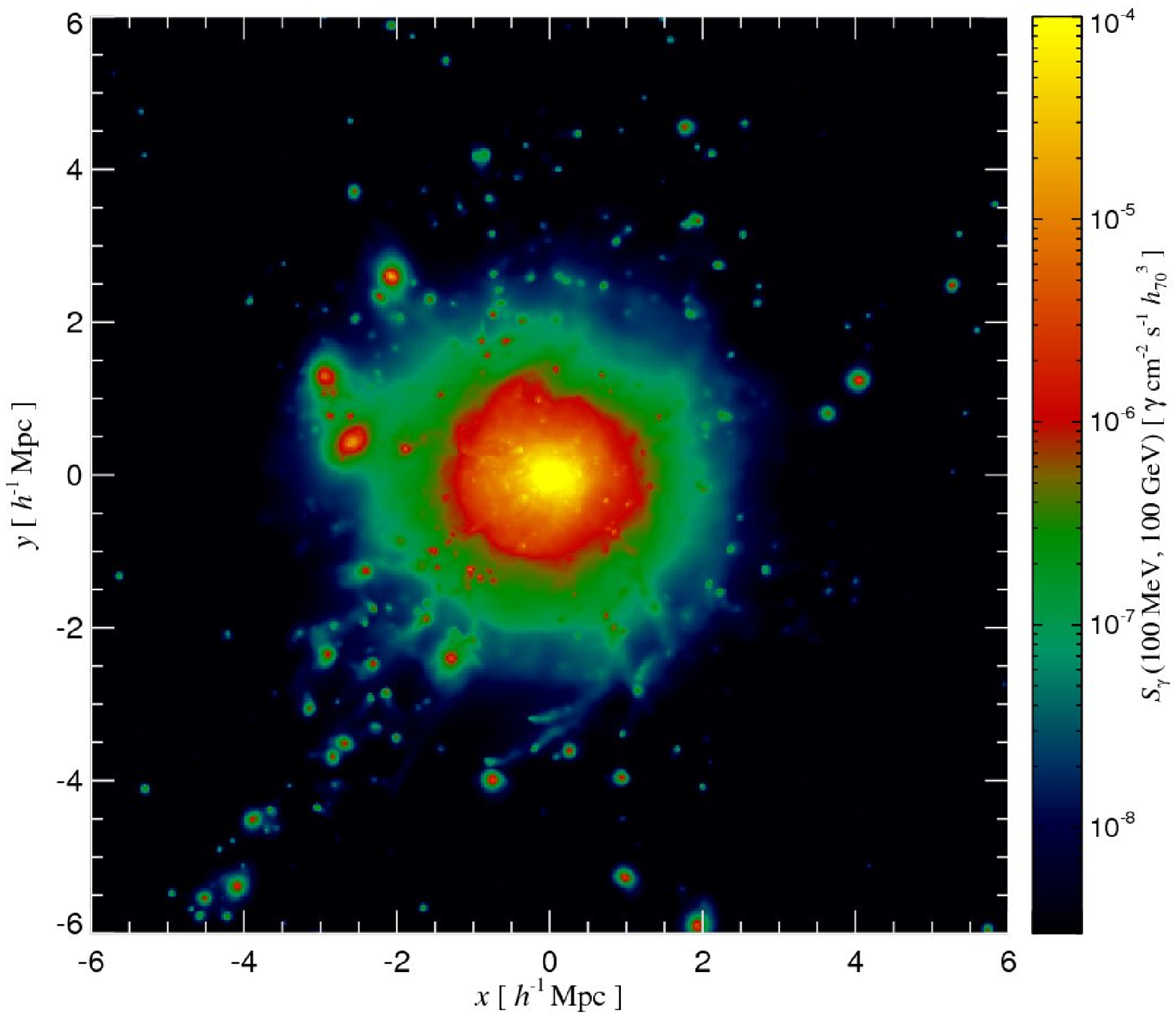}}\\
\end{center}
  \caption{Influence of simulated physics on pion decay $\gamma$-ray emission
  maps for $E_\gamma > 100\mbox{ MeV}$ of the large cool core cluster g8a. The
  peaked emission of each galaxy in the radiative simulation (left panel,
  model S2) contrasts the smoother and fluffier $\gamma$-ray emission of our
  non-radiative simulation with CR protons injected at structure formation
  shocks (right side, model S1).}
  \label{fig:gamma_physics_maps}
\end{figure*}

Figures~\ref{fig:gamma_physics_maps} and \ref{fig:gamma_physics} study the
influence of different simulated physics on $\gamma$-ray emission at $E_\gamma
> 100\mbox{ MeV}$.  The $\gamma$-ray emission is more inhomogeneous in
radiative simulations compared to non-radiative simulations. This is due to the
CR pressure equipartition within each galaxy. The overall $\gamma$-ray
luminosity, however, is very similar as can be inferred from the azimuthally
averaged emission profiles.  This confirms our finding for the synchrotron
emission in Fig.~\ref{fig:synchro_profiles} and confirms our explanation that
this is indeed a self-regulated effect of CR feedback and not biased due to
the magnetic weighting of the synchrotron emission. The complete CR model that
also accounts for CR from SNe (dashed lines, model S3) shows a slightly
enhanced $\gamma$-ray emission level since we opened up a new acceleration
channel compared to our standard model S2.

\subsection{Correlations between thermal and non-thermal emission}
\label{sec:correlations}

Compressing intrinsically non-spherical emission purely into surface
brightness profiles causes loss of information and might yield biased results.
In addition to emission maps and surface brightness profiles, we complete our
analysis using pixel-to-pixel correlations of the thermal X-ray surface
brightness with non-thermal cluster emission processes.  To this end we
compare these correlations of our post-merging cluster simulation g72
(Fig.~\ref{fig:correlations}) to the cool core cluster simulation g8
(Fig.~\ref{fig:correlations-g8a}).  Each of these figures shows the
correlation space density of the radio surface brightness (top panels) and the
$\gamma$-ray surface brightness for $E_\gamma > 100\mbox{ MeV}$ (bottom
panels), as well as that of the hadronically induced non-thermal emission
(left side, red colour scale) and the non-thermal emission of {\em primary CR
  electrons} that were accelerated directly at structure formation shocks
(right side, blue colour scale).

While the hadronically induced non-thermal emission is tightly correlated with
the thermal bremsstrahlung emission, the correlation is much weaker and the
scatter is increased in the case of primary non-thermal emission where
structures in the correlation space density correspond to individual structure
formation shock waves. We can see preferably tangential shocks that are
characterised by a varying non-thermal emission for a constant X-ray surface
brightness and to a smaller extend radial shocks where the role of thermal and
non-thermal emission is interchanged.

\subsubsection{Correlations of the synchrotron emission}
\label{sec:correlations_synch}

\begin{figure*}
\begin{center}
  \begin{minipage}[t]{0.495\textwidth}
    \centering{\it \large Secondary synchrotron emission (1.4 GHz):}
  \end{minipage}
  \hfill
  \begin{minipage}[t]{0.495\textwidth}
    \centering{\it \large Primary synchrotron emission (1.4 GHz):}
  \end{minipage}
\resizebox{0.5\hsize}{!}
{\includegraphics{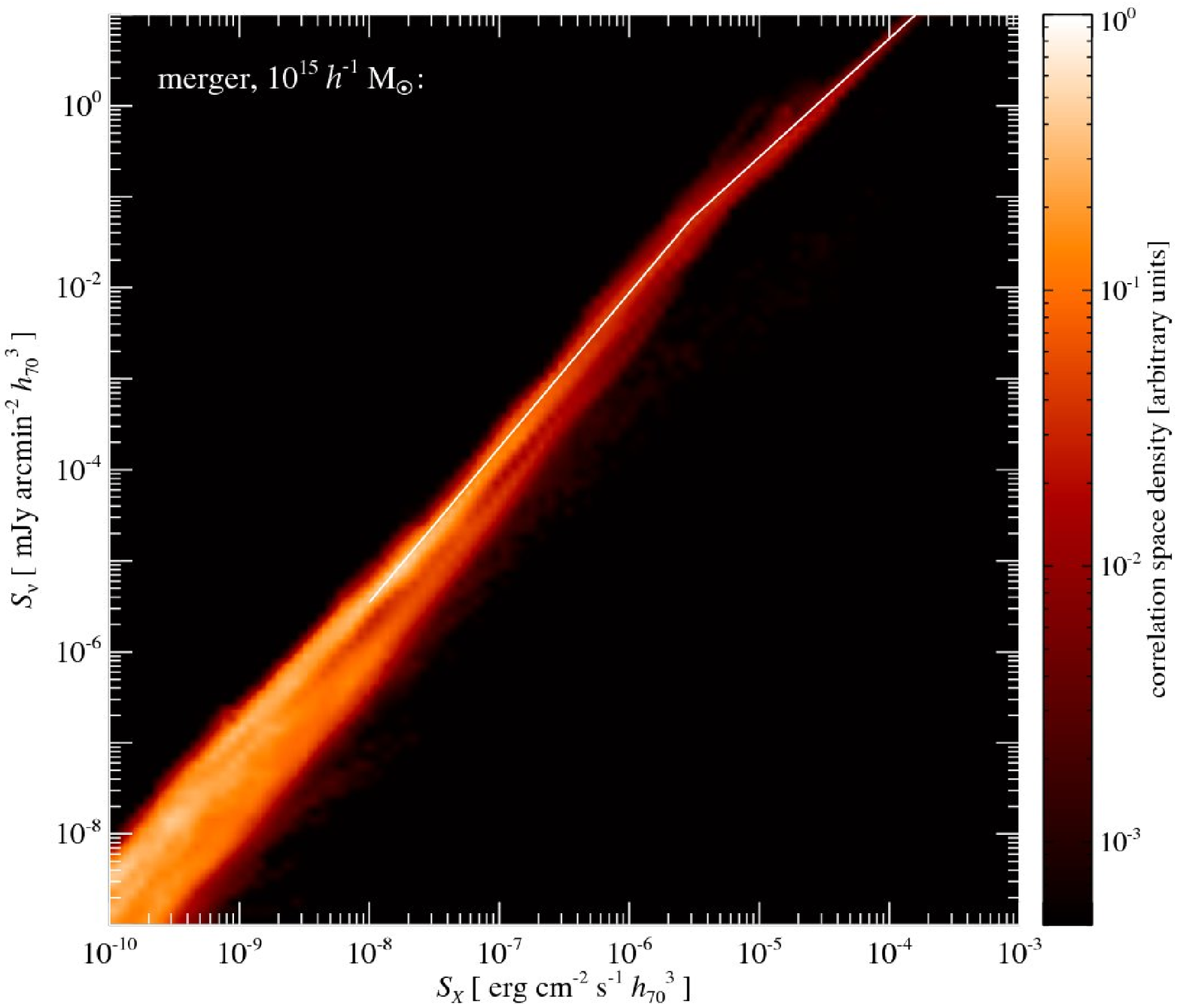}}%
\resizebox{0.5\hsize}{!}
{\includegraphics{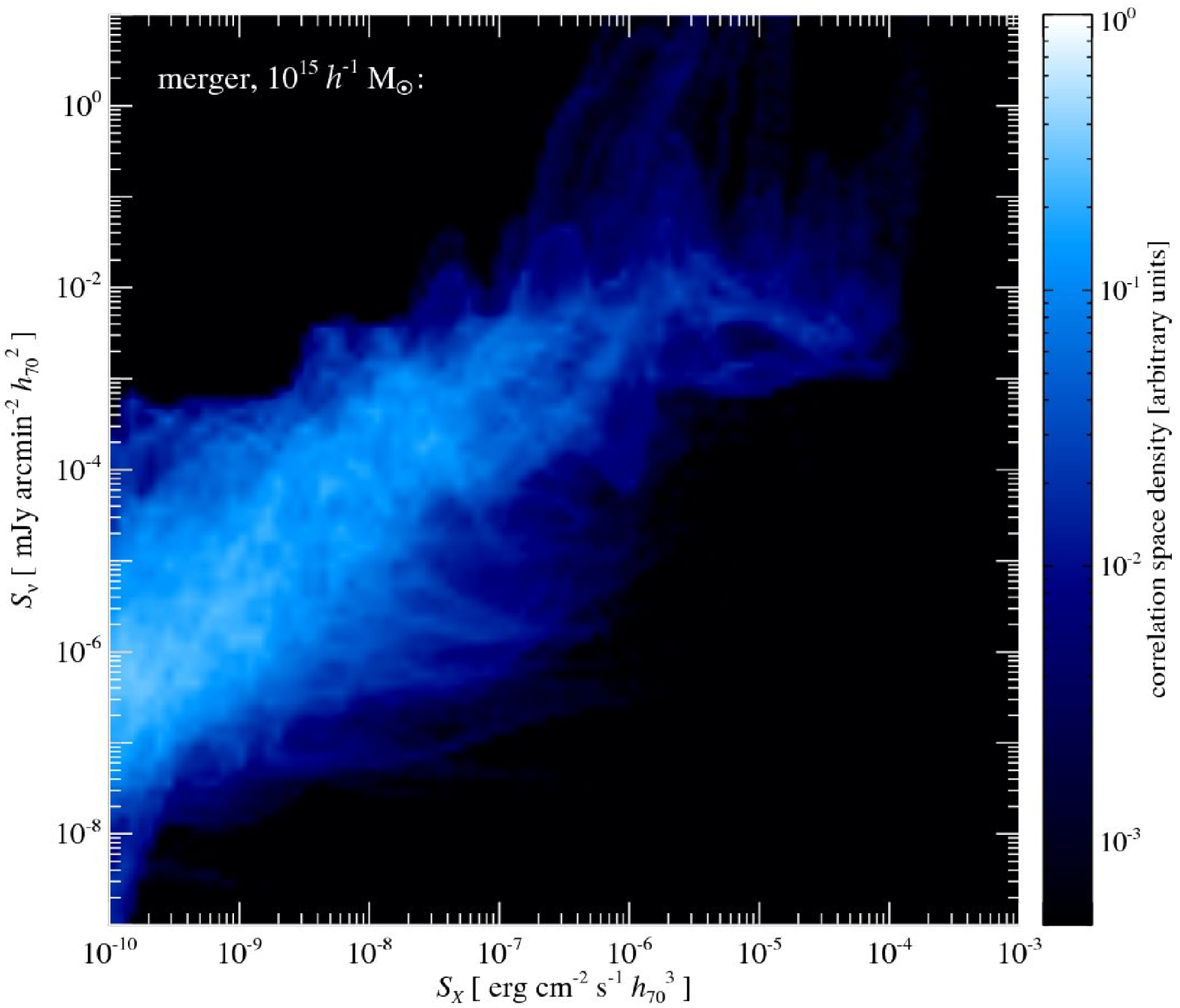}}\\
  \begin{minipage}[t]{0.495\textwidth}
    \centering{\it \large Pion decay \& secondary IC emission ($E_\gamma>100$ MeV):}
  \end{minipage}
  \hfill
  \begin{minipage}[t]{0.495\textwidth}
    \centering{\it \large Primary IC emission ($E_\gamma>100$ MeV):}
  \end{minipage}
\resizebox{0.5\hsize}{!}
{\includegraphics{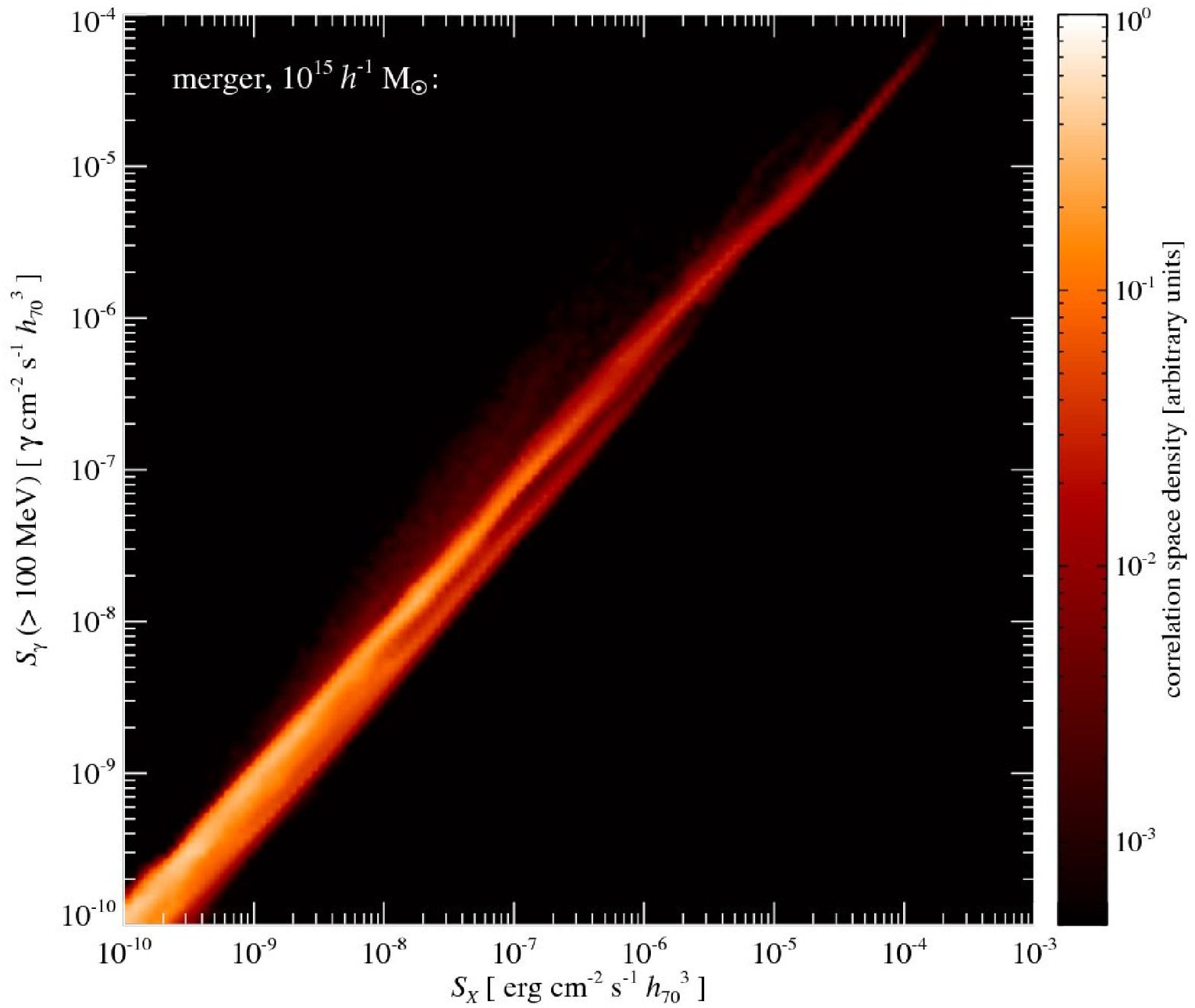}}%
\resizebox{0.5\hsize}{!}
{\includegraphics{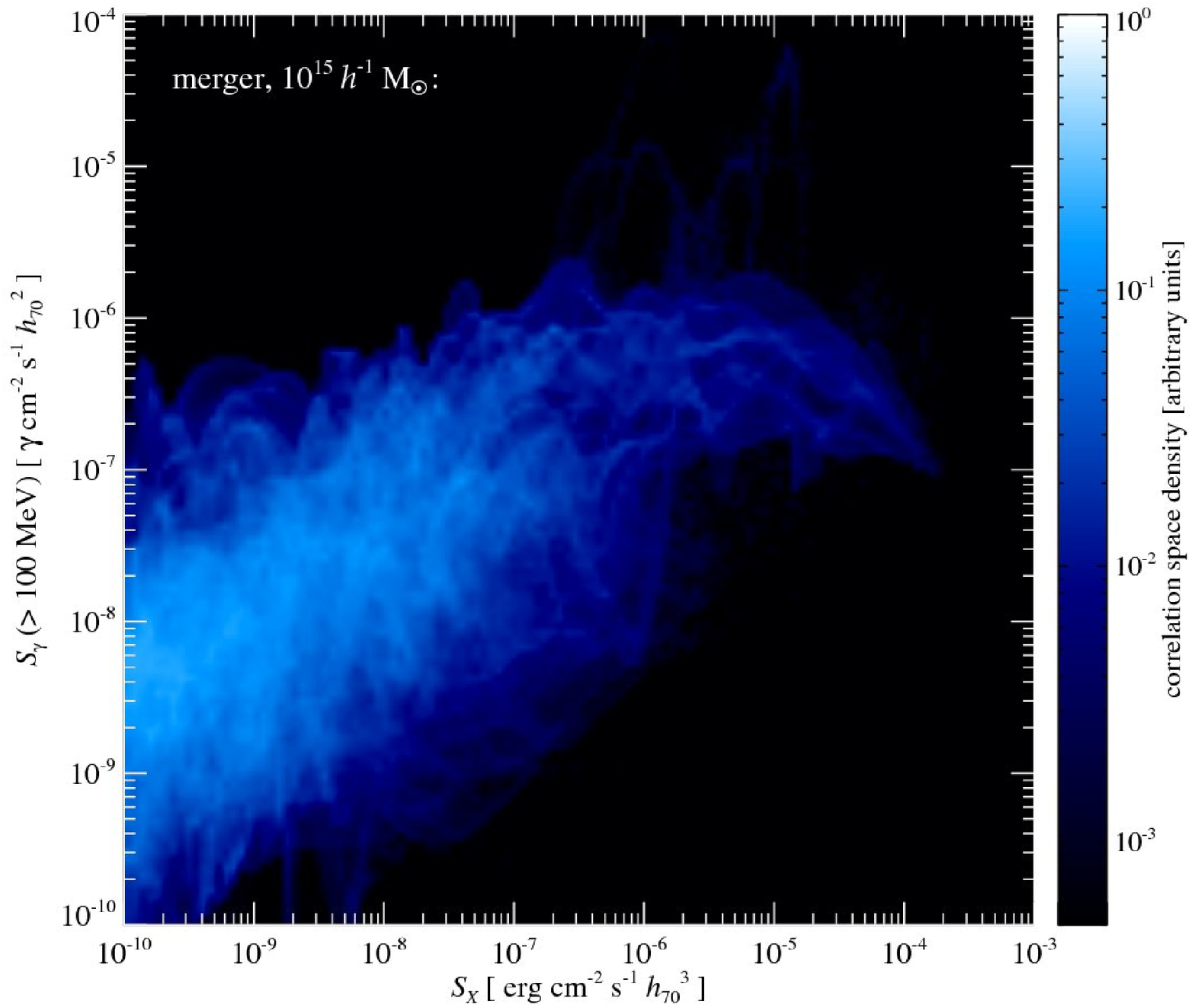}}\\
\end{center}
  \caption{Pixel-to-pixel correlation of the thermal X-ray surface brightness
  with both the radio surface brightness (top panels) and the $\gamma$-ray
  surface brightness for $E_\gamma > 100\mbox{ MeV}$ (bottom panels) for our
  post-merging cluster simulation g72 using our model S2. Shown is the
  correlation space density of the hadronically induced non-thermal emission
  (left side, red colour scale) and the non-thermal emission of {\em
  primary CR electrons} that were accelerated directly at structure formation
  shocks (right side, blue colour scale). The bottom left panel
  includes pion decay $\gamma$-ray emission as well as IC emission from
  secondary CR electrons. While the hadronically induced non-thermal emission
  is tightly correlated with the thermal bremsstrahlung emission, the
  correlation is much weaker and the scatter is increased in the case of
  primary non-thermal emission where the structures in the correlation space
  density correspond to individual structure formation shock waves. The line in
  the top left panel is a fit to the correlation where the slope flattens
  from 1.7 to 1.3 towards high luminosities.  }
  \label{fig:correlations}
\end{figure*}
\begin{figure*}
\begin{center}
  \begin{minipage}[t]{0.495\textwidth}
    \centering{\it \large Secondary synchrotron emission (1.4 GHz):}
  \end{minipage}
  \hfill
  \begin{minipage}[t]{0.495\textwidth}
    \centering{\it \large Primary synchrotron emission (1.4 GHz):}
  \end{minipage}
\resizebox{0.5\hsize}{!}
{\includegraphics{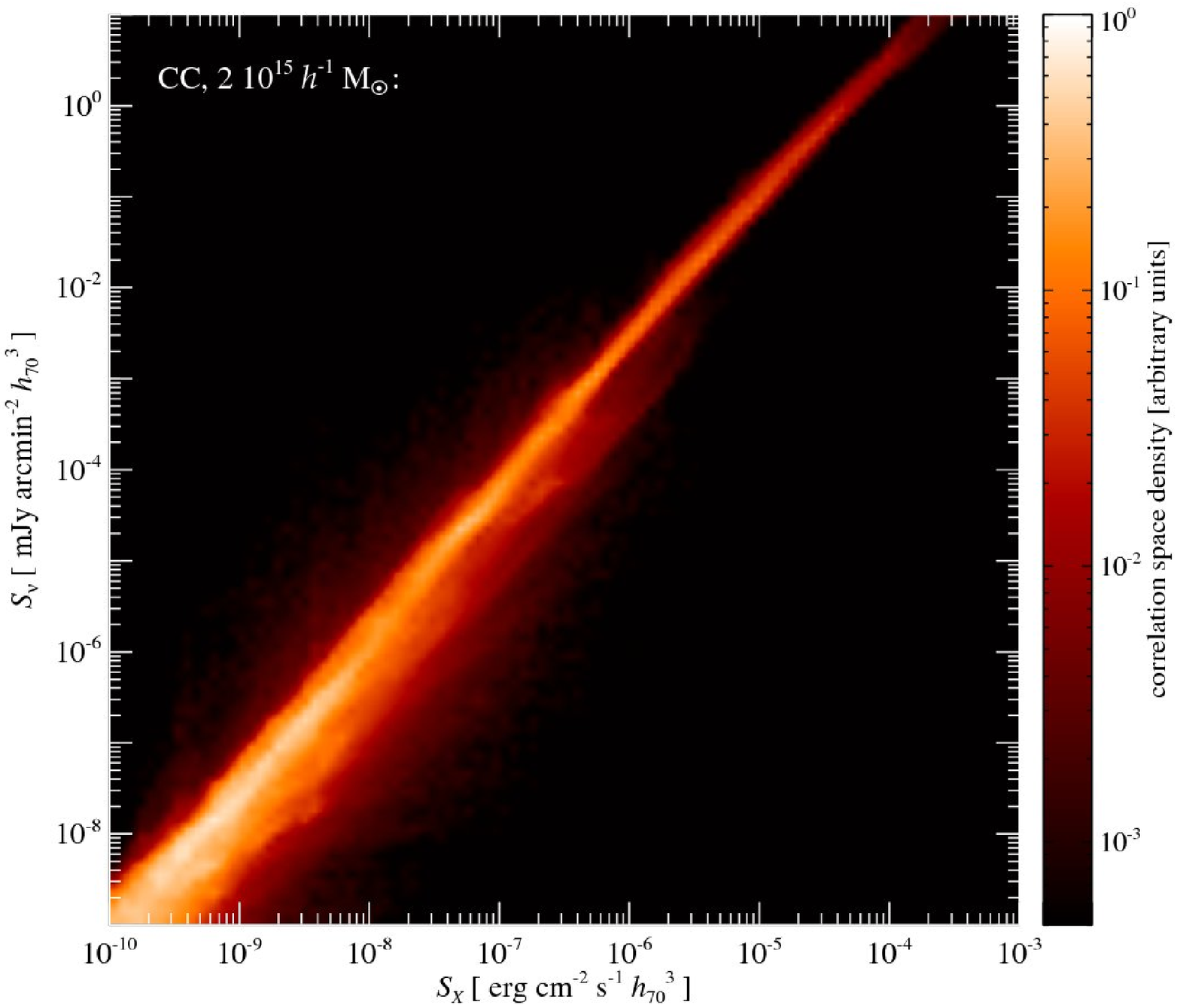}}%
\resizebox{0.5\hsize}{!}
{\includegraphics{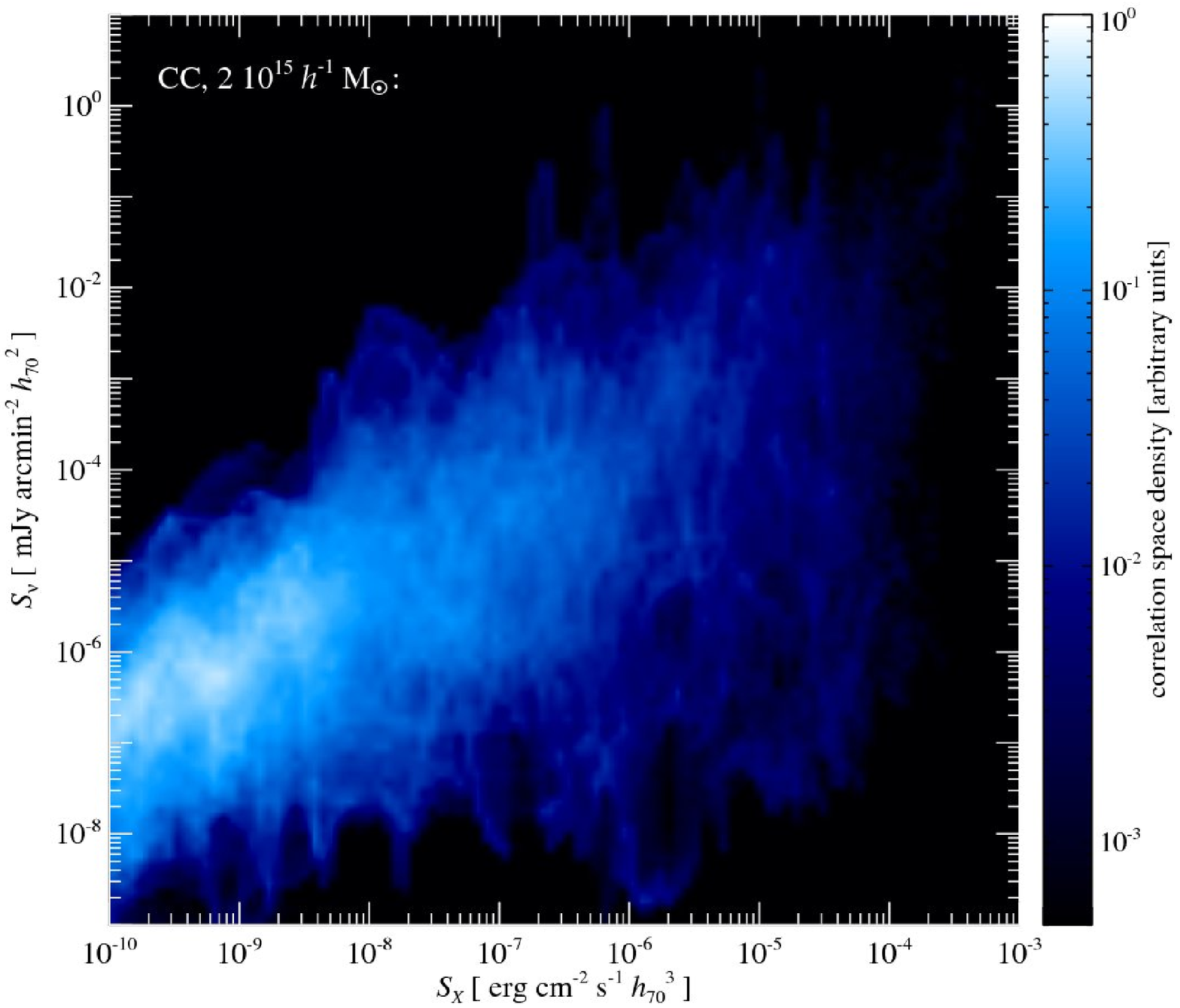}}\\
  \begin{minipage}[t]{0.495\textwidth}
    \centering{\it \large Pion decay \& secondary IC emission ($E_\gamma>100$ MeV):}
  \end{minipage}
  \hfill
  \begin{minipage}[t]{0.495\textwidth}
    \centering{\it \large Primary IC emission ($E_\gamma>100$ MeV):}
  \end{minipage}
\resizebox{0.5\hsize}{!}
{\includegraphics{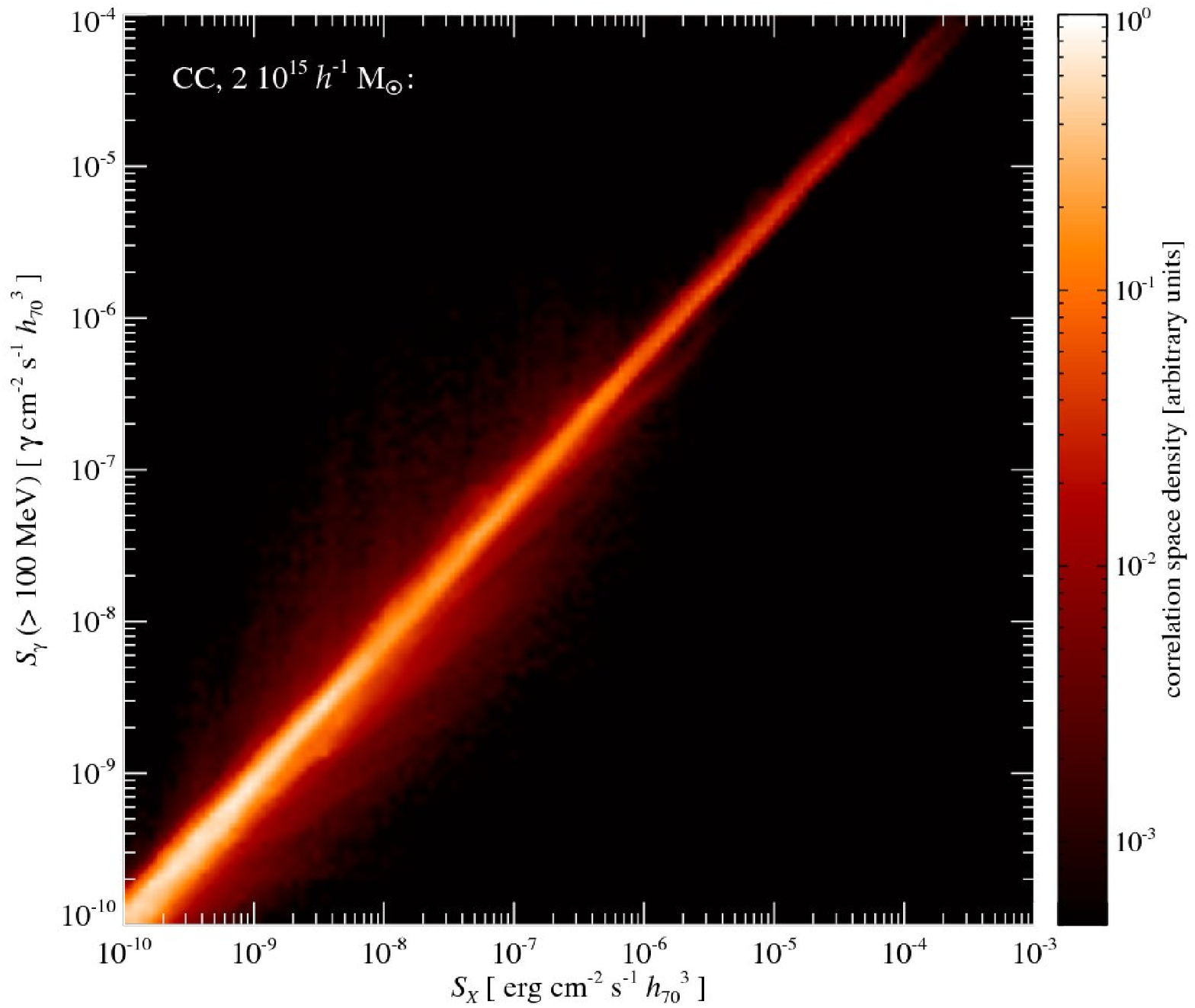}}%
\resizebox{0.5\hsize}{!}
{\includegraphics{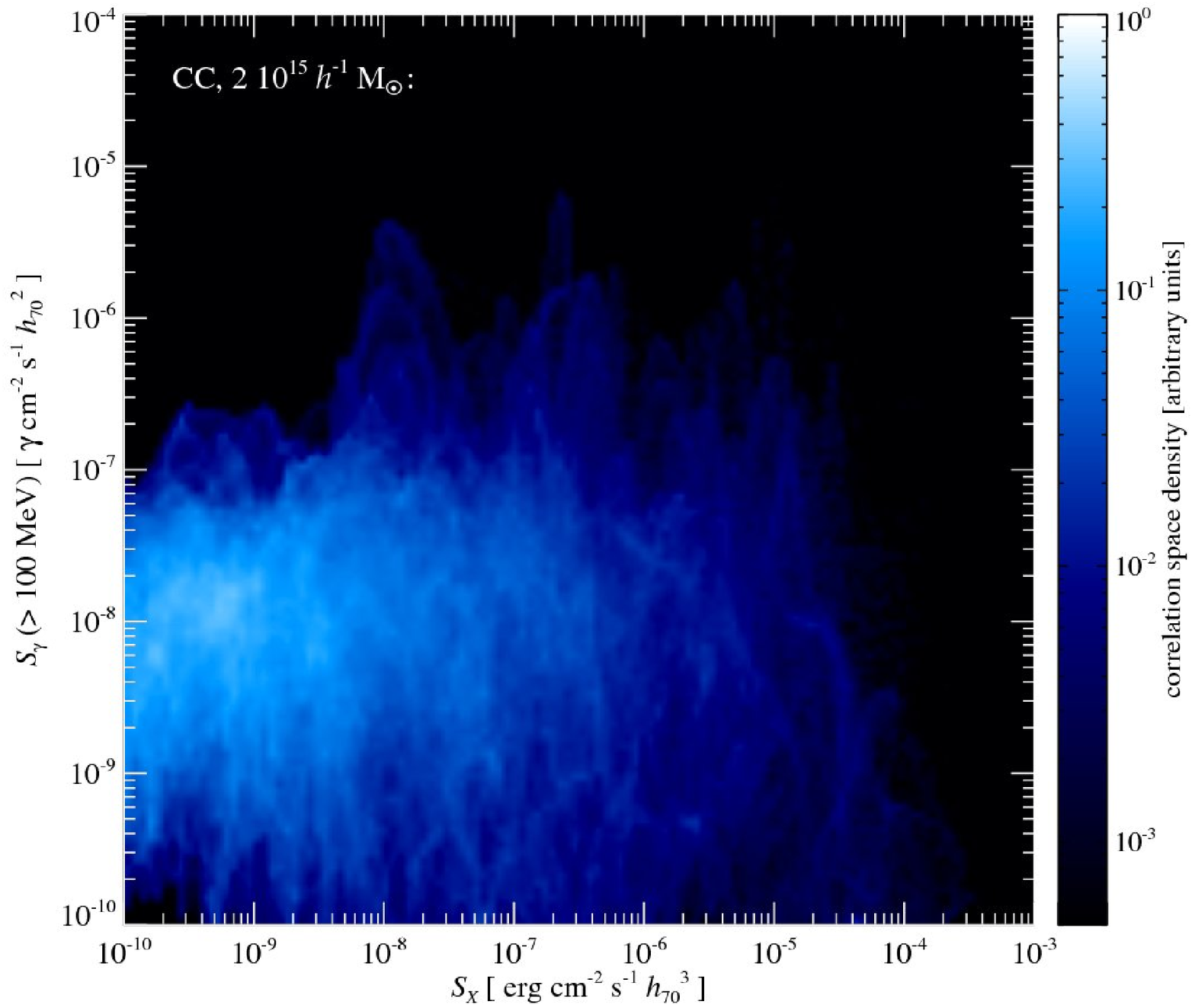}}\\
\end{center}
  \caption{Same as Fig.~\ref{fig:correlations} but for our cool core cluster
  simulation g8. Note the absence of any correlation in the bottom right
  panel showing the primary IC surface brightness for $E_\gamma > 100\mbox{
  MeV}$ which suggests that the correlation of the primary radio emission (top
  right panel) is solely due to our assumed model of the magnetic field,
  $\eps_B \propto \eps_\th$.}
  \label{fig:correlations-g8a}
\end{figure*}

Closer inspection of the {\em secondary synchrotron emission} (top left panel
in Fig.~\ref{fig:correlations}) shows flattening of the correlation 
\begin{equation}
  \label{eq:Snu-SX}
  S_\nu =  S_{\nu,0} \left(\frac{S_X}{S_{X,0}}\right)^{a},
\end{equation}
where the power-law index changes from $a=1.7$ to $a=1.3$ above $S_{X,0} =
3\times 10^{-6} \mbox{ erg cm}^{-2} \mbox{ s}^{-1}\, h_{70}^{3}$, with a
normalisation $S_{\nu,0}=0.057\mbox{ mJy arcmin}^{-2}\, h_{70}^3$.  This is due
to a combination of the following two effects. (1) Towards higher X-ray surface
brightness, the radio emission experiences a transition from IC to synchrotron
regime with a weaker dependence on magnetic field that goes along with weaker
density dependence.  (2) The merger displaces the central cool core and
disturbs the ICM by means of merging shock waves that dissipate the
gravitational binding energy associated with the merger. This yields an
increased CR proton pressure and number density relative to the thermal gas
within the central X-ray luminous regions and causes the change in the
correlation. Since the thermal energy distribution is equally effected by these
re-distribution of energy, this merger induced effect can be furthermore
amplified by the transition from the IC to synchrotron regime as explained
above.  The change of the power-law index of the correlation is more pronounced
for our post-merging cluster simulation g72 compared to our cool core cluster
g8 suggesting the importance of the second effect.  Note that the explicit
values of the correlation, in particular $S_{\nu,0}$, depend on the assumed
model for the magnetic field.  There is a second branch in the correlation of
the hadronically induced non-thermal emission visible that is due to the other
smaller cluster forming in that simulation.

The correlation of the {\em primary synchrotron emission} (top right side)
shows a large scatter especially at large surface brightness such that the
correlation of the {\em total synchrotron brightness} is expected to broaden
and to become flatter towards lower surface brightness to yield a roughly
linear correlation.  The exact realisation of the correlation at a surface
brightness that is substantially supported by primary synchrotron emission
sensitively depends on the mass ratio, geometry, and the advanced state of the
merger. Thus our simulations can only provide quantitative predictions for the
statistical behaviour rather than deterministic predictions for the
correlations.  Our correlations are strikingly similar to the ones found in
observed radio halos \citep[cf.][]{2001A&A...369..441G}.  They find a linear
relation between the radio and X-ray surface brightness that is tight at high
surface brightness while it broadens and flattens towards dimmer brightness.
This behaviour is one of the strongest arguments in favour of our new model for
radio halos that should be dominated in the centre by secondary emission with a
transition to the primary synchrotron emission in the cluster periphery.  In
accordance with our findings, their radio halo emission is slightly more
extended compared to the thermal X-ray emission. Varying spectral index
distributions preferably in the cluster periphery \citep{2004JKAS...37..315F}
support this picture. In particular, our model supports a strong link between
radio halos and cluster mergers for which there is a strong evidence in the
literature \citep[][ and references therein]{2004JKAS...37..315F}.

\subsubsection{Correlations of the $\gamma$-ray emission}

The bottom left panels in Figs.~\ref{fig:correlations} and
\ref{fig:correlations-g8a} include pion decay $\gamma$-ray emission as well as
IC emission from secondary CR electrons. These tight correlations are
characterised by a sub-linear power-law relation. This is due to the shallower
decay of the CR number density $n_\CR$ compared to that of the thermal gas
leading to an increase of the CR fraction $n_\CR / n_\th$
(cf.~Fig.~\ref{fig:fCR}).  Our post-merging cluster g72a shows a small
variation of the power-law index of the $\gamma$-ray to X-ray correlation at
high surface brightness albeit not as pronounced as in the case of radio
synchrotron emission. This confirms that both the merger induced boost of the
CR pressure and the transition from IC to synchrotron regime is responsible for
the flattening of the correlation between radio and X-ray surface brightness.
The primary IC emission of the cool core cluster simulation g8a (lower
right panel of Fig.~\ref{fig:correlations-g8a}) shows a complete absence
of any correlation.  The influence of the merger activity of a cluster on
enhancing the non-thermal cluster emission can thus unchallengeably be studied
with the primary IC $\gamma$-ray emission (lower right panel of
Fig.~\ref{fig:correlations}). Comparing this primary IC $\gamma$-ray emission
to its counterpart radio emission indicates that a large part of that
correlation with the X-ray emission is indeed owed to the merger induced CR
enhancement and only in parts by the density dependence of the magnetic field.

\section{Discussion and future work}
\label{sec:discussion}

\subsection{Comparison to previous literature}

There have been a series of pioneering papers simulating the non-thermal
emission from clusters by numerically modelling discretised CR energy
spectra on top of Eulerian grid-based cosmological simulations
\citep{2001CoPhC.141...17M, 2001ApJ...559...59M, 2001ApJ...562..233M,
  2002MNRAS.337..199M, 2003MNRAS.342.1009M}. In contrast to our approach,
these models neglected the hydrodynamic pressure of the CR component, were
quite limited in their adaptive resolution capability, and they neglected
dissipative gas physics including radiative cooling, star formation, and
supernova feedback. Comparing the non-thermal emission characteristics of
primary CR electrons, hadronically generated secondary CR electrons, and pion
decay $\gamma$-rays, we confirm the general picture put forward by these
authors while we find important differences on smaller scales especially in
cluster cores. Our inhomogeneous, peripheral radio relic emission resembles
their findings. However, the hadronic component of our simulated radio halos
is more centrally concentrated \citep[cf.][]{2001ApJ...562..233M}. Our
simulations both agree that the predicted level of hard X-ray inverse Compton
emission falls short of the claimed detection in Coma and Perseus albeit the
discrepancy is more dramatic in our simulations. We confirm that the
high-energy $\gamma$-ray emission ($E_\gamma > 100$~MeV) from cluster cores is
dominated by pion decays while at lower energies, the IC emission of secondary
CR electrons takes over \citep{2003MNRAS.342.1009M} -- at least for
non-merging clusters. We reproduce their finding that the $\gamma$-ray
emission in the virial regions of clusters and beyond in super-cluster regions
stems from IC emission of primary shock accelerated electrons. Contrarily to
these authors, we find that the surface brightness of this emission component
remains sub-dominant in projection compared to the hadronically induced
emission components in the cluster core and that the pion decay completely
dominates the high-energy $\gamma$-ray emission of clusters. We note that our
$\gamma$-ray fluxes from clusters are typically a factor of two smaller than
the estimates given in \citet{2001ApJ...559...59M}. As worked out in
\citetalias{2007PfrommerIII}, this has important implications for the number
of detectable $\gamma$-ray clusters by GLAST.

All the discrepancies can be understood by two main effects that lead to an
overestimation of the CR pressure inside the clusters simulated by
\citet{2001ApJ...559...59M} and thus overproduced the resulting non-thermal
emission: (1) \citet{2000ApJ...542..608M} identified shocks with Mach numbers
in the range $4 \lesssim \M \lesssim 5$ as the most important in thermalizing
the plasma. In contrast, \citet{2003ApJ...593..599R} and
\citet{2006MNRAS.367..113P} found that the Mach number distribution peaks in
the range $1 \lesssim \M \lesssim 3$.  Since diffusive shock acceleration of
CRs depends sensitively on the Mach number, this implies a more efficient CR
injection in the simulations by \citet{2001ApJ...559...59M}. (2) The
grid-based cosmological simulations have been performed in a cosmological box
of side-length $50\,h^{-1}$~Mpc with a spatial resolution of
$200\,h^{-1}$~kpc, assuming an Einstein-de Sitter cosmological model
\citep{2001ApJ...559...59M}.  The lack of resolution in the observationally
accessible, dense central regions of clusters in the grid-based approach
underestimates CR cooling processes such as Coulomb and hadronic losses.
Secondly, these simulations are unable to resolve the adiabatic
compression of a composite of CRs and thermal gas, an effect that disfavours
the CR pressure relative to the thermal pressure.  To summarise, their modest
resolution in non-radiative simulations anticipates some of the results that
we obtained using high-resolution simulations with {\em radiative
  hydrodynamics and star formation}, however for different reasons.

\subsection{Limitations and future work}

An accurate description of CRs should follow the evolution of the spectral
energy distribution of CRs as a function of time and space, and keep track of
their dynamical, non-linear coupling with the hydrodynamics. We made several
simplifying assumptions to enable the task of following CR physics
self-consistently in cosmological simulations of smoothed particle
hydrodynamics (SPH). In the following, we outline the possibly most severe
limitations of our approach for computing the non-thermal emission processes
that will be addressed in future work \citep[cf.][ for a more complete list of
the assumptions of our CR formalism]{2007A&A...473...41E}.
\begin{enumerate}
\item We assumed a simple scaling of the magnetic energy density with the
  thermal energy density that allows us to effectively scan the observationally
  allowed parameter space for the magnetic field. Note that current SPH
  implementations that are capable of following magneto-{hy\-dro\-dy\-nam\-ics}
  (MHD) are presently still fraught with numerical and physical difficulties,
  in particular when following dissipative gas physics
  \citep{1999A&A...348..351D, 2005JCAP...01..009D, 2004MNRAS.348..139P,
  2005MNRAS.364..384P}. While the inverse Compton and pion decay emission are
  mostly independent of the magnetic field, our synchrotron maps might be
  modified when the magneto-{hy\-dro\-dy\-nam\-ics} is properly accounted for.
\item We neglected the population of re-accelerated electrons throughout this
  work: strong merger shocks and shear motions at the cluster periphery might
  inject hydrodynamic turbulence that cascades to smaller scales, feeds the MHD
  turbulence and eventually might be able to re-accelerate an aged CR electron
  population. Due to non-locality and intermittency of turbulence, this could
  partly smooth the very inhomogeneous primary emission component predominantly
  in the virial regions of clusters where simulations indicate a higher energy
  density in random motions. However, to study these effects, high-resolution
  AMR simulations are required that refine not only on the mass but also on
  some tracer for turbulence such as the dimensionless vorticity parameter.
\item In our model, the emphasis is given to the dynamical impact of CRs on
  hydrodynamics, and not on an accurate spectral representation of the CRs.
  The pion decay emission is almost independent on the spectral CR properties.
  However, the secondary CR component starts to be affected by this
  simplification since the dimensionless CR momentum $q \simeq 16
  \gamma_\e\,m_\e/m_\p \simeq 100$, that gives rise to synchrotron/IC emitting
  electrons with a Lorentz factor of $\gamma_\e \simeq 10^4$ is already quite
  high. Improving the spectral description of CR physics will not only allow
  us to study the spectral variations of the CR proton component but also
  enable reliable predictions for the TeV $\gamma$-ray emission. This is of
  great interest for imaging air \v{C}erenkov telescopes.
\item We neglected microscopic CR diffusion in our simulations. The
  diffusivity can be rewritten into a macroscopic advection term that we
  fully resolve in our Lagrangian SPH simulations by construction and a
  microscopic diffusivity.  The advection term dominates over microscopic
    term, in particular for merging clusters that are relevant for radio halos
    as the following estimate for the diffusivities shows: $\kappa_\rmn{adv} =
    100 \mbox{ kpc}\, 1000 \mbox{ km/s} = 10^{31.5} \mbox{ cm}^2/\mbox{s} \gg
    \kappa_\rmn{diff} = 10^{29} \mbox{ cm}^2/\mbox{s}$.
\item Our model of the diffusive shock acceleration mechanism assumes a
  featureless power-law for both, the proton and the electron acceleration,
  that is injected from the thermal distribution.  The complete theoretical
  understanding of this mechanism is currently an active research topic that
  includes thermalization processes of the time evolution of the kinetic
  distribution of particles \citep{2006ApJ...638..125W} as well as non-linear
  effects and magnetic field amplification \citep{2006ApJ...652.1246V}.
  Phenomenologically however, we believe that there are strong indications for
  the diffusive shock acceleration mechanism to be at work which come from
  observations of supernova remnants over a wide range of wavelengths from the
  radio, X-rays into the TeV $\gamma$-rays \citep[e.g.,][]{2000AIPC..528..383E,
    2000ApJ...543L..61H, 2005ICRC....3..261E, 2005ApJ...634..376W,
    2004Natur.432...75A, 2006Natur.439..695A} as well as the bow shock of the
  Earth \citep{1990ApJ...352..376E, 1999Ap&SS.264..481S}.  Future work will be
  dedicated on improving our model to incorporate more elaborate plasma
  physical models and to study the uncertainty of our results with respect to
  the saturated value of our CR acceleration efficiency
  \citep[e.g.,][]{2007APh....28..232K, 2007arXiv0706.0587E}.  Varying the
  physics in our simulations (non-radiative versus radiative) results in a very
  different Mach number distribution and changes the injection efficiency
  dramatically \citepalias{2007MNRAS...378..385P}. However, the resulting
  non-thermal emission is almost independent of the simulated physics. For this
  reason, we are confident that our model produces reliable results due to the
  self-regulated nature of CR proton feedback.
\item In this work, we did not account for feedback processes by AGN despite
  their importance for understanding the nature of the very X--ray luminous
  cool cores found in many clusters of galaxies. In particular we neglected an
  additional CR population that diffuses out of AGN-inflated bubbles and
  postpone their study to future work \citep{2007Sijacki}.
\item We furthermore neglected a hypothetical source of secondary electrons
  that are produced ``in-situ'' in dark matter (DM) neutralino annihilations
  \citep{2006A&A...455...21C}. However, we see two main challenges associated
  with the DM model for giant radio halos. (1) Only a sub-class of clusters
  that seems to be associated with past or present merging events exhibit radio
  halos. This fact makes it hard to believe that the dark matter with its
  nearly universal density profile in halos \citep{1996ApJ...462..563N,
    1997ApJ...490..493N} could be responsible for such an infrequent cluster
  property, especially given the stringent upper limit on the diffuse radio
  emission in some massive clusters \citep{2007ApJ...670L...5B}. (2) In order
  to explain the extended radio halo emission, the DM model needs to invoke a
  profile for the magnetic energy density, that increases by two orders of
  magnitude beyond the thermal core causing the plasma beta parameter to
  decrease by a factor of 220 to a value of 15. This behaviour is not only in
  contrast to the magnetic profile predicted by numerical MHD simulations of
  galaxy clusters \citet{2001A&A...378..777D} but also in contrast to turbulent
  dynamo models for the growth of the magnetic fields that will saturate on a
  level which is determined by the strength of the magnetic back-reaction
  \citep[e.g.,][]{2003PhRvL..90x5003S, 2006PhPl...13e6501S} and is typically a
  fraction of the turbulent energy density. Observationally, it is clear that
  the rotation measure values towards radio galaxies show a larger dispersion
  for smaller projected separation from the cluster centre in a large rotation
  measure sample \citep{2001ApJ...547L.111C}. This is mainly due to the
  distribution of these sources within the cluster atmosphere causing some
  foreground objects to experiencing a small Faraday rotation depth which is
  however no argument in favour of a decreasing magnetic profile towards the
  cluster centre.
\end{enumerate}

\section{Conclusions}
\label{sec:conclusions}

We find that the cosmic ray (CR) proton pressure traces the time integrated
non-equilibrium activities of clusters and is only modulated by the recent
dynamical activities.  In contrast, the pressure of primary shock accelerated
CR electrons resembles the current dynamical, non-equilibrium activity of
forming structure and results in an inhomogeneous and aspherical spatial
distribution. This is due to the different cooling time scales of CR electrons
and protons and reflects their large mass difference.  Hence, the radio
synchrotron and inverse Compton emission of primary electrons provides a
snapshot of violent non-equilibrium processes that are responsible for
dissipating gravitational energy associated with structure formation such as
merger shock waves. Signatures of this emission component are irregular
morphologies, large spectral variations, and a high degree of synchrotron
polarisation. On the other hand, non-thermal emission processes of pions and
secondary CR electrons produced in hadronic CR proton interactions trace the
comparably smooth CR proton distribution centred on the cluster that the CR
protons accumulate over the Hubble time.

\begin{figure}
\resizebox{\hsize}{!}{\includegraphics{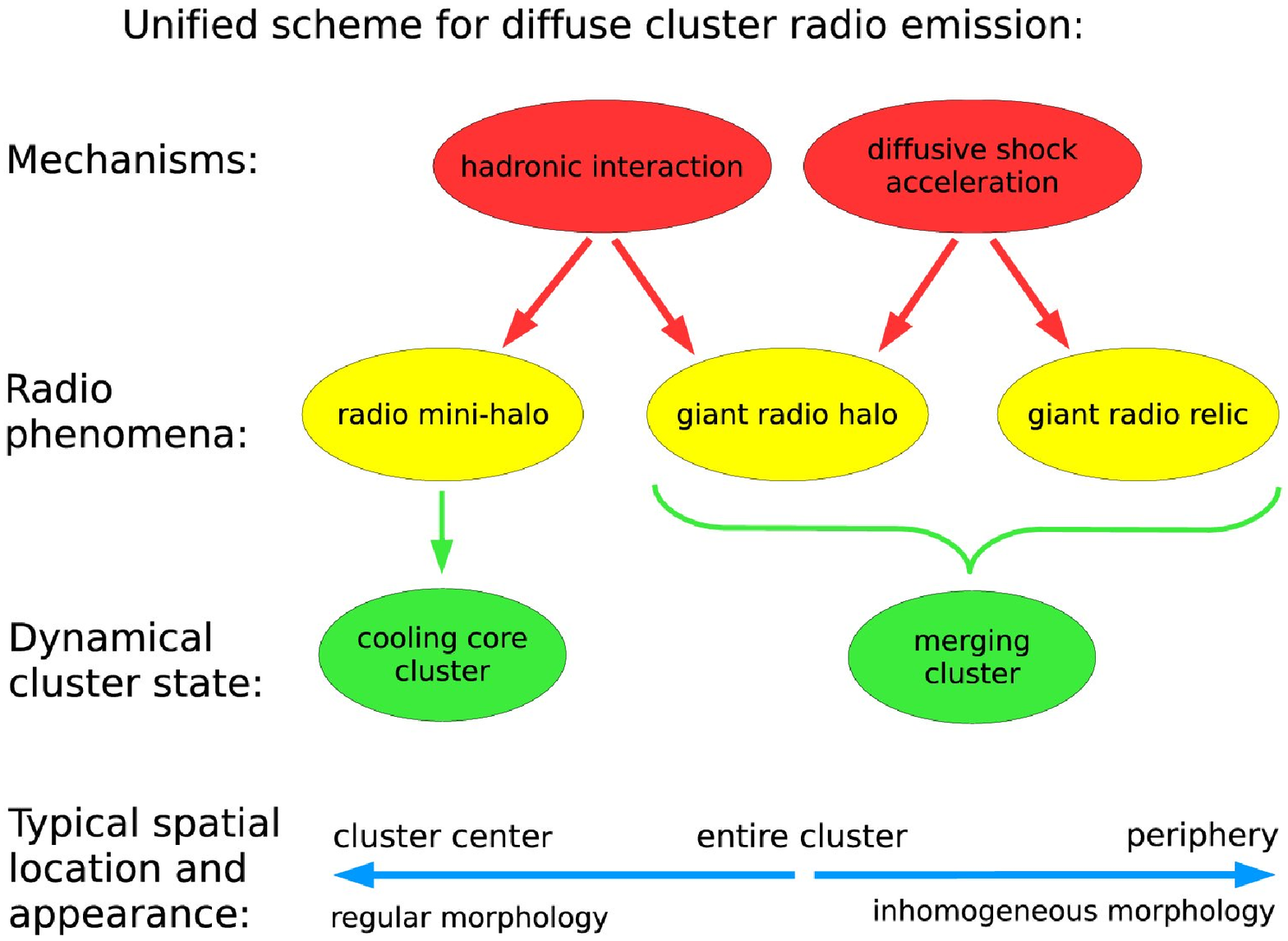}}
\caption{A cartoon showing our unified scheme for the diffuse cluster
    emission.  Hadronic cosmic ray proton interactions with ambient protons of
    the thermal ICM are thought to be responsible for radio mini-halos in
    cooling core clusters and the central parts of giant cluster radio halos in
    merging clusters. This emission mechanism produces a regular cluster-wide
    morphology resembling the thermal bremsstrahlung emission. In contrast, the
    radio synchrotron emission of shock accelerated cosmic ray electrons
    (through the Fermi 1 mechanism) is believed to be responsible for giant
    radio relics that have an inhomogeneous morphology and are primarily
    located at the cluster periphery.}
\label{fig:Unified_scheme}
\end{figure}

\subsection{Radio synchrotron emission}

{\bf Unified scheme:} we propose a {\em unified scheme for the generation of
  giant radio halos}, {\em radio mini-halos}, and {\em radio relics} that
naturally arises from our simulated radio synchrotron maps and emission
profiles. It is schematically shown in a cartoon in
  Fig.~\ref{fig:Unified_scheme}. It predicts that the diffuse radio emission
from a cluster varies with its dynamical stage as follows:
\begin{enumerate}
\item Once a cluster relaxes and develops cool core, a {\em radio mini-halo}
  develops due to synchrotron emission of hadronically produced CR electrons.
  Adiabatic compression of magnetic fields during the formation of the cool
  core should confine the observable radio synchrotron emission to the cooling
  region of the cluster. Since the cooling gas accretes onto the central black
  hole, this triggers the radio mode feedback of the AGN. Radio emission from
  the jet typically outshines the diffuse mini-halo which implies a high
  dynamic flux range. This leads to a negative selection effect that disfavours
  the detection of radio mini-halos or makes it very challenging.
\item If a cluster experiences a major merger, two leading shock waves are
  produced at first core passage that move outwards and become stronger as they
  break at the shallow peripheral cluster potential. Relativistic electrons are
  efficiently accelerated by means of diffusive shock acceleration and magnetic
  fields are amplified by shock compression and MHD turbulence at these
  shocks. Due to the short cooling times of $\tau \sim 10^8$~yrs, the
  synchrotron radiating electrons are confined to a narrow emission volume
  around the shock wave.  In combination with the preferred magnetic field
  direction in the shock surface, this implies a high degree of synchrotron
  polarisation.  The observer will typically observe one or two large-scale
  {\em radio relics}, depending on the merger characteristics such as the
  relative cluster masses, concentrations, and gas fractions, the merger
  geometry with respect to the line-of-sight, as well as the time dependent
  merger stage. Our cosmological simulations supports the picture put forward
  in isolated cluster merging simulations \citep{1996ApJ...473..651R}.
\item Simultaneously, virialisation of the gravitational energy, that is
  associated with the merger, generates a morphologically complex network of
  shock waves. The lower sound speed in the cluster outskirts imply stronger
  shocks that accelerate a spatially irregular distribution of CR electrons in
  these regions.  The injected MHD turbulence amplifies magnetic fields
  through strong shear motions and turbulent dynamo processes. The induced
  radio synchrotron emission traces these non-equilibrium processes similarly
  as the water `gischt' traces breaking non-linear waves. A {\em giant radio
    halo} develops due to (1) boost of the hadronically generated radio
  emission in the centre and a transition to the (2) irregular radio `gischt'
  emission in the cluster outskirts that represents radio synchrotron
  radiation emitted from primary, shock-accelerated electrons. 
\end{enumerate}

\noindent
{\bf Predictions:} The observational consequences of our unified scheme can be
summarised as follows:
\begin{enumerate}
\item Clusters undergoing major mergers are expected to have a giant radio
  halo with an extended radio synchrotron emission region ($R_\rmn{halo}\simeq
  0.5\, R_{200}$) while relaxed cool core clusters should host a smaller
  radio-mini halo.
\item Our simulated radio luminosities reproduce observed luminosities of
  halos/relics for magnetic fields derived from Faraday rotation measurements
  \citepalias[cf.][]{2007PfrommerIII}.
\item The regular morphology of the central parts of giant radio halos is a
  consequence of the dominant contribution of hadronically produced electrons.
\item The morphology and the radio spectral index in the radio halo
  periphery is predicted to show large variations due to the dominant
  contribution from primary CR electrons generated by shock waves.
  Superposing in projection many causally disconnected synchrotron emitting
  shock regions with different shock strength and thus electron spectral
  indices leads to spectral variations.
\item The amount of the primary radio emission depends critically on the
  characteristics of the merger.  We thus expect a large scatter in the scaling
  relation of the radio halo luminosities with cluster mass as well as in the
  pixel-to-pixel correlation of the thermal X-ray brightness with radio
  surface brightness.
\item The central radio emission should be Faraday depolarised assuming
  statistically isotropic distribution of magnetic field while the external
  emission regions are expected to have a small degree of polarisation.  As a
  word of caution, in order to detect this polarisation one might be forced to
  go out to large impact parameters where the resulting synchrotron surface
  brightness is small and the emission is dominated by very few contributing
  emission regions along the line-of-sight.
\end{enumerate}

These predictions from our cosmological high-resolution simulations
successfully reproduce characteristics of observed radio relics, giant
radio halos, as well as radio mini halos \citep{2004JKAS...37..315F,
  2006MNRAS.369.1577C}.  In our approach, we choose the magnetic energy
density to scale with the thermal energy.  The formation of a cool core is
expected to compress the magnetic field adiabatically and should be
responsible for the peaked central radio mini-halo emission profile.  This
effect should reinforce our observed difference in emission size between giant
radio halos and mini-halos. 

The observed correlation between radio halos and merging clusters implies a
departure of these systems from hydrostatic equilibrium and leads to a
complicated non-spherical morphology.  The resulting X-ray mass estimates are
subject to large uncertainties which makes the analysis and theoretical model
building based on azimuthally averaged quantities questionable if not
impossible since it causes loss of information and might yield biased results.
For this reason, we analyse {\em pixel-to-pixel correlations} of the thermal
X-ray surface brightness with non-thermal cluster emission processes.  We find
that the hadronically induced non-thermal emission is tightly correlated with
the thermal bremsstrahlung emission with the slope depending on the
realisation of the magnetic field. In contrast, the correlation is much weaker
and the scatter is increased in the case of primary non-thermal emission where
structures in the correlation space density correspond to individual structure
formation shock waves. This implies that in general, simulations will only be
able to provide quantitative predictions for the statistical behaviour rather
than deterministic predictions for the correlations.  Our new radio halo model
matches qualitatively the observed tight correlation at high surface
brightness which broadens and flattens towards dimmer brightness
\citep[cf.][]{2001A&A...369..441G}. However, our correlation is slightly
steeper than observed ones.  Taking into account the uncertainty of the
magnetic field model we conclude that observed pixel-to-pixel correlations
support our model.  

\noindent
{\bf Future:} What can we learn from a future, large sample of clusters that
show diffuse radio emission?
\begin{enumerate}
\item Radio relics and giant radio halos occur in dynamically merging clusters
  and indicate a departure of these systems from hydrostatic equilibrium and
  spherical symmetry. This has to be taken into account in the derivation of
  the cluster mass.
\item The orthogonal information about the dynamical cluster activity, that in
  general can not be obtained from the thermal cluster observables such as
  X-ray emission and Sunyaev-Zel'dovich effect, will help us in constructing a
  `gold cluster sample' for cosmology.
\item The property of the spatially confined radio relic emission from shock
  accelerated electrons might be employed to solve the inversion problem of
  reconstructing the course of a merger event given the thermal and radio
  synchrotron observables.
\item Combining high-resolution X-ray, Sunyaev-Zel'dovich, and radio
  observations will allow us to probe fundamental plasma physics: diffusive
  shock acceleration, large scale magnetic fields, and turbulence.
\end{enumerate}

\subsection{Inverse Compton and $\gamma$-ray emission}

In principle, inverse Compton (IC) emission and high-energy $\gamma$-ray
emission from decaying pions, produced in hadronic CR interactions, is the
cleanest way of probing current structure formation shock waves as well as time
integrated non-equilibrium cluster activity. This is because these non-thermal
emission components are not weighted with the magnetic energy density as it is
the case for synchrotron emission.  Our main findings can be summarised as
follows.
\begin{enumerate}
\item We identify two main regions for the generation of non-thermal emission
  in clusters of galaxies: the core that is also emitting thermal X-rays and
  the virial regions where the accretion shocks reside and merging
  shock waves break at the shallower cluster potential.
\item In the {\em cluster core regions}, the emission for energies
  $E_\gamma>100$~MeV is dominated by pion decay $\gamma$-rays. At lower
  energies, the IC emission from secondary CR electrons dominate the
  emission. Only in merging clusters, the situation may be reversed for the {\em
  outer cluster regions} where the primary IC emission can attain a similar
  flux level as the pion decay emission and even exceed the secondary IC
  emission at lower energies.
\item While the total high-energy $\gamma$-ray emission is always dominated by
  the pion decay component irrespective of the clusters dynamical state, the
  total hard X-ray IC emission can be dominated by either primary or secondary
  emission components, depending whether a major merger takes place that boosts
  the primary IC emission.
\item A corollary of this is that the high-energy $\gamma$-ray emission can be
  reliably predicted for massive clusters using a scaling relation of
  non-thermal emission and the cluster mass. In contrast, the hard X-ray
  emission of even massive clusters is subject to large flux variations that
  depend sensitively on the dynamical state of the cluster.  
\item Due to larger variation of merging histories and the smaller
  gravitational potential in less massive systems, their CR pressure and the
  associated $\gamma$-ray emission level is subject to larger modulation and
  reflects more sensitively the current merging activity of the cluster than
  it is the case in large systems.
\item The morphology of the pion decay as well as the secondary IC component
  resemble the thermal X-ray emission albeit they decrease less steeply with
  growing radius and extend further out. This is due to the increasing CR
  number fraction $f_\CR = n_\CR / n_\rmn{th}$ with increasing radius and
  reflects the more efficient CR acceleration at stronger shocks in the
  cluster periphery.  The morphology of the primary IC emission is irregularly
  shaped and traces current non-equilibrium phenomena such as merger or
  accretion shock waves.
\item Possibly most surprising, we find that the dominant emission component
  at the centre (primary or secondary IC for $E_\gamma>10$~keV and pion decay
  $\gamma$-rays for $E_\gamma>100$~MeV) depends only weakly on whether
  radiative or non-radiative gas physics is simulated provided we consider in
  both cases only CRs from structure formation shocks. This is mainly due to
  self-regulating effects of the CR pressure.
\item Measuring the hard X-ray and $\gamma$-ray emission will have a huge
  astrophysical impact and teach us about: the CR pressure contribution to the
  intra-cluster medium, the generating mechanisms of radio halos (such that we
  can use them in addition to thermal observables to characterise clusters),
  the contribution of the pion decay emission as well as the primary and
  secondary IC radiation to the $\gamma$-ray background.
\item Detecting the non-thermal spectrum ranging from X-rays to $\gamma$-rays
  will enable us to probe fundamental plasma physics on large cluster scales
  such as inferring the energy conversion efficiency of diffusive shock
  acceleration of protons and electrons as well as probing the large scale
  magnetic fields.
\end{enumerate}

\section*{Acknowledgements}
All computations were performed on CITA's McKenzie cluster
\citep{2003...McKenzie} which was funded by the Canada Foundation for
Innovation and the Ontario Innovation Trust.

\bibliography{bibtex/gadget} \bibliographystyle{mn2e}

\appendix

\section{Relativistic electron populations}
\label{sec:electron_pop}

\subsection{Definitions}
\label{sec:def}
Throughout the paper we use the following definitions for the differential
source function $q(\vecbf{r},E)$, the emissivity $j(\vecbf{r},E)$ and the
volume integrated quantities, respectively:

\noindent
\resizebox{\hsize}{!}{
\begin{minipage}[t]{0.5\hsize}
\begin{eqnarray*}
q(\vecbf{r},E) &=& \frac{\dd^3 N}{\dd t\, \dd V\, \dd E}\,,\\
Q(E)         &=& \int \dd V\, q(\vecbf{r},E)\,,
\end{eqnarray*}
\end{minipage}\hfill
\begin{minipage}[t]{0.5\hsize}
\begin{eqnarray}
\rule{0mm}{5mm}   j(\vecbf{r},E) &=& E\, q(\vecbf{r},E)\,,\\
\rule{0mm}{6.2mm} J(E)         &=& E\, Q(E)\,,\label{Q}
\end{eqnarray}
\vspace{0.1cm}
\end{minipage}}
where $N$ denotes the integrated number of particles.
From the source function the integrated number density production rate of
particles $\lambda(\vecbf{r})$, the number of particles produced per unit time
interval within a certain volume, $\mathcal{L}$, and the particle flux
$\mathcal{F}$ can be derived. 
The definitions of the energy weighted quantities are denoted on the right hand
side, respectively,

\noindent
\resizebox{\hsize}{!}{
\begin{minipage}[t]{0.5\hsize}
\begin{eqnarray*}
\lambda(\vecbf{r})&=& \int\dd E\,q(\vecbf{r},E)\,,\\
\mathcal{L}     &=& \int\dd V\,\lambda(\vecbf{r})\,,\\
\mathcal{F}     &=& \frac{\mathcal{L}}{4\,\pi\,D^2}\,,
\end{eqnarray*}
\end{minipage}\hfill
\begin{minipage}[t]{0.5\hsize}
\begin{eqnarray}
\Lambda(\vecbf{r})&=& \int\dd E\,E\,q(\vecbf{r},E)\,,\\
L               &=& \int\dd V\,\Lambda(\vecbf{r})\,,\label{luminosity}\\
F               &=& \frac{L}{4\,\pi\,D^2}\,.\label{flux}
\end{eqnarray}
\vspace{0.1cm}
\end{minipage}}

\subsection{Timescales}
\label{sec:timescales}

This section presents general considerations for deriving the characteristic
electron momentum scales of the distribution function. To this end, we compare
the energy loss timescales $\tau_\rmn{loss} = -E / \dot{E}_\rmn{loss}$ to the
acceleration timescales as a function of particle kinetic energy for the most
important processes in this context.  The detailed calculations for the
equipartition distribution of CR electrons can be found in the next two
sections.  The particle kinetic energy $E / (m_\e c^2)= \gamma-1 =
(1-\beta^2)^{-1/2}-1$ is defined in terms of the Lorentz factor $\gamma$ and
the velocity of the electron, $\beta c$.  Energy loss processes due to Coulomb
interactions \citep{1972Phy....60..145G}, inverse Compton (IC) and synchrotron
emission \citep{1979rpa..book.....R} are defined as follows:
\begin{eqnarray}
  \label{eq:Coul}
  -\dot{E}_\rmn{Coul} &=& 
  \frac{3\, \sigma_\T\, m_\e c^3\, n_\e}{2 \beta}
  \left[\ln\left(\frac{m_\e c^2\, \beta \sqrt{\gamma-1}}
      {\hbar \omega_\rmn{pl}} \right)\right.\nonumber \\
      &&- \left.
      \left(\frac{\beta^2}{2} + \frac{1}{\gamma}\right)\,\ln 2 + 
      \frac{1}{2} + 
      \left(\frac{\gamma - 1}{4\gamma}\right)^2\right], \\
  \label{eq:ICcooling}
  -\dot{E}_\rmn{IC,synch} &=&
  \frac{4}{3}\,\sigma_\T\, c\, \left(\eps_\rmn{ph} + \eps_B\right)
  \,\gamma^2\beta^2.
\end{eqnarray}
Here $\sigma_\rmn{T} = 8\upi r_\e^2 / 3$ is the Thomson cross section,
$r_\e = e^2 / (m_\e c^2)$ the classical electron radius, $\omega_{\rm
  pl} = \sqrt{4 \upi e^2 n_{\rm e} /m_{\rm e}}$ is the plasma
frequency, and $n_\e$ is the number density of free electrons. IC
losses of electrons depend on the energy density of the cosmic
microwave background (CMB) and the starlight photon field,
$\eps_\rmn{ph} = \eps_\rmn{CMB} + \eps_\rmn{stars}$, where we neglect
the latter one for simplicity and express $\eps_\rmn{CMB} =
B^2_\rmn{CMB}/(8\upi)$ by an equivalent field strength $B_\rmn{CMB} =
3.24\, (1+z)^2\umu\mbox{G}$.  Synchrotron losses of an isotropic pitch
angle distribution of electrons depend on the energy density of the
local magnetic field, $\eps_B = \bra B^2 \ket/(8\upi)$, where $B =
\sqrt{\bra \vecbf{B}^2\ket}$ is the {\em rms} of the magnetic vector
field $\vecbf{B}$.  Comparing these two loss processes, we obtain a
synchrotron dominated cooling regime for $B > B_\rmn{CMB}$ and an IC
dominated regime in the weak field limit.  The timescale for CR
electron injection by means of hadronic interactions of CR protons
with ambient protons of the thermal plasma is given by:
\begin{equation}
  \label{eq:pp}
  -\dot{E}_\rmn{pp} = E \,c\, \bar{\sigma}_\rmn{pp}\, n_\rmn{N},
\end{equation}
where $\bar{\sigma}_\rmn{pp} = 32 \mbox{ mbarn}$ is the average inelastic
cross section for proton-proton interactions, $n_\rmn{N} = n_\rmn{H} +
4n_\rmn{He} = \rho / m_\p$ is the number density of target nucleons. 

\begin{figure}
\resizebox{\hsize}{!}{\includegraphics{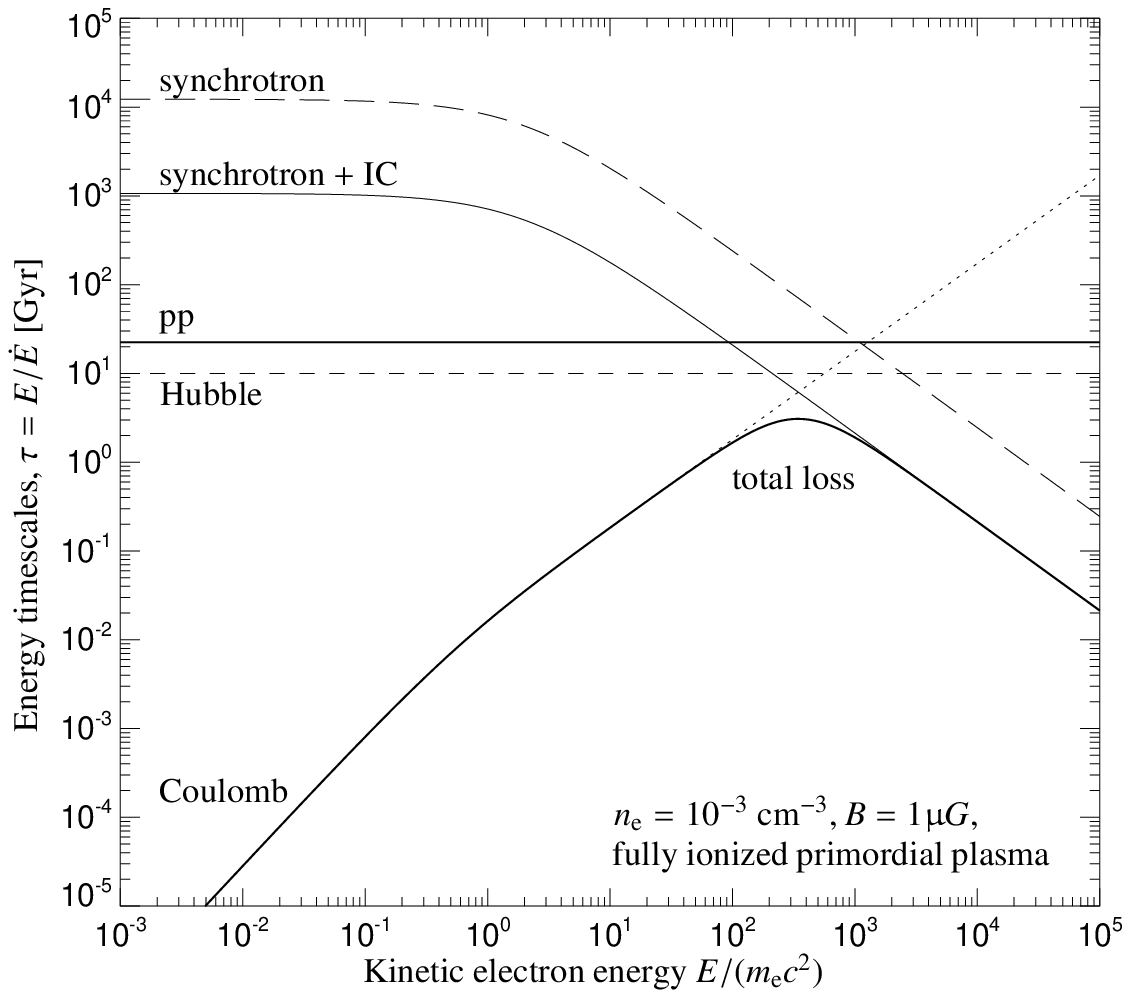}}
  \caption{Energy gain and loss timescales as a function of the kinetic energy
    of electrons for typical conditions of the ICM. The three solid lines from
    the bottom to the top denote the total loss timescale for electrons, the
    timescale due to hadronic injection of secondary electrons (pp), and the
    combined inverse Compton (IC)/synchrotron cooling timescale. The dotted
    line shows the Coulomb timescale, the long dashed one the synchrotron
    timescale, and the dashed one the Hubble time. Note that the relative
    normalisation of the hadronic injection timescale compared to the total
    loss timescale is subject to different density dependencies and the shock
    acceleration timescale depends crucially on the properties of structure
    formation shocks. }
  \label{fig:timescales}
\end{figure}

Comparing the different energy gain and loss rates of relativistic electrons
yields characteristic momentum scales that are responsible for spectral breaks
or cutoffs in the CR electron distribution function. Conveniently, we denote
these characteristic momentum scales with the dimensionless electron momentum
$q = \beta\gamma = P_\e\, (m_\e c)^{-1}$ and label these with the competing
processes considered. Equating the timescale for Coulomb interactions and
IC/synchrotron losses yields
\begin{eqnarray}
  \label{eq:q_Coulomb,IC}
  q_\rmn{Coul,IC/synch} &=& 
  \left\{
    \frac{9\,m_\e c^2\,n_\e}{8\, \left(\eps_\rmn{ph} + \eps_B\right)}
    \left[
      \ln\left(\frac{m_\e c^2 \bra \gamma^{1/2} \ket}{\hbar \omega_\rmn{pl}}\right)
      + 0.216
    \right]
  \right\}^{1/2} \\
  &\simeq& 
  300 \left(\frac{n_\e}{10^{-3}\mbox{cm}^{-3}}\right)^{1/2}.\nonumber
\end{eqnarray}
In the last step and the following numerical examples, we assume the IC cooling
regime for simplicity.  This energy scale provides a bottleneck through which
high-energy electrons have to pass when they age and one expects a
characteristic pile-up at this energy scale in their distribution function for
an integration over the energy injection history of different CR electron
populations \citep{1999ApJ...520..529S}.

The timescale for diffusive shock acceleration of electrons at cosmological
shock waves is much shorter than cosmologically relevant timescales,
$\tau_\rmn{shock} \ll 1$~Gyr. This implies that $\tau_\rmn{shock}$ intersects
the total loss timescale in the low-energy Coulomb regime, $\tau_\rmn{Coul}$,
as well as in the high-energy IC/synchrotron regime, $\tau_\rmn{IC/synch}$
(cf.~Fig.~\ref{fig:timescales}). We can thus identify two characteristic
momenta of the {\em primary population of electrons} that are obtained by
equating $\tau_\rmn{shock}$ with the Coulomb and the IC/synchrotron timescale,
\begin{eqnarray}
\label{eq:inj,IC/synch}
q_\rmn{inj,IC/synch} &=&  \frac{3\, m_\e c}
{4\,\sigma_\rmn{T} \tau_\inj\,(\eps_B + \eps_\rmn{ph})},\\
\label{eq:inj,Coul}
q_\rmn{inj,Coul} &=&  \frac{3}{2}\,\sigma_\rmn{T} c\, n_\e\,\tau_\inj\, 
\left[
  \ln\left(\frac{m_\e c^2\, \bra\gamma^{1/2}\ket}{\hbar \omega_\rmn{pl}}\right)
  + 0.216
\right].
\end{eqnarray}
At high energies, we expect to have an IC/synchrotron cooled electron spectrum
that joins at lower energies into the shock injection spectrum of CR electrons
that have had no time to cool radiatively yet. The low momentum cutoff of the
CR electron distribution function is determined by Coulomb losses.

\begin{figure}
\resizebox{\hsize}{!}{\includegraphics{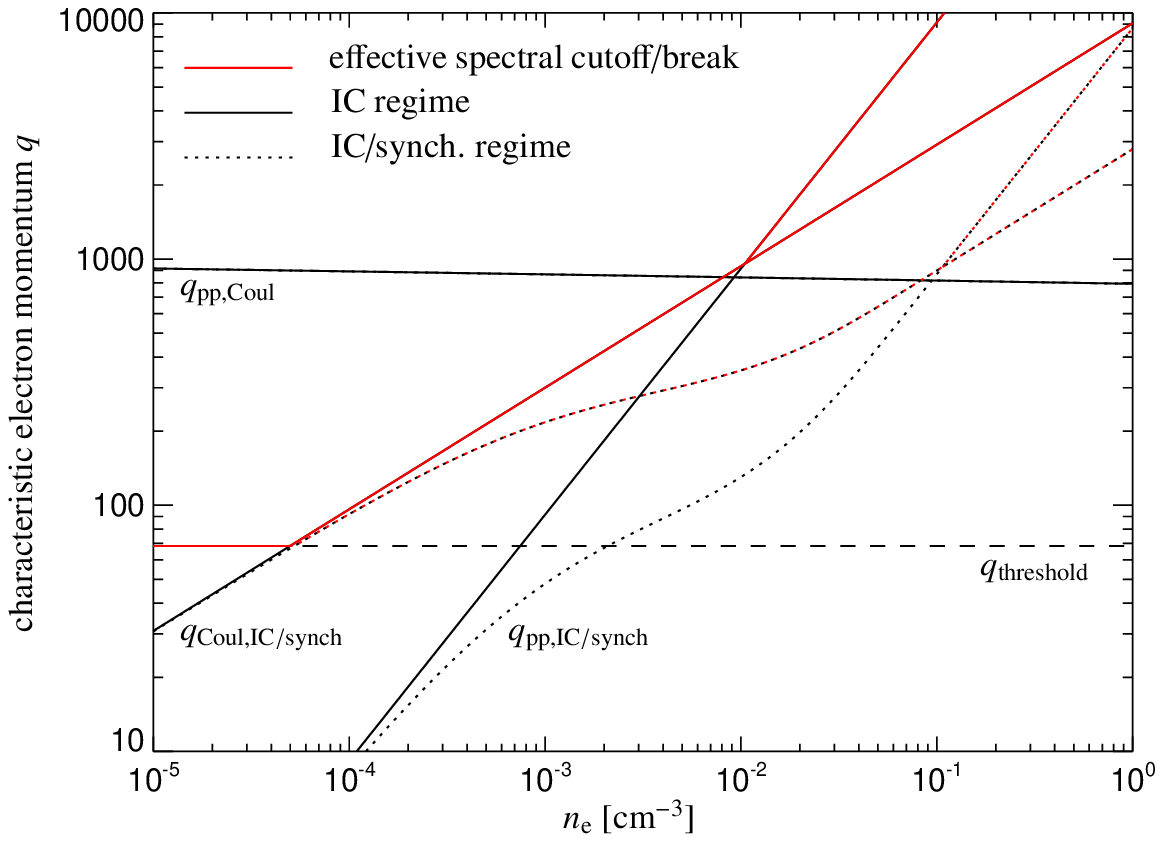}}
  \caption{Density dependence of characteristic momentum scales $q= P_\e\,
    (m_\e c)^{-1}$ for a hadronically injected {\em secondary CR electron
      population}.  Comparing the different energy gain and loss rates of
    competing processes such as hadronic injection of secondary electrons
    (pp), Coulomb cooling, and IC/synchrotron cooling yields characteristic
    momentum scales that are responsible for spectral breaks or cutoffs in the
    CR electron distribution function. The solid red line denotes the
    effective spectral cutoff: for $n_\e \lesssim 10^{-2}\mbox{ cm}^{-3}$, it
    denotes the lower cutoff of the power-law distribution function. At higher
    densities, it denotes the lower cutoff $q_\rmn{Coul,IC}$ and the spectral
    break ($q_\rmn{pp,IC}$) of the broken power-law spectrum. The dotted lines
    include synchrotron losses and assume a scaling of the magnetic field of
    $\eps_B\propto \eps_\th$ with a saturation value for $B$ at $B_\rmn{max} =
    10\,\umu$G.}
  \label{fig:q_ne}
\end{figure}

Characteristic momenta of the {\em secondary population of electrons} are obtained
by equating the energy injection rate through hadronic proton interactions
with the energy loss rates,
\begin{eqnarray}
  \label{eq:q_pp}
  q_\rmn{threshold} &=& \frac{m_{\pi^\pm}}{4\, m_\e} \simeq 70, \\
  \label{eq:q_pp,IC}
  q_\rmn{pp,IC/synch} &=& \frac{3\, \bar{\sigma}_{pp}\, 
    n_\rmn{N}\, m_\e\, c^2}
  {4\, \sigma_\T\, \left(\eps_\rmn{ph} + \eps_B\right)} \simeq 
  70\, \left(\frac{n_\e}{10^{-3}\mbox{cm}^{-3}}\right), \\
  \label{eq:q_pp,Coulomb}
  q_\rmn{pp,Coul} &=& 
  \frac{3\, \sigma_\T\, n_\e}{2\,\bar{\sigma}_\rmn{pp}\, n_\rmn{N}}
    \left[
      \ln\left(\frac{m_\e c^2\, \bra\gamma^{1/2}\ket}{\hbar \omega_\rmn{pl}}\right)
      +0.216
    \right] \\
  &\simeq& 
  1300 - \ln\left(\frac{n_\e}{10^{-3}\mbox{cm}^{-3}}\right)^{1/2}.\nonumber
\end{eqnarray}
$q_\rmn{threshold}$ reflects the threshold momentum for the inelastic
proton-proton reaction. The shortest equipartition timescale at these
characteristic momenta dominates the resulting electron equilibrium
distribution. In the case of a double-valued solution for $q$, i.e.{\ }when we
obtain two equipartition cutoffs with a similar timescale, we choose the
larger one which is in equilibrium with the IC/synchrotron cooling.  All these
momentum scales have different density dependencies which are visualised in
Fig.~\ref{fig:q_ne}.  At typical densities of the ICM for $n_\e\lesssim
10^{-2}\mbox{ cm}^{-3}$, the two momentum scales $q_\rmn{pp,IC/synch}$ and
$q_\rmn{pp,Coul}$ have an associated timescale that is much longer than the
timescale at $q_\rmn{Coul, IC/synch}$ as can be readily inferred from
Fig.~\ref{fig:timescales}.  This implies that the equilibrium distribution
function of secondary CR electrons has a low-energy cutoff at $q_\rmn{Coul,
  IC/synch}$.

The situation is reversed for the dense ICM or the inter-stellar medium with
electron densities $n_\e \gtrsim 10^{-2}\mbox{ cm}^{-3}$ (assuming the IC
dominated cooling regime), and the momentum scale $q_\rmn{Coul, IC/synch}$
drops out of the problem due to its long timescale. The equilibrium
distribution function of secondary CR electrons develops a break at
$q_\rmn{pp,IC/synch}$ above which the secondary electron injection is in
equilibrium with IC/synchrotron cooling and below which the electron injection
spectrum remains unchanged (similar to the case of the primary CR electron
population). The lower cutoff of the distribution function is given by
$q_\rmn{pp,Coul}$, provided it exceeds the threshold $q_\rmn{threshold}$ for
the hadronic reaction and provided the CR proton distribution extends down to
these low energies. The effective spectral cutoff $q_\rmn{eff}$ of the
electron distribution function is visualised in Fig.~\ref{fig:q_ne} as a {\em
  solid red line}. For $n_\e \lesssim 10^{-2}\mbox{ cm}^{-3}$, we have a
simple power-law with a lower cutoff given by $q_\rmn{eff}$. At higher
densities, the lower solid red line denotes again the cutoff of the
distribution function, while the upper red line (that coincides with
$q_\rmn{pp,IC}$) denotes the spectral break above which the hadronic injection
and IC/synchrotron cooling established a steady state spectrum.

So far, we only considered regime of weak magnetic fields where IC cooling
dominates. The complete picture including synchrotron cooling is however only
slightly changed due to the following line of arguments.  Assuming a simple
scaling model for the magnetic energy density, $\eps_B \propto n_\e$, in
Eqns.~(\ref{eq:q_Coulomb,IC}) and (\ref{eq:q_pp,IC}) will cause these momentum
scales to become independent of density, as modelled in Fig.~\ref{fig:q_ne}
with {\em dotted lines}. Eventually, the enhancement of the magnetic field
strength through turbulent dynamo processes will saturate on a level which is
determined by the strength of the magnetic back-reaction
\citep[e.g.,][]{2003PhRvL..90x5003S}. For illustrative purposes in
Fig.~\ref{fig:q_ne}, we model such a saturation effect with a simple model,
$\eps_B = \eps_{B_0}\,(1 - \exp(n_\e / n_{\e_0}))$, where we chose
$B_\rmn{max} = 10\, \umu$G and $n_{\e_0}=10^{-2}\mbox{ cm}^{-3}$. This causes
the transition to the broken power-law CR electron spectrum to occur at the
higher critical electron density
\begin{equation}
\label{eq:ne_crit}
n_\rmn{e,crit} = 10^{-2}\mbox{ cm}^{-3}\, 
\left(\frac{\eps_{B_\rmn{max}} + \eps_\rmn{CMB}}{\eps_\rmn{CMB}}\right),
\end{equation}
where $\eps_{B_\rmn{max}} = B^2_\rmn{max}/(8\upi)$ and we assumed a plasma of
primordial composition with a hydrogen mass fraction of $X_\rmn{H}=0.76$ and
full ionisation. These modifications due to synchrotron losses cause a curvature
of the straight lines in the log-log representation in Fig.~\ref{fig:q_ne} of
the original power-law dependencies on $n_\e$. As before, shown in red is the
spectral cutoff and spectral break, respectively, above which the hadronic
injection and IC/synchrotron cooling established a steady state spectrum.

\subsection{Shock-accelerated electron population}
\label{sec:fe}

\subsubsection{Injection spectrum}
\label{sec:fe_inj}

In this section, we discuss electron acceleration processes at shock
waves due to gas accretion and galaxy mergers, using the framework of
{\em diffusive shock acceleration} using the thermal leakage model
  originally proposed by \citet{1981JGZG...50..110E}.  Our
description follows the approach of \citet{2007A&A...473...41E} for
the acceleration of CR protons.  The shock surface separates two
regions: the {\em upstream regime} defines the region in front of the
shock which is causally unconnected for super sonic shock waves,
whereas the wake of the shock wave defines the {\em downstream
  regime}.  The shock front itself is the region in which the mean
plasma velocity changes rapidly on a scale of the order of the plasma
skin depth. In the rest frame of the shock, particles are impinging
onto the shock surface at a rate per unit area of $\rho_2 \vel_2 =
\rho_1 \vel_1$. Here $\vel_1$ and $\vel_2$ give the plasma velocities
(relative to the shock's rest frame) in the upstream and downstream
regimes of the shock, respectively. The corresponding mass densities
are denoted by $\rho_1$ and $\rho_2$.

We assume that after passing though the shock front most of the electron gas
thermalizes to a Maxwell-Boltzmann distribution with characteristic post-shock
temperature $T_2$ and the dimensionless electron momentum $p = P_\e\, (m_\e
c)^{-1}$:
\begin{equation}
  \label{eq:MBdistibution}
  f_\rmn{e}(p) = 4\upi\, n_\rmn{e}\,
  \left(\frac{m_\e c^2}{2 \upi\, k T_2}\right)^{3/2}\! p^2
  \exp\left(-\frac{m_\e c^2\,p^2 }{2 \,k T_2}\right),
\end{equation}
where the number density of electrons of the thermal distribution in the
downstream regime, $n_\rmn{e}=n_2$, as well as $T_2$ can be inferred by
means of the mass, momentum, and energy conservation laws at the shock surface
for a gas composed of relativistic particles and thermal constituents.  Note
that we use an effective one-dimensional distribution function $f(p)\equiv
4\upi p^2 f^{(3)}(p)$.  In our simulations, we follow the spatial and temporal
evolution of the hydrodynamic quantities such as temperature and density
(although for brevity we suppress this in our notation).  For cosmological
applications, we have to consider the primordial composition of the
cosmological fluid, i.e.~the ionised electron number density is given by
$n_\rmn{e} = X_\rmn{H} x_\e \rho / m_\p$, where $X_\rmn{H}=0.76$ is the
primordial hydrogen mass fraction, and $x_\e$ is the ratio of electron and
hydrogen number densities which we dynamically track in our radiative
simulations.  Assuming that a fraction of the thermalized particles experience
stochastic shock acceleration by diffusing back and forth over the shock front,
the test particle theory of diffusive shock acceleration predicts a resulting
CR power-law distribution in momentum space.  Within our model, this CR
injection mechanism can be treated as an instantaneous process.

For a particle in the downstream region of the shock to return upstream it is
necessary to meet two requirements. The particle's effective velocity component
parallel to the shock normal has to be larger than the velocity of the shock
wave, and secondly, its energy has to be large enough to escape the
``trapping'' process by Alfv\'en waves that are generated in the downstream
turbulence \citep{1995A&A...300..605M,1998AdSpR..21..551M}.  Thus, only
particles of the high-energy tail of the distribution are able to return to the
upstream shock regime in order to become accelerated. The complicated detailed
physical processes of the specific underlying acceleration mechanism are
conveniently compressed into a few parameters
\citep{1993ApJ...402..560J,1994APh.....2..215B,1995ApJ...447..944K}, one of
which defines the momentum threshold for the particles of the thermal
distribution to be accelerated,
\begin{equation}
  \label{eq:qinj}
  q_\rmn{inj} = x_\rmn{inj} p_\rmn{e} = 
  x_\rmn{inj} \sqrt{\frac{2 \,k T_2}{m_\e c^2}}.
\end{equation}

In the linear regime of CR electron acceleration, the thermal distribution
joins in a smooth manner into the resulting CR electron power-law distribution
at $q_\rmn{inj}$ so that $x_\rmn{inj}$ represents the only parameter in our
simplified diffusive shock acceleration model,
\begin{equation}
  \label{eq:finj}
  f_\rmn{lin}(p) = 
  f_\rmn{e}(q_\rmn{inj}) \left(\frac{p}{q_\rmn{inj}}\right)^{-\alpha_\rmn{inj}} 
  \theta(p-q_\rmn{inj}).
\end{equation}
The slope of the injected CR electron spectrum is given by
\begin{equation}
  \label{eq:ainj}
  \alpha_\rmn{inj} = \frac{r + 2}{r - 1}, \quad\mbox{where}\quad
  r = \frac{\rho_2}{\rho_1} = \frac{\vel_1}{\vel_2} 
\end{equation}
denotes the shock compression ratio \citep{1978MNRAS.182..147B,
  1978MNRAS.182..443B, 1983SSRv...36...57D}.  In combination with the
slope $\alpha_\rmn{inj}$, the value of $x_\rmn{inj}$ regulates the
amount of kinetic energy which is transferred to the CR electrons.
Theoretical and observational studies of shock acceleration of CR
protons at galactic supernova remnants suggest a range of $x_\rmn{inj}
\simeq 3.3$ to $3.6$, implying a particle injection efficiency of
$\eta_\rmn{lin} \simeq 10^{-4}$ to $10^{-3}$
\citep{1989A&A...225..179D, 1993ApJ...402..560J, 1994APh.....2..215B,
  1995ApJ...447..944K, 1995A&A...300..605M, 2000AIPC..528..383E,
  2000ApJ...543L..61H, 2005ApJ...634..376W}.  In the linear regime,
the number density of injected electrons is given by
\begin{equation}
  \label{eq:ninj}
  \Delta n_\rmn{lin} = \int_0^\infty \! \dd p\, f_\rmn{lin}(p)
  = f_\rmn{e}(q_\rmn{inj})\, \frac{q_\rmn{inj}}{\alpha_\rmn{inj}-1}.
\end{equation}
This enables us to infer the particle injection efficiency which is a measure
of the fraction of downstream thermal gas particles which experience diffusive
shock acceleration,
\begin{equation}
  \label{eq:eta}
  \eta_\rmn{lin}\equiv\frac{\Delta n_\rmn{lin}}{n_\rmn{e}} = 
  \frac{4}{\sqrt{\upi}}\,\frac{x_\rmn{inj}^3}{\alpha_\rmn{inj}-1}\,
  \rmn{e}^{-x_\rmn{inj}^2}.
\end{equation}
The particle injection efficiency is independent of the downstream post-shock
temperature $T_2$.  These considerations allow us to infer the injected
electron energy density in the linear regime:
\begin{equation}
  \Delta\eps_\rmn{lin}  =
  \eta_\rmn{lin}E_\e^\rmn{inj}(\alpha_\rmn{inj},q_\rmn{inj})\,n_\rmn{e}(T_2). 
\end{equation}

The average kinetic energy of $E_\e^\rmn{inj}(\alpha_\rmn{inj}, q_\rmn{inj})$ of an
injection power-law spectrum with CR electron spectral index $\alpha_\rmn{inj}$ and
lower momentum cutoff $q_{\rm inj}$ is given by
\begin{eqnarray}
\label{eq:Te}
E_\e^\rmn{inj} &=& 
\frac{\eps_\rmn{inj}}{n_\rmn{inj}} \quad = \quad
n_\rmn{inj}^{-1} \int_0^\infty  \dd p\, f_\rmn{inj}(p) \,E(p)   \nonumber \\
 &=&\me\,c^2\left[ \frac{q^{\alpha-1}}{2}\, 
\B_\frac{1}{1+q^2} \left(\frac{\alpha-2}{2},
\frac{3-\alpha}{2} \right) + \sqrt{1+q^2} -1\right],
\end{eqnarray}
where $E(p) = (\sqrt{1+p^2} -1)\, \me\,c^2$ is the kinetic energy of an
electron with momentum $p$, the electron distribution function $f_\rmn{inj}$ is
given by Eqn.~(\ref{eq:f_inj}), and we used the abbreviations $\alpha =
\alpha_{\rm inj}$ and $q = q_\rmn{inj}$. $\B_x(a,b)$ denotes the incomplete
beta-function \citep{1965hmfw.book.....A}, assuming $\alpha>2$.  In our
description, the CR electron energy injection efficiency in the linear regime
is defined to be the energy density ratio of freshly injected CR electrons to
the total dissipated energy density in the downstream regime,
\begin{equation}
  \zeta_\rmn{lin} =
  \frac{\Delta\eps_\rmn{lin}}{\Delta\eps_\rmn{diss}},
   \quad\mbox{where}\quad
  \Delta\eps_\rmn{diss} = \eps_\rmn{e2} - \eps_\rmn{e1} r^\gamma.
\end{equation}
The dissipated energy density in the downstream regime,
$\Delta\eps_\rmn{diss}$, is given by the difference of the thermal energy
densities in the pre- and post-shock regimes, corrected for the
contribution of the adiabatic part of the energy increase due to the
compression of the gas over the shock.

In order to obey energy conservation as well as the saturation effect for
strong shocks, we propose the following modification of the electron injection
efficiency at high values of the Mach number:
\begin{equation}
\label{eq:saturation}
\zeta_\rmn{inj} = \left[1 - \exp\left(-\frac{\zeta_\rmn{lin}}
  {\zeta_\rmn{max}}\right)\right]\,\zeta_\rmn{max}.
\end{equation}
\citet{2003ApJ...585..128K} suggest a value of $\zeta_\rmn{max} \simeq 0.05$
for the limiting case of the electron energy injection efficiency.  One can
then infer the injected CR electron energy density in terms of the energy injection
efficiency of diffusive shock acceleration processes,
\begin{equation}
  \label{eq:zeta}
\Delta \eps_\rmn{inj} = \zeta_\rmn{inj} \Delta \eps_\rmn{diss}.
\end{equation}

Putting these considerations together, one arrives at the injection spectrum
for the electrons, 
\begin{eqnarray}
\label{eq:f_inj}
f_\inj(p)\dd p &=& C_\inj\, p^{-\alpha_\inj}\, \theta(p - q_\rmn{inj})\dd p,\\
C_\inj &=& \left( 1 - \e^{-\delta}\right)\, \delta^{-1}\, f_\e(q_\inj)\,
q_\inj^{\alpha_\inj}, \\ 
\label{eq:delta_inj}
\delta &=&\frac{\Delta \eps_{\e,\lin}}{\zeta_\rmn{max}\,\Delta\eps_\rmn{diss}}
\simeq \frac{\eta_{\e,\lin}\, E_\e^\rmn{inj}(\alpha_\inj, q_\inj)}{\zeta_\rmn{max}\,
  \dot{E}_\rmn{diss}\tau_\rmn{shock}}, \\
f_\e(q_\inj) &=& \frac{4}{\upi}\, n_\e\, x_\rmn{inj}^3\, q_\rmn{inj}^{-1}
\rmn{e}^{-x_\rmn{inj}^2},
\end{eqnarray}
where $\dot{E}_\rmn{diss} = \dot{E}_\rmn{diss,~SPH} \,m_\p /
(M_\rmn{SPH}\,X_\rmn{H}\,x_\e)$ denotes the dissipated energy per timestep and
per electron and $\tau_\rmn{shock} = f_h h/\vel$ is the time it takes the
particle to pass through the broadened shock front. The front has a
characteristic length scale that is a multiple of the SPH smoothing length $h$
(with $f_h = 2$), and one may approximate $\vel$ with the pre-shock velocity
$\vel_1 = \M_1 c_\rmn{s1}$.

\subsubsection{Equilibrium spectrum of shock accelerated electrons}
\label{sec:fe_eq}

This section describes the steady-state approximation for the equilibrium CR
electron spectrum.  This is only justified if the dynamical and diffusive
timescales are long compared to the shock injection or IC/synchrotron
timescale. This may well be the case in clusters of galaxies, however,
probably not in our own Galaxy.  Moreover, this section neglects possible
re-acceleration processes of CR electrons like continuous in-situ acceleration
via resonant pitch angle scattering by compressible MHD modes.

The steady-state CR electron spectrum at high energies $p = \beta \gamma >
\mbox{GeV}/c$ is governed by the injection of shock-accelerated CR electrons,
denoted by the source function $s_\e$, and their cooling processes so that it
can be described by the continuity equation
\begin{equation}
\frac{\partial }{\partial p} \left[ \dot{p}(p) f_\e
(p) \right] = s_\e( p)\,.
\end{equation}
For $\dot{p}(p) < 0$, this equation is solved by
\begin{equation}
\label{eq:fe}
f_\e (p) = \frac{1}{|\dot{p}(p)|} \int_{p}^\infty
\! \dd p'  s_\e( p')\,.
\end{equation}
For the energy range of interest, the cooling of the radio emitting CR
electrons is dominated by synchrotron and inverse Compton losses,
$\dot{p}_\rmn{IC,synch} = \dot{E}_\rmn{IC,synch}/(m_\e c^2)$ where
$\dot{E}_\rmn{IC,synch}$ is given by Eqn.~(\ref{eq:ICcooling}).  The source
function of the shock-accelerated CR electrons for the energy range of
interest is given by
\begin{equation}
s_\e(p) = \frac{f_\inj(p)}{\tau_\inj}
\end{equation}
In our formalism, we set $\tau_\rmn{inj} = \min(\tau_\rmn{shock},
\tau_\rmn{Hubble})$ using $\tau_\rmn{shock}$ of Eqn.~(\ref{eq:delta_inj}) due
to the following line of arguments: the freshly accelerated relativistic
electron population in post-shock regions cools and finally diminishes as a
result of loss processes.  In the interesting observational bands such as
inverse Compton (IC) $\gamma$-rays and radio synchrotron emission the electron
population diminishes on such a short timescale that we could describe this by
instantaneous cooling. In this approximation, there is no steady-state
electron population and we would have to convert the energy from the electrons
to IC and synchrotron radiation. However, we can introduce a virtual electron
population that lives in the SPH broadened shock volume only which is defined
to be the volume of energy dissipation. Within this volume that is comoving
with the shock, we can indeed use the steady state solution for the
distribution function of relativistic electrons and we assume no relativistic
electrons in the post-shock volume where there is no energy dissipation.
Thus, the cooled CR electron equilibrium spectrum can be derived from
Eqn.~(\ref{eq:fe}) yielding
\begin{eqnarray}
\label{eq:f_eq}
f_\e(p)\,\dd p &=& C_\e\, p^{-\alpha_\e}\,\dd p, \\
\label{eq:C_e_prim}
C_\e &=& \frac{3\, C_\rmn{inj}\, m_\e c}
{4\,(\alpha_\e - 2)\,\sigma_\rmn{T}\,\tau_\inj\,(\eps_B + \eps_\rmn{ph})}
\end{eqnarray}
Here, $\alpha_\e = \alpha_\inj + 1$ is the spectral index of the equilibrium
electron spectrum. The normalisation scales linearly with the gas density
$C_\e \propto \rho$ which we evolve dynamically in our simulations and depends
indirectly on $\alpha_\inj$ and $\dot{E}_\rmn{diss}$ through the variable
$C_\rmn{inj}$.

At high energies, we have the IC/synchrotron cooled power-law electron spectrum
that joins at lower energies into the shock injection spectrum which has had no
time to cool radiatively yet. The low-energy regime of the CR electron
distribution function is determined by Coulomb losses. It turns out that the
timescale associated with the momentum scale $q_\rmn{Coul,IC/synch}$ is always
larger than the injection timescale $\tau_\rmn{inj}$ such that the transition
from the injection spectrum (\ref{eq:f_inj}) to the cooled equipartition
spectrum (\ref{eq:f_eq}) takes place at the characteristic momentum
\begin{equation}
\label{eq:break}
q_\rmn{break,prim} =  \frac{3\, m_\e c}
{4\,(\alpha_\e - 2)\,\sigma_\rmn{T} \tau_\inj\,(\eps_B + \eps_\rmn{ph})}.
\end{equation}
The injection spectrum extends down to the lower cutoff $q_\rmn{min,prim} =
\max(q_\rmn{inj}, q_\rmn{inj,Coul})$ where $q_\rmn{inj}$ and
$q_\rmn{inj,Coul}$ are given by Eqns.~(\ref{eq:qinj}) and (\ref{eq:inj,Coul}),
respectively. 

The pressure of a CR electron or proton power-law population as e.g. described
by Eqn.~(\ref{eq:f_eq}), that is characterised by two momentum cutoffs $p_1$
and $p_2$ is given by
\begin{eqnarray}
\label{eq:Pcr}
P_\CR &=& \frac{m c^2}{3}\,\int_0^\infty \dd p\, f(p)\,\beta\,p\\
  &=& \frac{C\,m c^2}{6} \, 
\left[
  \B_{\frac{1}{1+p^2}} \left( \frac{\alpha-2}{2},\frac{3-\alpha}{2}
  \right)
\right]_{p_2}^{p_1} ,
\end{eqnarray}
where $\beta := \vel/c = p/\sqrt{1+p^2}$ is the dimensionless velocity of the
CR particle.  The CR population can hydrodynamically be described by an
isotropic pressure component as long as the CRs are coupled to the thermal gas
by small scale chaotic magnetic fields. Note, that for $2<\alpha<3$ the kinetic
energy density and pressure of the CR populations are well defined for the
limit $q\rightarrow 0$, although the total CR number density diverges.

\subsection{Hadronically produced electron population}
\label{sec:fe_had}

Considering CR protons, which are at least in our Galaxy the dominant CR
species, it is convenient to introduce the dimensionless proton momentum $p_\p
= P_\p/(\mp\,c)$.  We assume that the differential particle momentum spectrum
per volume element can be approximated by a single power-law:
\begin{equation}
\label{eq:fp}
f_\p(p_\p) = \frac{\dd N}{\dd p_\p\,\dd V} = C_\p \, p_\p^{-\alpha_\p}\,
\theta(p_\p- q_\p) ,
\end{equation}
where $\theta(x)$ denotes the Heaviside step function, $C_\p = C_\p
(\vecbf{x}, t)$ denotes the normalisation, $q_\p = q_\p(\vecbf{x},t)$ is the
lower cutoff of the distribution function, and $\alpha_\p$ is the CR spectral
index that is taken to be constant in space and time for simplicity.  In our
simulation, we dynamically evolve the quantities $C_\p$ and $q_\p$ according
to the dominant gain and loss processes in the intra-cluster medium.  The
modelling of the cosmic ray physics includes adiabatic CR transport processes,
injection by supernovae and cosmological structure formation shocks, as well
as CR thermalization by Coulomb interaction and catastrophic losses by
hadronic interactions. As already laid out in the introduction, the hadronic
reaction of CR protons with ambient thermal protons produces pions which decay
into secondary electrons, positrons, neutrinos and $\gamma$-rays.

There are two analytical models in the literature that describe the hadronic
proton-proton reaction while assuming isospin symmetry.
\citet{1950PTP...5.570F} proposed the {\em fireball model} which assumes a
state of hot quark-gluon plasma in thermal equilibrium after the hadronic
interaction that subsequently ablates pions with energy dependent
multiplicities. Since this model is only valid in the high-energy limit for CR
protons, we use the analytic formalism by \citet{2004A&A...413...17P} that
parametrises important effects near the pion threshold and is based on an
approximate description developed by \citet{1986ApJ...307...47D,
  1986A&A...157..223D}, which combines isobaric \citep{1970Ap&SS...6..377S}
and scaling models \citep{1977PhRvD..15..820B,1981Ap&SS..76..213S} of the
hadronic reaction.

\subsubsection{Injection spectrum}
\label{sec:fe_had,inj}

The pion production spectrum can be derived from general considerations
including branching ratios and multiplicities of the hadronic reaction
\citep{1971cgr..book.....S}. The pion production spectrum describes the
produced number of pions per unit time, volume and momentum intervals, $\dd N
/ (\dd t\, \dd V\, \dd p_\pi\, \dd p_\p)$, and reads in this context
\begin{equation}
  \label{eq:pionsource}
  s_\pi(p_\pi,p_\p) = c n_\rmn{N} \xi(p_\p) \sigma_\rmn{pp}^\pi(p_\p)
  \delta_\rmn{D}(p_\pi-\bra p_\pi\ket) \theta(p_\p - p_\rmn{p,th}),
\end{equation}
where $n_\rmn{N} = n_\rmn{H} + 4 n_\rmn{He} = \rho / m_\p$ is the target
density of nucleons in a fluid of primordial element composition,
$\sigma_\rmn{pp}^\pi$ the inelastic p-p cross section, $\bra p_\pi\ket$ the
average momentum of a single produced pion, and $p_\rmn{p,th} = 0.78$ denotes
the threshold momentum for pion production.  For a differential CR proton
distribution, the pion source function can be marginalised over the proton
energy, yielding
\begin{equation}
  \label{eq:pionsourcepm}
  s_{\pi^\pm}(p_\pi) = 
  \frac{2}{3}q_\pi(p_\pi) = 
  \frac{2}{3} \int_{-\infty}^\infty \dd p_\p f_\p (p_\p)s_\pi(p_\pi,p_\p),
\end{equation}
where the CR proton population is given by Eqn.~(\ref{eq:fp}). The scaling
behaviour in the high-energy limit of Dermer's model can be described by a
constant pion multiplicity $\xi(p_\p)\simeq \xi=2$ and the dependence of the
mean pion momentum is given by $\bra p_{\pi^\pm}\ket = m_\p p_\p / (2
m_{\pi^\pm}\xi)$.  The weak energy dependencies of the pion multiplicity and
the inelastic cross section can be absorbed in a semi-analytical
parametrisation of the cross section, $\sigma_\rmn{pp}^\pi(\alpha_\p)$
\citep[for details see][]{2004A&A...413...17P}.  The mean energy of the
produced secondary electrons ($\pi^\pm \to e^\pm + 3 \nu$) in the laboratory
frame is given by $\bra E_\e\ket= \frac{1}{4}\bra E_{\pi^\pm}\ket$.  Employing
the transformation law for distribution functions and using the mean value of
the electron momentum in the relativistic limit allows us to approximate the
electron source function by
\begin{eqnarray}
  \label{eq:se}
  s_\e(p)\,\dd p &=& s_{\pi^\pm}[p_\pi(p)]\,
  \frac{\dd p_\pi}{\dd p}\,\dd p =
  \frac{4 m_\e}{m_{\pi^\pm}} s_{\pi^\pm}
  \left(\frac{4 m_\e}{m_{\pi^\pm}}\,p\right)\,\dd p\\
\label{eq:se_sec}
 &=&\frac{4}{3}\, 16^{1-\alpha_\p} c\,\sigma_\rmn{pp}\,
n_\rmn{N}\, C_\p\, \left(\frac{m_\e}{m_\p}\right)^{1-\alpha_\p}
p^{-\alpha_\p}\, \dd p,
\end{eqnarray}
where the effective cross section $\sigma_\rmn{pp}$ depends in our model on
the spectral index of the CRp spectrum $\alpha_\p$ according to
\begin{equation}
\label{eq:sigmapp}
\sigma_\rmn{pp} \simeq 32 \cdot
\left(0.96 + \rmn{e}^{4.4 \,-\, 2.4\,\alpha_\p}\right)\mbox{ mbarn}. 
\end{equation}
Thus, we can write down the injection spectrum for CR electrons resulting
from hadronic reactions of CR protons with ambient gas protons,
\begin{eqnarray}
\label{eq:finj_sec}
f_\rmn{inj,pp}\, \dd p &=& 
C_\rmn{inj,pp} p^{-\alpha_\p}\, \dd p \\
C_\rmn{inj,pp} &=& 
\frac{4}{3}\, 16^{2-\alpha_\e} c\,\tau_\rmn{pp}\,\sigma_\rmn{pp}\,
n_\rmn{N}\, C_\p\, \left(\frac{m_\e}{m_\p}\right)^{2-\alpha_\e},
\end{eqnarray}
and $\tau_\rmn{pp} = \min[(c\, \sigma_\rmn{pp} n_\rmn{N})^{-1},
\tau_\rmn{Hubble}]$, and $\alpha_\e = \alpha_\p + 1$.

\subsubsection{Equilibrium spectrum of secondary electrons} 
\label{sec:fe_had,eq}

The same line of arguments presented in Sec.~\ref{sec:fe_eq} allows us to derive 
the equilibrium distribution of secondary CR electrons above a GeV due to IC
and synchrotron cooling,
\begin{eqnarray}
\label{eq:fe_hadr}
f_\e (p)\, \dd p &=& C_\e p^{-\alpha_\e} \,\dd p \\
\label{eq:C_e}
C_\e &=& \frac{16^{2-\alpha_\e}
  \sigma_\rmn{pp}\, n_\rmn{N} C_\p\,m_\e\,c^2}
     {(\alpha_\e - 2)\,\sigma_\rmn{T} \,(\eps_B + \eps_\rmn{ph})}
\left(\frac{m_\p}{m_\e}\right)^{\alpha_\e-2},
\end{eqnarray}
where the effective CR-proton cross section $\sigma_\rmn{pp}$ is given by
Eqn.~(\ref{eq:sigmapp}), and $n_\rmn{N} = n_\rmn{H} + 4 n_\rmn{He} = \rho /
m_\p$ is the target density of nucleons in a fluid of primordial element
composition. As discussed in Sect.~\ref{sec:timescales}, the equilibrium
spectrum of secondary electrons looks different depending on the ambient
electron density relative to the critical electron density (\ref{eq:ne_crit}).
At average ICM densities below $n_\rmn{e,crit}$, the equilibrium spectrum is
given by the IC/synchrotron cooled spectrum (\ref{eq:fe_hadr}) with the lower
cutoff
\begin{equation}
  \label{eq:qmin_sec}
  q_\rmn{min,sec} = \max(q_\rmn{Coul, IC/synch}, q_\rmn{threshold},
  q_\rmn{pp}),
\end{equation}
where $q_\rmn{pp}=q_\p m_\p /(16 m_\e)$ is the lower cutoff of the injected
electron population that is inherited from the lower proton cutoff $q_\p$,
while $q_\rmn{Coul, IC/synch}$ and $q_\rmn{threshold}$ are given by
Eqns.~(\ref{eq:q_Coulomb,IC}) and (\ref{eq:q_pp}), respectively.  

In the high ICM densities/ISM-regime above $n_\rmn{e,crit}$, the equilibrium
spectrum is given by the injection spectrum (\ref{eq:finj_sec}) at low
energies between $q_\rmn{min,sec}$ and
\begin{equation}
  \label{eq:qbreak_sec}
  q_\rmn{break,sec} = \frac{q_\rmn{pp,IC/synch}}{\alpha_\e-2}.
\end{equation}
Above $q_\rmn{break,sec}$, the equipartition spectrum steepens and and joins
continuously into the IC/synchrotron cooled spectrum (\ref{eq:fe_hadr}).


\section{Radiative processes}
\label{sec:rad_proc_formulae}

The non-thermal radio and hard X-ray emission is generated by CR electrons
with energies $E_\e > \mbox{GeV}$ (cf.~Eqns.~(\ref{eq:synchrotron}) and
(\ref{eq:ICphoton})).  For convenience, we rescale the cooled CR electron
equilibrium spectra of Eqns.~(\ref{eq:f_eq}) and (\ref{eq:fe_hadr}) to the
energy scale of a GeV,
\begin{eqnarray}
\label{eq:f_GeV}
f_\e(E_\e) \dd E_\e &=&
\frac{\tilde{C}_\e}{\mbox{GeV}}
\left( \frac{E_\e}{\mbox{GeV}}\right)^{-\alpha_\e} \dd E_\e,\\
\label{eq:Ctilde}
\tilde{C}_\e &=& 
C_\e \left(\frac{m_\e c^2}{\mbox{GeV}}\right)^{\alpha_\e-1}\!,  
\end{eqnarray}
and $C_\e$ is given by Eqn.~(\ref{eq:C_e_prim}) respectively (\ref{eq:C_e}),
depending on the electron population.

\subsection{Cluster magnetic fields}
\label{sec:magfield}

In principle, cosmological structure formation calculations with SPH are
capable of following magneto-hydrodynamics \citep{1999A&A...348..351D,
  2005JCAP...01..009D, 2004MNRAS.348..139P, 2005MNRAS.364..384P}, although
this is presently still fraught with numerical and physical difficulties.
Secondly, the origin of cluster magnetic fields is still an open question
\citep[][ and references therein]{2002RvMP...74..775W}. There are studies of
the Faraday rotation measure (RM) as a function of cluster impact parameter
using the position of a sample of radio lobes in different clusters
\citep{2001ApJ...547L.111C} which hints at a magnetic profile centred on the
cluster with $\umu$G field strengths. Field reversals along the line-of-sight
lead to cancellations in RM, since $RM \propto \int n_\e \vecbf{B} \cdot \dd
\vecbf{l}$.  The unknown behaviour of the characteristic length scale of the
magnetic field with cluster radius leaves us with some degree of freedom for
the magnetic profile that is unconstrained by current observations.  Assuming
primordial origin, and amplification of magnetic fields in the process of
structure formation would still require scanning the parameter space of the
field strengths in the initial conditions \citep{1999A&A...348..351D}.

Thus, we refrain from running self-consistent MHD simulations on top of the
radiative gas and CR physics and postpone a detailed analysis of the influence
of MHD on the radio emission to future work.  We chose the following simple
model for the magnetic energy density:
\begin{equation}
  \label{eq:magnetic_scaling_app}
  \eps_B = \eps_{B,0} \left(\frac{\eps_\rmn{th}}{\eps_\rmn{th,0}}\right)^{2
  \alpha_B},
\end{equation}
where $\eps_{B,0}$ and $\alpha_B$ are free parameters in our model.
Rather than applying a scaling with the gas density as non-radiative
MHD simulations by \citet{1999A&A...348..351D, 2001A&A...378..777D}
suggest, we chose the energy density of the thermal gas.  This
quantity is well behaved in the centres of clusters where current
cosmological radiative simulations, that do not include feedback from
AGN, have an over-cooling problem which results in an overproduction
of the amount of stars, enhanced central gas densities, too small
central temperatures, and too strong central entropy plateaus compared
to X-ray observations.  Theoretically, the growth of magnetic field
strength is determined through turbulent dynamo processes that will
saturate on a level which is determined by the strength of the
magnetic back-reaction \citep[e.g.,][]{2003PhRvL..90x5003S,
  2006PhPl...13e6501S} and is typically a fraction of the turbulent
energy density that itself should be related to the thermal energy
density, thus motivating our model theoretically.

\subsection{Synchrotron radiation}
\label{sec:synchro}

The synchrotron emissivity $j_\nu$ at frequency $\nu$ and per steradian of a
CR electron population described by Eqn.~(\ref{eq:f_GeV}), which is located in
an isotropic distribution of magnetic fields \citep[Eqn.~(6.36)
in][]{1979rpa..book.....R}, is obtained after averaging over an isotropic
distribution of electron pitch angles, yielding
\begin{eqnarray}
\label{eq:jnu}
j_\nu&=& A_{E_\rmn{synch}}(\alpha_\e)\, \tilde{C}_\e\,
\left[\frac{\eps_B}{\eps_{B_\rmn{c}}}\right]^{(\alpha_\nu+1)/2}
\propto \eps_\cre\, B^{\alpha_\nu+1} \nu^{-\alpha_\nu},
\\
\label{eq:Bc}
B_\rmn{c} &=& \sqrt{8\upi\, \eps_{B_\rmn{c}}} 
   = \frac{2\upi\, m_\e^3\,c^5\, \nu}{3\,e \mbox{ GeV}^2}
   \simeq 31 \left(\frac{\nu}{\mbox{GHz}} \right)~\umu\rmn{G},
\\
A_{E_\rmn{synch}} &=& \frac{\sqrt{3\upi}}{32 \upi}
   \frac{B_\rmn{c}\, e^3}{m_\e c^2}
   \frac{\alpha_\e + \frac{7}{3}}{\alpha_\e + 1}
   \frac{\Gamma\left( \frac{3\alpha_\e-1}{12}\right)
         \Gamma\left( \frac{3\alpha_\e+7}{12}\right)
         \Gamma\left( \frac{\alpha_\e+5}{4}\right)}
        {\Gamma\left( \frac{\alpha_\e+7}{4}\right)},
\end{eqnarray}
where $\Gamma(a)$ denotes the $\Gamma$-function \citep{1965hmfw.book.....A},
$\alpha_\nu = (\alpha_\e-1)/2 = \alpha_\inj/2$, $\tilde{C}_\e$ is given by
Eqn.~(\ref{eq:Ctilde}), and $B_\rmn{c}$ denotes a (frequency dependent)
characteristic magnetic field strength which implies a characteristic magnetic
energy density $\eps_{B_\rmn{c}}$.  Line-of-sight integration of the radio
emissivity $j_\nu $ yields the surface brightness of the radio emission
$S_\nu$.

For later convenience, we calculate the radio luminosity per unit frequency
interval of a steady state population of hadronically generated electrons
(\ref{eq:fe_hadr}),
\begin{eqnarray}
  L_\nu &=& 4\upi\int\dd V\, j_\nu  =
  A_\nu \int\dd V\,C_\p n_\rmn{N}\, \frac{\eps_B}{\eps_B+\eps_\rmn{ph}}
  \left(\frac{\eps_B}{\eps_{B_\rmn{c}}}\right)^{(\alpha_\nu-1)/2}\nonumber\\
  \label{eq:hadronic_radio}
  &\simeq& A_\nu \int\dd V\,C_\p n_\rmn{N}, 
  \mbox{ for }\eps_B\gg \eps_\rmn{ph}, \\
  A_\nu &=&  4\upi\, A_{E_\rmn{synch}}
  \frac{16^{2-\alpha_\e} \sigma_\rmn{pp}\, m_\e\,c^2}
   {(\alpha_\e - 2)\,\sigma_\rmn{T}\,\eps_{B_\rmn{c}}}
   \left(\frac{m_\p}{m_\e}\right)^{\alpha_\e-2}
   \left(\frac{m_\e c^2}{\rmn{GeV}}\right)^{\alpha_\e-1},
\end{eqnarray}
where we introduced the abbreviation $A_\nu$ with the dimensions $[A_\nu] =
\mbox{erg cm}^3 \mbox{ s}^{-1} \mbox{ Hz}^{-1}$ and the volume integral extends
over the entire cluster.  In the last step of Eqn.~(\ref{eq:hadronic_radio}),
we assumed typical radio spectral indices of cluster radio halos of
$\alpha_\nu\sim 1$ such that the radio luminosity of the equilibrium
distribution of CR electrons becomes independent of the magnetic field in the
synchrotron dominated regime for $\eps_B\gg \eps_\rmn{ph}$
(cf.~Fig.~\ref{fig:IC-synchro_B}).

\subsection{Inverse Compton radiation}

Inverse Compton (IC) scattering of cosmic microwave background (CMB) photons
off ultra-relativistic electrons with Lorentz factors of $\gamma_\e\sim 10^4$
redistributes these photons into the hard X-ray regime according to
Eqn.~(\ref{eq:ICphoton}).  The integrated IC source density $\lambda_\IC$ for
an isotropic power law distribution of CR electrons as described by
Eqn.~(\ref{eq:f_eq}) or (\ref{eq:fe_hadr}), can be obtained by integrating the
IC source function $s_\gamma(E_\gamma)$ in Eqn.~(43) of
\citet{2004A&A...413...17P} (in the case of Thomson scattering) over an energy
interval between observed photon energies $E_1$ and $E_2$ yielding
\begin{eqnarray}
\label{eq:IC}
\lambda_\IC(E_1, E_2) &=& \int_{E_1}^{E_2} \dd E_\IC\,s_\IC(E_\IC) \\
&=& \tilde{\lambda}_0\,
f_\IC (\alpha_e)\, 
\left(\frac{m_\e\, c^2}{\mbox{GeV}}\right)^{1 - \alpha_\e}
\left[\left(\frac{E_\IC}{k T_\mathrm{CMB}}\right)^{-\alpha_\nu}
  \right]_{E_2}^{E_1}, \\
f_\mathrm{IC} (\alpha_e) &=& \frac{2^{\alpha_\e+3}\, 
  (\alpha_\e^2 + 4\, \alpha_\e + 11)}
  {(\alpha_\e + 3)^2\, (\alpha_e + 5)\, (\alpha_e + 1)} \nonumber\\
&&\times\,\Gamma\left(\frac{\alpha_e + 5}{2}\right)\,
  \zeta\left(\frac{\alpha_e + 5}{2}\right)\,,\\
\mbox{and }\tilde{\lambda}_0 &=& 
\frac{16\, \pi^2\, r_\e^2\, \tilde{C}_\e\,
      \left(k T_\mathrm{CMB}\right)^3\,}{(\alpha_\e-1)\,h^3\, c^2}\,, 
\end{eqnarray}
where $\alpha_\nu = (\alpha_\e-1)/2$ denotes the spectral index, $r_\e = e^2
/(m_\e\, c^2)$ the classical electron radius, $\zeta(a)$ the Riemann
$\zeta$-function \citep{1965hmfw.book.....A}, and $\tilde{C}_\e$ is given by
Eqn.~(\ref{eq:Ctilde}). The IC photon number flux $\mathcal{F}_\gamma$ is
derived by means of volume integration over the emission region and correct
accounting for the growth of the area of the emission sphere on which the
photons are distributed:
\begin{equation}
  \label{eq:flux_IC}
  \mathcal{F}_\gamma (E_1, E_2) = \frac{ 1+z}{4\upi \,D^2} \int \dd V\,
  \lambda_\IC[(1+z) E_1, (1+z) E_2].
\end{equation}
Here $D$ denotes the luminosity distance and the additional factors of $1+z$
account for the cosmological redshift of the photons.

\subsection{$\bgamma$-ray emission from decaying pions}

Provided the CR population has a power-law spectrum, the integrated
$\gamma$-ray source density $\lambda_\gamma$ for pion decay induced
$\gamma$-rays can be obtained by integrating the $\gamma$-ray source
function $s_\gamma(E_\gamma)$ \citep[cf.][]{2007A&A...473...41E},
\begin{eqnarray}
\label{eq:lambda_gamma}
\lambda_\gamma &=& \lambda_\gamma(E_1, E_2)  = \int_{E_1}^{E_2} \dd E_\gamma\,
s_\gamma(E_\gamma) \\
\label{eq:lambda_gamma2}
&=& \frac{4\, C_\p}{3\, \alpha\delta_\gamma}
\frac{m_{\pi^0} c\, \sigma_{\rmn{pp}}n_\rmn{N}}{m_\p}
\left( \frac{m_\p}{2 m_{\pi^0}}\right)^{\alpha}\,
\left[\mathcal{B}_x\left(\frac{\alpha + 1}{2\,\delta_\gamma},
    \frac{\alpha - 1}{2\,\delta_\gamma}\right)\right]_{x_1}^{x_2},\\
x_i &=& \left[1+\left(\frac{m_{\pi^0}c^2}{2\,E_i}
      \right)^{2\,\delta_\gamma}\right]^{-1} \mbox{~for~~}
      i \in \{1,2\},
\end{eqnarray}
where we used the abbreviation $\alpha = \alpha_\gamma$. $C_\p$ is the
normalisation of the proton distribution function which we follow dynamically
in our simulations (cf.~Eqn.~(\ref{eq:fp})), and the rest mass of a neutral
pion is $m_{\pi^0} c^2 \simeq 135 \,\mbox{MeV}$.  The shape parameter
$\delta_\gamma$ depends on the spectral index of the $\gamma$-ray spectrum
$\alpha$ according to
\begin{equation}
\label{eq:delta}
\delta_\gamma \simeq 0.14 \,\alpha_\gamma^{-1.6} + 0.44.
\end{equation}
There is a detailed discussion in \citet{2004A&A...413...17P} how the
$\gamma$-ray spectral index $\alpha_\gamma$ relates to the spectral index of
the parent CR population $\alpha_\p$.  In Dermer's model, the pion multiplicity
is independent of energy yielding the relation $\alpha_\gamma = \alpha$
\citep{1986ApJ...307...47D, 1986A&A...157..223D}.  The formalism underlying
Eqns.~(\ref{eq:lambda_gamma}) and (\ref{eq:lambda_gamma2}) includes the
detailed physical processes at the threshold of pion production like the
velocity distribution of CRs, momentum dependent inelastic CR-proton cross
section, and kaon decay channels.  The $\gamma$-ray luminosity is defined by
\begin{eqnarray}
  \mathcal{L}_\gamma &=& \int\dd V\, \lambda_\gamma  =
  A_\gamma \int\dd V\,C_\p n_\rmn{N},
  \label{eq:hadronic_gamma}
\end{eqnarray}
where we introduced the constant $A_\gamma$ with the dimensions $[A_\gamma] =
\gamma \mbox{ cm}^3 \mbox{ s}^{-1}$ that is given by $A_\gamma =
\lambda_\gamma / (C_\p n_\rmn{N})=\mbox{const}$, according to
Eqn.~(\ref{eq:lambda_gamma2}). $\mathcal{F}_\gamma$ is derived by
Eqn.~(\ref{eq:flux_IC}) substituting $\lambda_\rmn{IC}$ by $\lambda_\gamma$.

\subsection{SPH projections and Hubble scaling}

We produced projected maps of the density, Mach number of shocks, relative CR
pressure of protons and electrons, and non-thermal cluster observables in the
radio, hard X-ray, and $\gamma$-ray regime.  Generally, a three-dimensional
scalar field $a(\vecbf{r})$ along any ray was calculated by distributing the
product of $a(\vecbf{r})$ and the specific volume $M_\alpha / \rho_\alpha$ of
the gas particles over a grid comoving with the cosmic expansion. This yields
the projected quantity $A (\vecbf{r}_{\bot})$:
\begin{equation}
  \label{eq:projection}
  A (\vecbf{r}_{\bot,\, ij}) = \frac{1}{L_\rmn{pix}^2}
  \sum_\alpha a_\alpha \frac{M_\alpha}{\rho_\alpha}\,
  W_{\alpha,\,ij}(\vecbf{r}_{\bot,\, ij} - \vecbf{r}_\alpha),
\end{equation}
where $W_{\alpha,\,ij}$ is the value of the projected smoothing kernel
(normalised to unity for the pixels covered) of an SPH particle $\alpha$ at
comoving grid position $\vecbf{r}_{\bot,\, ij}$, and $L_\rmn{pix}^2$ is the
comoving area of the pixel. In order to obtain a line-of-sight average of some
mass-weighted quantity, say temperature, we project the quantity $a_\alpha =
T_\alpha\, \rho_\alpha$ divided each pixel by the mass projection (e.g. setting
$a_\alpha = \rho_\alpha$).

Combining primary and secondary non-thermal emissivities requires the knowledge
of the scaling with the Hubble constant. It turns out that the primary
synchrotron/IC emissivities scale as $j_{\nu/IC,\rmn{prim}} \propto h^3$
leading to a scaling of the surface brightness of $S_{\nu/IC,\rmn{prim}}
\propto h^2$. In contrast, the secondary synchrotron/IC/$\gamma$-ray
emissivities scale as $j_{\nu/IC,\rmn{sec}} \propto h^4$ which results in a
scaling of the surface brightness of $S_{\nu/IC,\rmn{prim}} \propto h^3$. The
different scaling of the primary and secondary non-thermal emission components
with the Hubble constant is the reason why we choose to show all non-thermal
luminosities in units of the currently favoured Hubble constant, $h_{70}$,
where $H_0 = 70\, h_{70}\mbox{ km s}^{-1} \mbox{ Mpc}^{-1}$.

\bsp

\label{lastpage}

\end{document}